\documentclass{aa}  

\usepackage{color, colortbl}
\usepackage{graphicx}
\usepackage{tikz}
\usepackage{float}
\usepackage{subcaption}
\usepackage{multicol}
\usepackage{longtable}
\usepackage{placeins}
\usepackage{hyperref}
\usepackage{txfonts}
\begin{document}

   \title{A dynamic view of V Hydrae}

   \subtitle{Monitoring of a spectroscopic-binary AGB star with an alkaline jet\thanks{Based on observations made with the Mercator Telescope, operated on the island of La Palma by the Flemish Community, at the Spanish Observatorio del Roque de los Muchachos of the Instituto de Astrof\'isica de Canarias. Based on observations obtained with the HERMES spectrograph, which is supported by the Research Foundation - Flanders (FWO), Belgium, the Research Council of KU Leuven, Belgium, the Fonds National de la Recherche Scientifique (F.R.S.-FNRS), Belgium, the Royal Observatory of Belgium, the Observatoire de Genève, Switzerland and the Thüringer Landessternwarte Tautenburg, Germany. }}

\author{L. Planquart
          \inst{1}\fnmsep\thanks{Research fellow, FNRS, Belgium.}
          \and
          A. Jorissen \inst{1}
          \and A. Escorza \inst{2, 3, 4}
        \and
        O. Verhamme \inst{5}   
        \and
        H. Van Winckel \inst{5}
          }

   \institute{Institut d’Astronomie et d’Astrophysique, Université Libre de Bruxelles, CP 226, Boulevard du Triomphe, 1050 Brussels,
Belgium\\
              \email{lea.planquart@ulb.be}
            \and
            European Southern Observatory, Alonso de Córdova, 3107 Vitacura, Santiago, Chile
         \and
        Instituto de Astrofísica de Canarias, C. Vía Láctea, 38205 La Laguna, Tenerife, Spain
\and
Universidad de La Laguna, Departamento de Astrofísica, Av. Astrofísico Francisco Sánchez, 38206 La Laguna, Tenerife, Spain
         \and
             Instituut voor Sterrenkunde (IvS), KU Leuven, Celestijnenlaan 200D, B-3001 Leuven, Belgium \\
             }

   \date{Received 12 September 2023 / Accepted 30 October 2023}

 
  \abstract
   {The well studied carbon star V Hydrae is known to exhibit a complex asymmetric environment made of a dense equatorial wind and high-velocity outflows, hinting at its transition from the AGB phase to the asymmetric planetary nebula phase. In addition, V Hydrae also exhibits a long secondary period of 17 years in its light curve, suggesting the presence of a binary companion that could shape the circumstellar environment. } 
   {In this paper, we aim to confirm the binary nature of V Hydrae by deriving its orbital parameters and investigating the effect of the orbital motion on the circumbinary environment. }
   {In a first step, we used a radial-velocity monitoring performed with the HERMES spectrograph to disentangle the pulsation signal of the AGB from its orbital motion and to obtain the spectroscopic orbit. We combined the spectroscopic results with astrometric information to get the complete set of orbital parameters, including the system inclination. Next, we reported  the time variations of the sodium and potassium resonance doublets. Finally, following the methods used for post-AGB stars, we carried out spatio-kinematic modelling of a conical jet to reproduce the observed spectral-line modulation. }
   {We found the orbital solution of V Hydrae for a period of 17 years. We correlated the companion passage across the line of sight with the obscuration event and the blue-shifted absorption of alkaline resonant lines. Those variations were modelled by a conical jet emitted from the companion, whose opening angle is wide and whose sky-projected orientation is found to be consistent with the axis of the large-scale bipolar outflow previously detected in the radio-emission lines of CO.}
   {We show that the periodic variation seen for V Hydrae is likely to be due to orbital motion. The presence of a conical jet offers a coherent model to explain the various features of V Hydrae environment.}

   \keywords{stars: AGB and post-AGB -- Stars: variables: general
 -- technique: spectroscopic -- stars: individual: V Hya -- binaries: spectroscopic -- ISM: jets and outflows}

   \maketitle
%

\section{Introduction}
\label{sect:introduction}

A star with an initial mass ranging from 0.8 to 8 $\rm M_{\odot}$ will eventually evolve through the asymptotic giant branch (AGB) stage. During this phase, the star undergoes a process of significant mass loss ($10^{-7}$ to $10^{-4}$ $\rm M_{\odot}$/yr)  through a slow (5 to 20 $\rm km\,s^{-1}$) wind that enriches the interstellar medium \citep{Habing_AGB_2004agbs.book....1H}. 
When leaving the AGB phase, its effective temperature increases and goes through a short-lived transition of post-AGB and pre-planetary nebula (lasting about a few thousand years) phase. The majority of planetary nebulae (PN) show a variety of aspherical (bi- or multi-polar) structures \citep{Planetary_nebulae_2002ARA&A..40..439B}. Those bipolar structures are expected to arise from binary interaction with a close-by companion \citep{PN_2017NatAs...1E.117J}. In the preceding phase, the post-AGB phase, binarity has been widely detected and studied (\citealt{Oomen_2018A&A...620A..85O} and references therein). Common building blocks of post-AGB binaries have been identified (e.g. \citealt{Kluska_2022A&A...658A..36K}), consisting of a central binary surrounded by a stable circumbinary disk of gas and dust. They often exhibit a bipolar outflow which is likely launched by an accretion disc around the companion (e.g. \citealt{Dylan_2022A&A...666A..40B} and references therein). Thanks to the near-infrared (NIR) interferometry, these building blocks can also be spatially resolved (e.g. \citealt{Kluska_2019A&A...631A.108K}; \citealt{Corporaal_2023A&A...671A..15C}). While these post-AGB binaries represent about a third of all known Galactic post-AGB stars \citep{Kluska_2022A&A...658A..36K}, they are poorly connected to the AGB progenitor sample, as evidence for binaries among these progenitors is lacking. 
Due to the intrinsic properties of AGB stars, detecting binaries is not a straightforward task. Given their photosphere is bright and dynamic, the detection of orbital motion is possible only if it exceeds the amplitude of the intrinsic pulsation of the stellar envelope \citep{Chapitre9}.

The late-type carbon star V Hydrae (V Hya) is among the best candidates to study the impact of a companion on the circumstellar environment of AGB stars. 
Secular photometric monitoring indicates that the star exhibits two photometric periods: a long period of 6160~$\pm$~400 d with an amplitude of 3 mag and a short period of 530 $\pm$ 30 d with an amplitude of 1.5~mag \citep{Knapp_LC_1999}. 
While the 530 d period is commonly associated with the Mira pulsation of the AGB star, the origin of the 17~y period is still under debate. Several scenarios have been suggested, many of which invoke the presence of a binary companion: eclipse by a dusty cloud surrounding a companion \citep{Knapp_LC_1999}, extinction by the plasma ejection of the companion \citep{Sahai1}, a triple system with an inner eccentric orbit of 8 y \citep{Salas}, along with possible episodic dust formation \citep{Lloyd_Evans}. 
In addition, the presence of a binary companion has been inferred on the basis of the near-UV (NUV) excesses \citep{GALEX}, detections of blue emission lines in the spectrum (\citealt{LloydEvansJet}, \citealt{Lloyd_Evans_VHyaB_2009AIPC.1094..963L}), and the exceptional structure of the large-scale environment of the star, as we describe below.

Several radio observations in the CO bands highlight the presence of an environment shaped by different velocity regimes: a slow (15 $\rm km\,s^{-1}$) dense equatorial wind, a medium-velocity (45 $\rm km\,s^{-1}$) spherical wind, and a fast (150-200 $\rm km\,s^{-1}$) bipolar outflow \citep{Knapp_CO_1997}. \cite{KABM} mapped the fast wind component in the CO J=2-1 transition as a bipolar flow whose blue-shifted lobe is pointing toward the east with an inclination axis of about 30°.   
The presence of an eastward high-velocity knotty jet was also detected in the [\ion{S}{II}] emission by HST/STIS \citep{Sahai1} and its possible radio-equivalent by recent high-resolution ALMA observations together with the discovery of several concentric rings inside the dense equatorial structure \citep{Sahai_ALMA_2022ApJ...929...59S}.  
All the reported observations confirm the large-scale complexity of the system, although no model (so far) has been able to assemble all the pieces of the puzzle and link the large-scale bipolar structure with the expected binary  motion.

In this paper, we bring on new observations from the high-resolution HERMES spectrograph \citep{HERMES}. The long-term spectral monitoring of \object{V Hya} provides unprecedented information on the system dynamics and new constraints on its binary motion. We report radial-velocity and spectral-line variations in phase with the long-term photometric period and interpret them in the framework of a binary system with a period of 17 y.
The results are twofold: they provide the first spectroscopic evidence of a binary-AGB system and highlight the complex phenomena of the short-lived transition phase of V~Hya,  making it a case study for further investigation.

This paper is structured as follows. The observations and data reduction are presented in Sect. 2. The determination of the orbital parameters is described in Sect. 3. In Sect. 4, we compare the spectroscopic signal with the visual photometry. In Sect. 5, we analyze the temporal variation of selected spectral lines. The spatio-kinematic modeling of the \ion{Na}{I} resonance doublet is introduced in Sect. 6. 
In Sect. 7, we discuss the possible origin of the long obscuration event and put our new spectroscopic results in the context of previous observations. In Sect. 8, we outline our main findings.

\section{Observations}
\label{sect:observation}
\subsection{Spectroscopic data}
\label{sect:spectroscopic_data}
V Hya has been monitored with the HERMES spectrograph mounted on the 1.2m Mercator telescope at La Palma \citep{HERMES}. A total of 130 spectra were taken on different dates between 2011 and 2023, with an average exposure time of 900 seconds. The high-resolution mode was used, providing a resolution of R=86,000 over a wavelength range from 3800 to 9000 \AA. The dedicated python-based pipeline was used for the data reduction. The radial velocity (RV) was obtained by cross-correlating the object spectrum with a carbon-star template and fitting a Gaussian function to the mean line profile, as represented by the resulting cross-correlation function (CCF). The mask used to compute the CCF is built from a HERMES spectrum of the R-type carbon star BD +02$^\circ$4338, Doppler-corrected to zero velocity from a correlation with the master Arcturus template. All lines with a depth getting below the relative level of 0.8 with respect to the local continuum were kept in the mask.
This mask, covering the wavelength range 476.9 — 654.7 nm,  was selected because it provides the cleanest CCF for V~Hya (deepest and least asymmetric).
Thirty spectra were excluded from the orbital analysis because they bear the signature of shocks. Due to the presence of shocks linked to the stellar pulsation, the CCF profile at those epochs close to maximum light was distorted (asymmetric broadening or double-peaked; \citealt{Alvarez_2001A&A...379..305A}). Table \ref{tab:RV_table} summarises the log of the observations, including the observing dates of each spectra, the computed RVs and their formal uncertainties (due to the quality of the CCF fit). 
An uncertainty of 0.07~$\rm km\,s^{-1}$ expressing the long-term stability of the spectrograph is added quadratically to the formal uncertainty of the CCF fit.
This long-term accuracy is estimated from the average RV standard dispersion of
a sample of RV standard stars monitored along with the science targets
(see Fig. 10 in \citealt{HERMES}).

\subsection{Photometric data}
\label{sect:photometric_data}
For more than a century, V~Hya has been monitored by variable-star observers. We use the archival visual photometric data from the American Association of Variable Stars Observers, AAVSO \citep{AAVSO_ref}. The overall dataset (9000 measurements) covers a time interval from 1913 to 2023, allowing us to monitor six 17-yr cycles. 

\subsection{Astrometric data}
\label{sect:astrometric_data}

Astrometric measurements from both Hipparcos \citep{Hipparcos1997A&A...323L..49P} and Gaia \citep{Gaia2022A&A...667A.148G} ESA missions have been used to constrain additional parameters that are not accessible from spectroscopic data alone, such as the orbital inclination.
The difference between the instantaneous proper motion computed by the two missions and the average proper motion between the two reference epochs provides an estimate of the acceleration in the sky plane. That acceleration being non-null for V~Hya \citep{acceleration_2021ApJS..254...42B} is a signature of a long-trend motion and gives additional constraints on the orbital solution.

\section{Radial-velocity analysis} 
\label{sec:RV-analysis}
 
\subsection{Methods}  \label{RV-method}
In this section we describe the method used to retrieve the orbit from the RV curve. In a first step (hereafter, step 1), the contribution associated to the pulsation was fitted together with a Keplerian orbit. During a second step (hereafter, step 2), the cleaned (pulsation removed) RV curve was combined with the described astrometric data and fitted with the \textsc{orvara}\footnote{\url{https://github.com/t-brandt/orvara}} code based on a Markov chain Monte Carlo (MCMC) method \citep{orvara_2021AJ....162..186B}.

\subsubsection{Disentangling the pulsation (step 1)} 
\begin{figure}[t]
    \includegraphics[width = 0.5\textwidth]{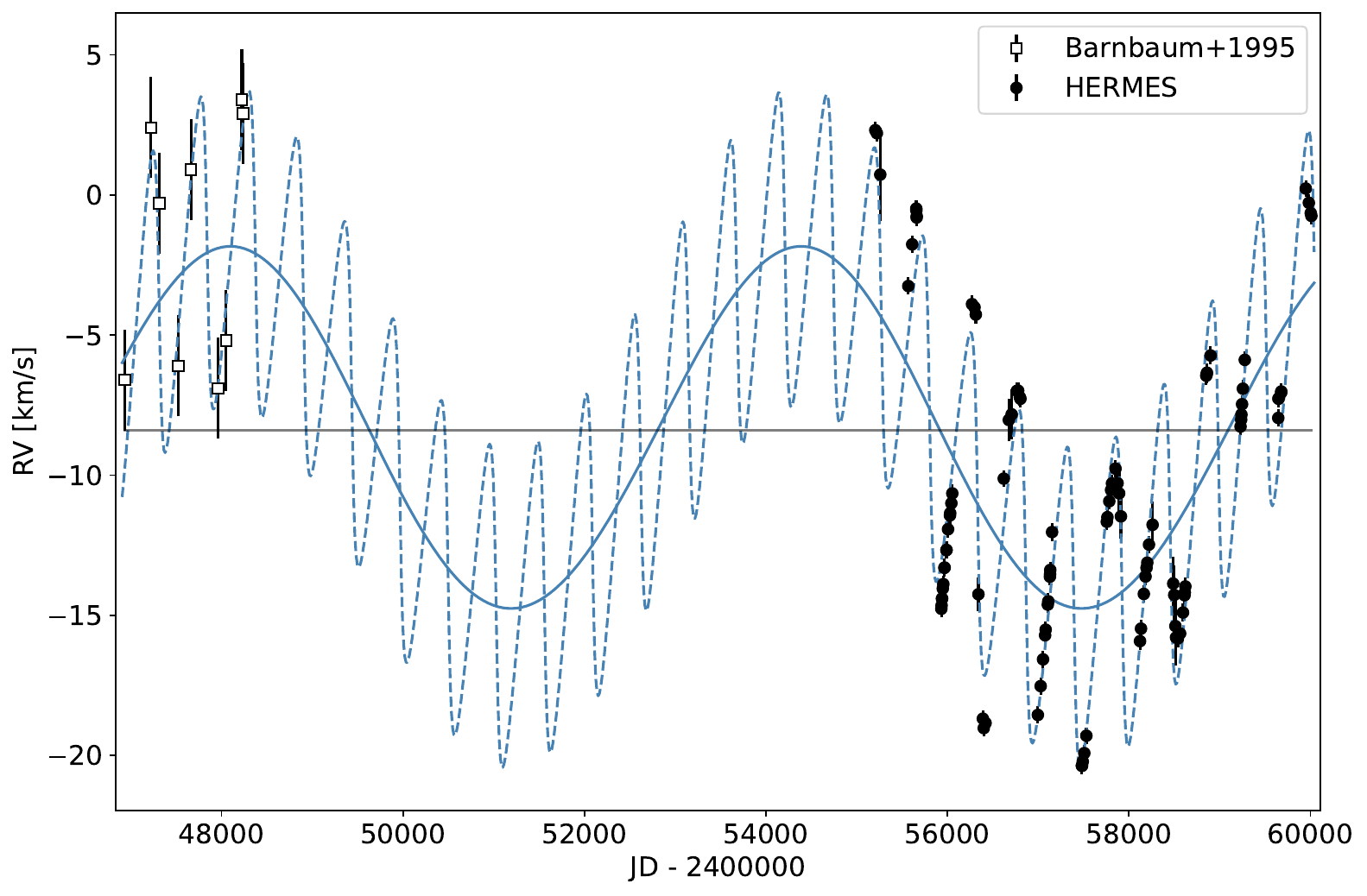}
    \includegraphics[width = 0.5\textwidth]{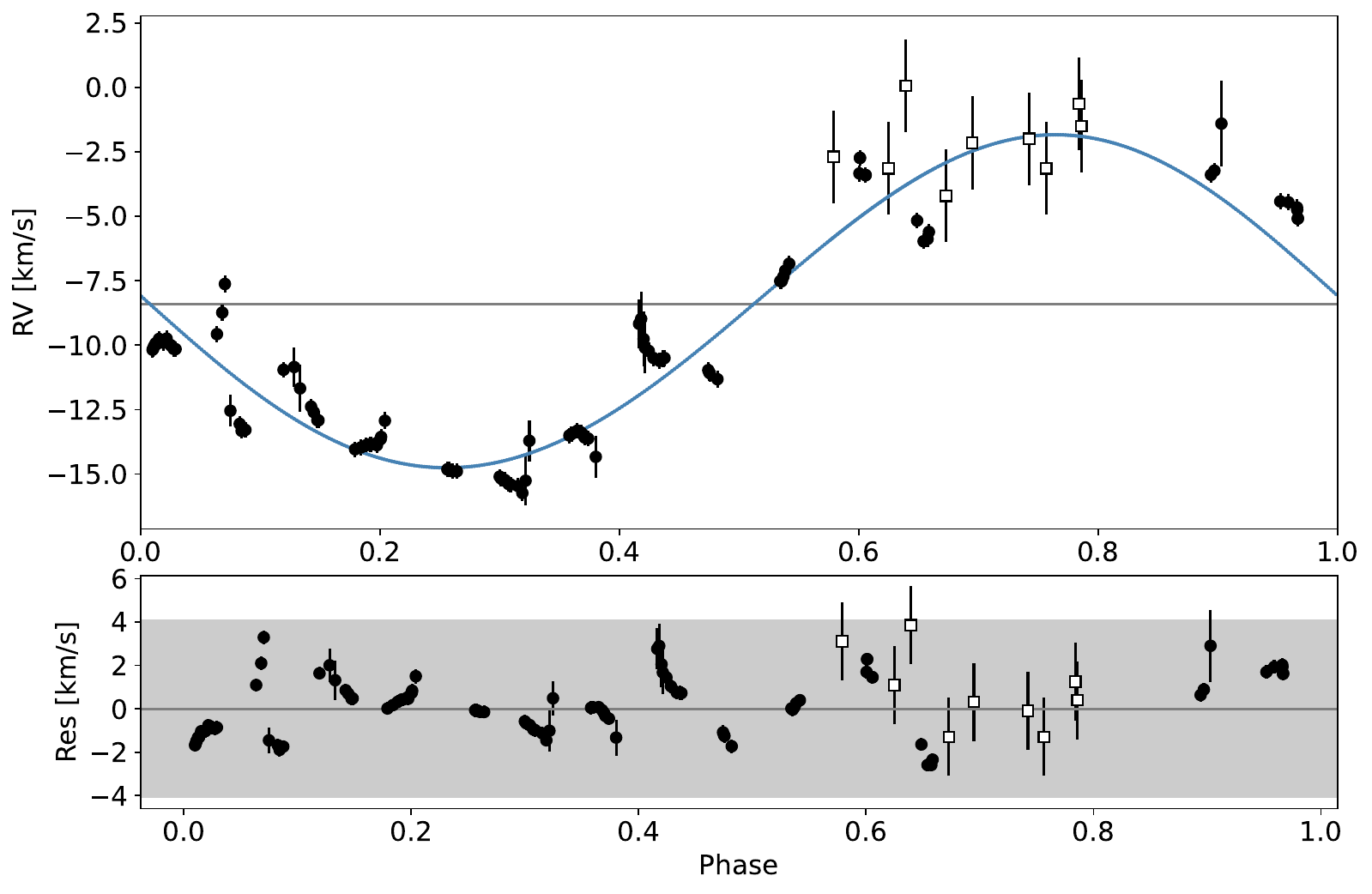}
    \caption{Radial velocities of V Hya. \textbf{Top}: RV curve as a function of time. The full line is the Keplerian orbit and the dashed line is the composite signal (orbit and pulsation). \textbf{Middle}: Phase-folded cleaned orbit of V~Hya. \textbf{Bottom} Residual values with the $\pm3\sigma$ from the mean is represented as the grey-shaded region.  }
    \label{fig:Fit_orbite}
\end{figure}

The RV curve exhibits two periodic variations: a short-period signal corresponding to the Mira pulsation (following the 530 d periodicity) as well as a long-term trend (see Fig. \ref{fig:Fit_orbite}). The long trend can be attributed to the 17~y modulation observed in the light curve \citep{Knapp_LC_1999} but, as this period is longer than the time spanned by the HERMES monitoring, an entire cycle is not visible. 
As mentioned in Sect. \ref{sect:spectroscopic_data}, the errors in the RV from HERMES consist of two terms: the error intrinsic to the CCF fit (photon noise and template mismatch) and a 0.07~$\rm km\,s^{-1}$ uncertainty to account for a possible offset of the wavelength zero-point. 
To better constrain the fit, nine additional RV values obtained by \cite{Barnbaum} have been added to the RV curve obtained with the HERMES spectrograph. Fortunately, those data, falling on the previous orbital cycle, correspond to different orbital phases as compared to the HERMES data, allowing us to cover almost the whole orbit. To account for an eventual offset of the RV zero point reference between the two instruments, an additional 0.2~$\rm km\,s^{-1}$ uncertainty is added to the RV data from HERMES and represents the minimum value required to increase the relative statistical weight of the oldest RV data from \cite{Barnbaum} to a significant level, given their higher uncertainties ($\pm$~1.5~$\rm km\,s^{-1}$).

The RV curve was fitted by a combination of two functions: a Keplerian orbit and an asymmetric sine, $\sin_a$. The asymmetric sine is expected to mimic the S-curve observed in pulsating stars \citep{Chapitre9} and is mathematically defined as:

\begin{equation}
    A \sin_a(\omega_0 (t+ t_0) ) = \frac{A}{\Gamma} \arctan \Bigg( \frac{\Gamma  \sin (\omega_0 (t+ t_0))}{1-\Gamma  \cos (\omega_0 (t+ t_0))}\Bigg),
    \label{eq:sin_asym}
\end{equation}
where $\omega_0 = 2\pi/P$ is the angular frequency, $A$ is the signal amplitude, $t_0$ the time reference and $\Gamma$ the tilt parameter satisfying  $|\Gamma| \leq 1$. $\Gamma = 0$ corresponds
to the usual sine function of period, $P,$ whereas $\Gamma = +1$ (resp. $- 1$) corresponds to the saw-tooth function tilted to the left (right, resp.). 

In the fitting process, the asymmetric sine period was fixed at  530 d and the Keplerian period was forced to remain in the range 6310$\pm$1220 days. Those values were obtained by computing the Lomb-Scargle periodogram of the light curve and taking the 3$\sigma$ uncertainty interval for the long-period signal (see. Fig. \ref{fig:Lombscragle}). In total, the fitting routine is composed of a nine-parameter function (3 for the asymmetric sine and 6 to describe the orbital motion). 

After a first fitting, the contribution of the pulsation is subtracted from the RV and the cleaned signal is fitted with a Keplerian orbit alone using a $\chi^2$-minimisation. The method allows us to get precise orbital parameters at the price of imposing the period. 
The standard deviations of the orbital parameters are obtained by computing the square root of the diagonal elements of the covariance matrix.

\subsubsection{Orbital parameters with \textsc{orvara} (step 2)}
\label{sec:Orbital_parameters_with_ORVARA_(step 2)}

As a complementary analysis, meant to confirm our results and to obtain the full orbital solution (including the inclination), another independent method was used on the RV curve cleaned from the pulsation. The software package \textsc{orvara}, developed by \cite{orvara_2021AJ....162..186B}, uses a parallel tempering MCMC method to simultaneously fit radial-velocity and astrometric measurements. Combining the measurements allows the companion mass and inclination for binary systems to be computed \citep{2023A&A...671A..97E}, which would otherwise be inaccessible with just the spectroscopic measurements. 
The code first fits the RV data then adds the absolute astrometric measurements: astrometric position ($\alpha$, $\delta$), proper motion ($\mu_{\delta}$, $\mu_{\alpha}$), and Gaia DR3 parallax for each MCMC step. It allows us to simultaneously fit the full orbital parameters (period, $P$; eccentricity, $e$; semi-major axis, $a$; inclination, $i$; argument of the periastron of the visible star, $\omega$; time of periastron passage ,$ T_0$; angle of the ascending node $\Omega$); the mass of the two stars; and an additional radial-velocity jitter, $\sigma_{vr}$. 
Details on the numerical implementation can be found in the reference paper \citep{orvara_2021AJ....162..186B}.

For V Hya, the primary mass was set as a Gaussian prior of 1.9 $\pm$1 $\rm M_{\odot}$ and the Gaia DR3 parallax was used as prior, corresponding to a distance of 434$\pm$21~pc.
As the parallax uncertainties for AGB stars is often underestimated, different prior values were tested, corresponding to the recomputed distances from \cite{Gaia_AGB_priors_2022A&A...667A..74A}: $484^{+107}_{-177}$~pc and $529^{+248}_{-131}$~pc. However, no significant difference was found in the final orbital parameters and, therefore, the Gaia DR3 parallax has been adopted in the remainder of the paper. 
We used 15 MCMC~temperatures and for each temperature, with 100 walkers run with 5~$\times10^6$ steps per walker. After a first try, the MCMC was oscillating between two orbital solutions: one with a high eccentricity for a period of 35 years and a mass ratio above 4 and another with a low eccentricity, a period of 17 years, and a smaller mass ratio. Such a high mass ratio would imply a companion that is more massive than V~Hya initial mass, which is inconsistent with its evolutionary stage. 
We therefore re-started the MCMC with an additional prior on the eccentricity ($e<0.15$) to favour the orbital solution with the shorter period, compatible with the observed modulation in the light curve.

\subsection{Results}
\label{sect:results}
The orbital parameters found with the two methods (step 1 and step 2) are listed in Table \ref{tab:orbital_parameters}.  
Despite their different implementation and assumptions, both methods produce a low-eccentricity orbit (compatible with a circular nature) with a period consistent with the light-curve modulation. 
The \textsc{orvara} solution has larger uncertainties, but it is expected to be more robust and does not require to impose the period. Below, we discuss the values obtained and their sensitivity regarding the astrometric measurement. 
\subsubsection{Spectroscopic fit (step 1)}
The pulsation was fitted with an asymmetric sine amplitude of 5.18~$\rm km\,s^{-1}$. Integrating the curve over half a period to obtain the corresponding radial distance covered leads to a value of $\Delta R$ = 0.25~au. For a stellar radius of 2.5 au (see Sect. \ref{section:best_fit_model}), this variation in radial distance corresponds to a variation of $\pm$ 10\% of $R_{*}$ over a pulsation cycle. 
The mass function, $f(m),$ obtained is equal to 0.18 $\rm M_\odot$. This value gives partial indication of the companion mass with RV data only depending on the inclination angle and the mass of the primary. 
The inclination angle cannot be obtained with the spectroscopic signal alone. It requires additional information from astrometric measurements and an estimation of the primary mass.

\begin{table}

    \caption{Orbital parameters from V~Hya for the two methods described in Sect. \ref{RV-method}. Left panel: parameters from step 1. Right panel: parameters from step 2, parameters with asterisk (*) are discussed in Sect.\ref{sect:sensitivity_hipparcos}. DoF stands for 'degrees of freedom'.}
    \label{tab:orbital_parameters}
\begin{tabular}{|lr|lr|}
\hline
  \multicolumn{2}{|c}{RV data only} & \multicolumn{2}{|c|}{RV and astrometry data}  \\
  \hline
  Parameter & Values & Parameter & Value  \\
\hline
$P$ [yr] &17.22$\pm$ 0.92 &$P$ [yr]      &   ${17.45}_{-0.29}^{+0.34}$ \\
$e$ &0.020 $\pm$ 0.020 & $e$      &   ${0.024}_{-0.017}^{+0.027}$\\
$T_0$ [JD] & 2455070$\pm$1665 &  $T_0$ [JD]      &   ${2458684}_{-2582}^{+2128}$\\
$\omega$ [°] & 40.62$\pm$98.66 & $\omega$ [°]    &     ${343}_{-122}^{+147}$\\

$ a_1 \sin{i}$ [au] & 3.73 $\pm$ 0.01&     $a$ [au]    &${11.2}_{-1.5}^{+1.2}$\\
$f(m)$ [M$_\odot$] &  0.176 $\pm$ 0.003& $i$ [°]*    &      ${37.7}_{-2.0}^{+2.2}$\\
$K$ [$\rm km\,s^{-1}$] & 6.46 $\pm$0.43  &$M_2$/$M_1$&  ${1.36}_{-0.29}^{+0.68}$\\
$\gamma$ [$\rm km\,s^{-1}$] & -8.40 $\pm$ 0.45 & $\Omega$ [°]*   &        ${159.7}_{-3.3}^{+43.0}$\\
 $A$ [$\rm km\,s^{-1}$] & 5.18 $\pm$ 0.18 & $M_2$ [$M_\odot$]   &      ${2.63}_{-0.69}^{+0.63}$\\
 $\Gamma$ & -0.68 $\pm$ 0.08 &$\sigma_{jit}$ [$\rm km\,s^{-1}$]& ${1.09}_{-0.8}^{+0.9}$\\
$t_0$ [JD] & 2449962 $\pm$ 4& &\\
$\chi_{red}^2$ &  13.65& $\chi_{red}^2$ & 14.0\\
DoF & 9&DoF & 9\\

\hline
\end{tabular}
\end{table}

\subsubsection{Astrometric-spectrometric orbit (step 2)}
\begin{figure*}[t]
    \includegraphics[clip, trim=0cm 0cm 2.5cm 0cm ,height = 6cm]{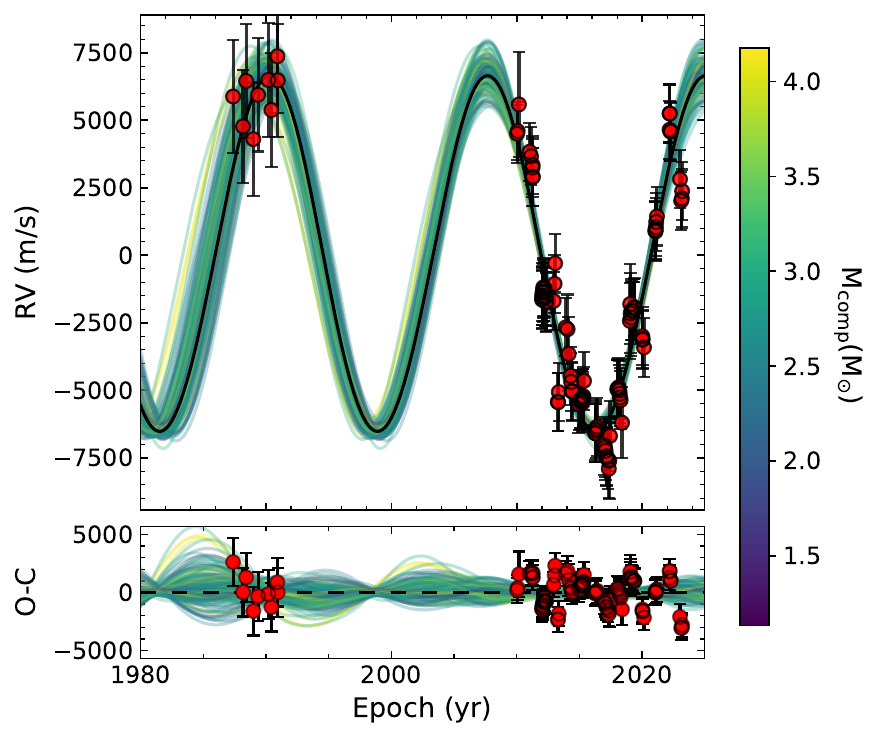}
    \includegraphics[height = 6cm]{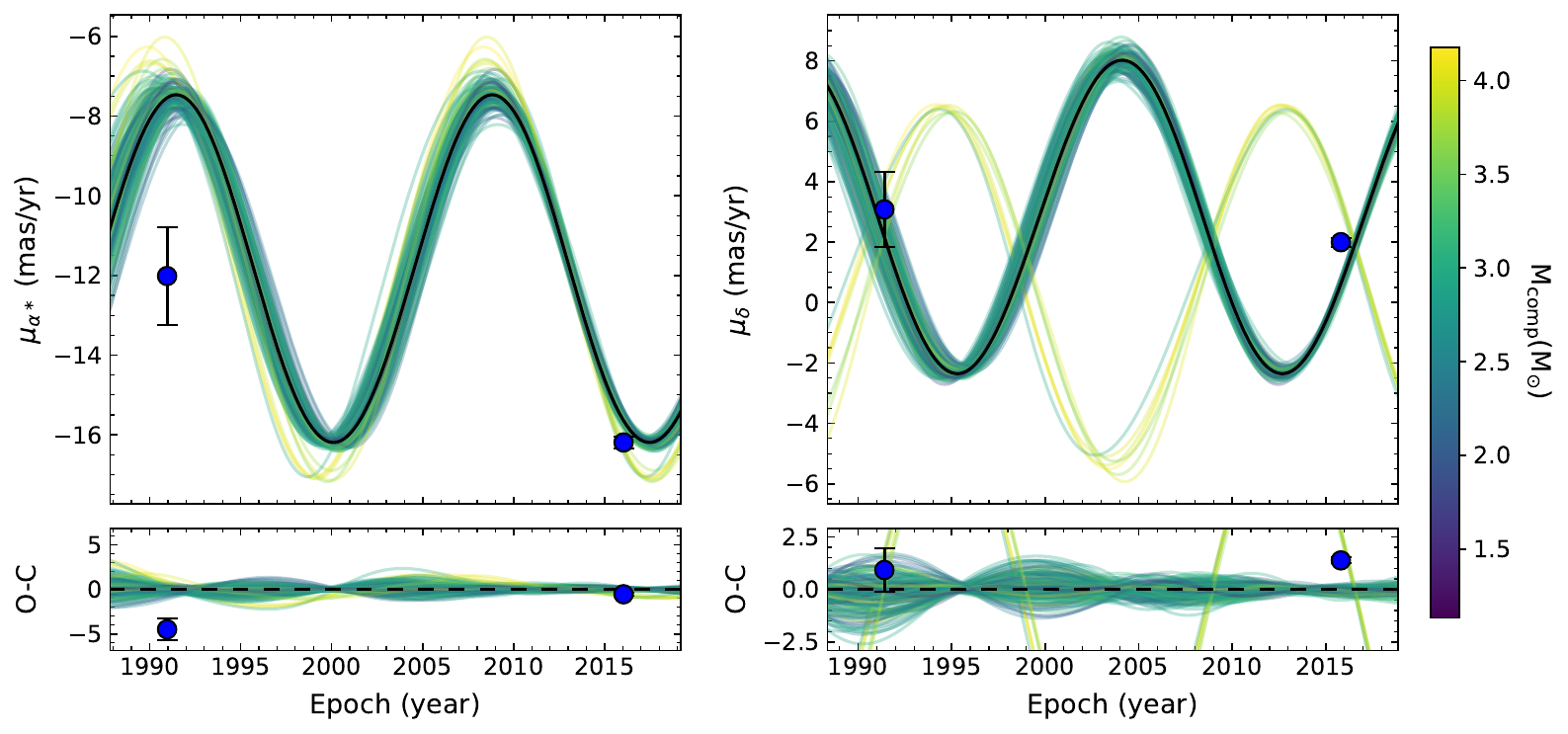}
    \caption{\textsc{orvara} results: RV curve (left), proper motions in right ascension (centre) and declination (right). In all plots, the black thick line represents the best-fitting orbit, while 40 other well-fitting orbits are included and colour-coded as a function of the companion mass.}
    \label{fig:orvara_fit}
\end{figure*}

The right panel of Table \ref{tab:orbital_parameters} gives the orbital parameters obtained when simultaneously fitting the RV curve and astrometric data. The $\chi^2_{red}$ is the overall reduced $\chi^2$, taken as the sum of the $\chi^2$ for the Hipparcos proper motion ($\chi^2_{H}$), the Gaia proper motion  ($\chi^2_{G}$), and the long-term Hipparcos-Gaia proper motion ($\chi^2_{HG}$). The different $\chi^2$ can be found in the central column of Table \ref{tab:chi2_orvara}. Figure \ref{fig:orvara_fit} displays the best fit for the RV curve and the proper motions from Hipparcos and Gaia. The corner plot of the MCMC chain is displayed in Appendix B (Fig. \ref{fig:orvara_corner}). The best-fitting value for the inclination is $37.7 \pm 2.2$°. This lies within the range of previous estimations: about 30° for \cite{Knapp_CO_1997}, and 45° for \cite{Sahai_ALMA_2022ApJ...929...59S}. 

\subsection{Sensitivity with respect to the Hipparcos data}
\label{sect:sensitivity_hipparcos}
In Fig. \ref{fig:orvara_fit}, the data point corresponding to the proper motion in right ascension at Hipparcos epoch, $\mu^H_{\alpha*}$, appears as an outlier in the best-fitted model. The instantaneous values of the proper motion at Hipparcos epoch are computed by \cite{acceleration_2021ApJS..254...42B}, as the 60/40 weighted average between the instantaneous proper motion for the original and the second reduction (\citealt{Hipparcos1997A&A...323L..49P}, \citealt{Hip_newdata_reduction_2008yCat.1311....0V}). For V~Hya, the values in right ascension between the two reductions (-14.21$\pm$1.41~mas/yr and -11.02$\pm$1.14~mas/yr, resp.) are not compatible within their uncertainties (as it is for the declination component, 2.72$\pm$1.14~mas/yr and 2.29$\pm$1.11~mas/yr resp.).
This discrepancy can be explained by the variability-induced mover (VIM) flag associated with V~Hya for the first reduction. Using an improved chromaticity
correction adapted for red stars, \cite{VIM_2003A&A...399.1167P} did not confirm the VIM solution; whereas in the second reduction, its recomputed solution is a standard five-parameter solution that does not bear any VIM processing. It indicates the uncertain nature of the $\mu^H_{\alpha*}$ datapoint. 
We thus decided to keep the 60/40 mean value computed in the catalog of acceleration \citep{acceleration_2021ApJS..254...42B}, but to assess the robustness of the obtained orbit regardless the $\mu^H_{\alpha*}$, a sensitivity analysis was performed. The $\mu^H_{\alpha*}$ value was artificially offset by +4, +2, -2, and -4 mas/yr and, for each new value, \textsc{orvara} was run again to fit the RV curve and the new astrometric indicators, using the same MCMC strategy described in Sect. \ref{sec:Orbital_parameters_with_ORVARA_(step 2)}, but without any prior on the eccentricity. The results are the following: for the positive offset all orbital parameters stay unchanged, except for a change in the inclination sign. The solution with an inclination of $i$=180-38=142° (and $\Omega$ = 180+20°) is favoured over the solution with $i$=38° (and $\Omega$ = 180-20°). It corresponds to a $\mu_{\delta}$ signal in anti-phase, compared to the best-fitting curve from Fig. \ref{fig:orvara_fit} (right panel), resulting in a change in the orbit motion direction  with respect to the previous solution.
For the negative offset, the obtained orbital solution is bimodal: either the previous estimated solution (period of 17~yr with $e$~<~0.1, $i$  =~38° and $a$ = 11.2~au) or the solution with a period that is twice larger (and, thus, a mass ratio greater than 5), implying a companion mass higher than 4~$\rm M_{\odot}$. This solution is more probable for higher shift of the $\mu_{\alpha*}$ parameter but corresponds to an important increase of the $\chi^2$ (see Table \ref{tab:chi2_orvara}). As discussed in Sect. \ref{sec:Orbital_parameters_with_ORVARA_(step 2)}, such a massive companion is considered as unlikely.
The fit of the PM and RV curves for the four artificial $\mu^H_{\alpha*}$ are displayed in the appendix, together with their respective corner plot. We conclude that the obtained orbital parameters are robust against the unreliable $\mu^H_{\alpha*}$ value. Therefore, we took the orbital parameters ($i$, $\Omega$, $M_1/M_2$, $a$) derived from the astrometry as indicative, rather than precise, and considered the orbital motion direction as undetermined: either counterclockwise rotation of the companion around the primary on the sky (i~=~40° solution) or clockwise rotation (i~=~180-40° solution)
Such a degeneracy could be lifted with additional constraints from astrometric data (forthcoming Gaia DR4 catalog) or through interferometric observations  \citep{VHya_MATISSE}.

\begin{table}[t]
    \centering
    \caption{\textsc{orvara} $\chi^2$  for the different offsets from $\mu_{\alpha*}$, expressed in mas/yr.}
    \begin{tabular}{|l|rrrrr|}
        \hline
         & +4& +2& 0 &-2& -4\\
        \hline
        $\chi^2_{H}$ & 0.48& 5.28& 13.84& 14.43& 25.26\\
        $\chi^2_{HG}$ & 0.08& 0.03& 0.05& 0.04&  0.07\\
        $\chi^2_{G}$ & 0.09& 0.13& 0.17& 0.15& 1.25\\
        $\chi^2_{tot}$ & 0.65& 5.44& 13.65& 14.62& 26.58\\
        \hline
    \end{tabular}
    
    \label{tab:chi2_orvara}
\end{table} 

\subsection{Nature of the companion}

The derived mass function (Table \ref{tab:orbital_parameters}) allows us to constrain the nature of the companion, hereafter referred to as 'V Hya B'. With the  inclination derived at step 2, V Hya B is expected to be more massive than V Hya A. With the masses of the two components being highly correlated (see corner plot in Fig. \ref{fig:orvara_corner}), no precise mass can be retrieved for the companion without imposing V Hya A mass. Taking a primary mass of 1~$\rm M_{\odot}$ as lower limit, the companion mass would be of 1.9~$\rm M_{\odot}$. Such a mass is too high to correspond to a white dwarf and therefore the companion is likely to belong to the main sequence phase. A main sequence companion with a mass above 2.0~$\rm M_{\odot}$ would corresponds to an A-type star. The main-sequence nature of V Hya B was previously inferred from the measured UV excess by GALEX \citep{GALEX}. Such UV excess would require a hot main sequence star. 

\begin{figure}[t]
    \centering
    \includegraphics[width = 0.5\textwidth]{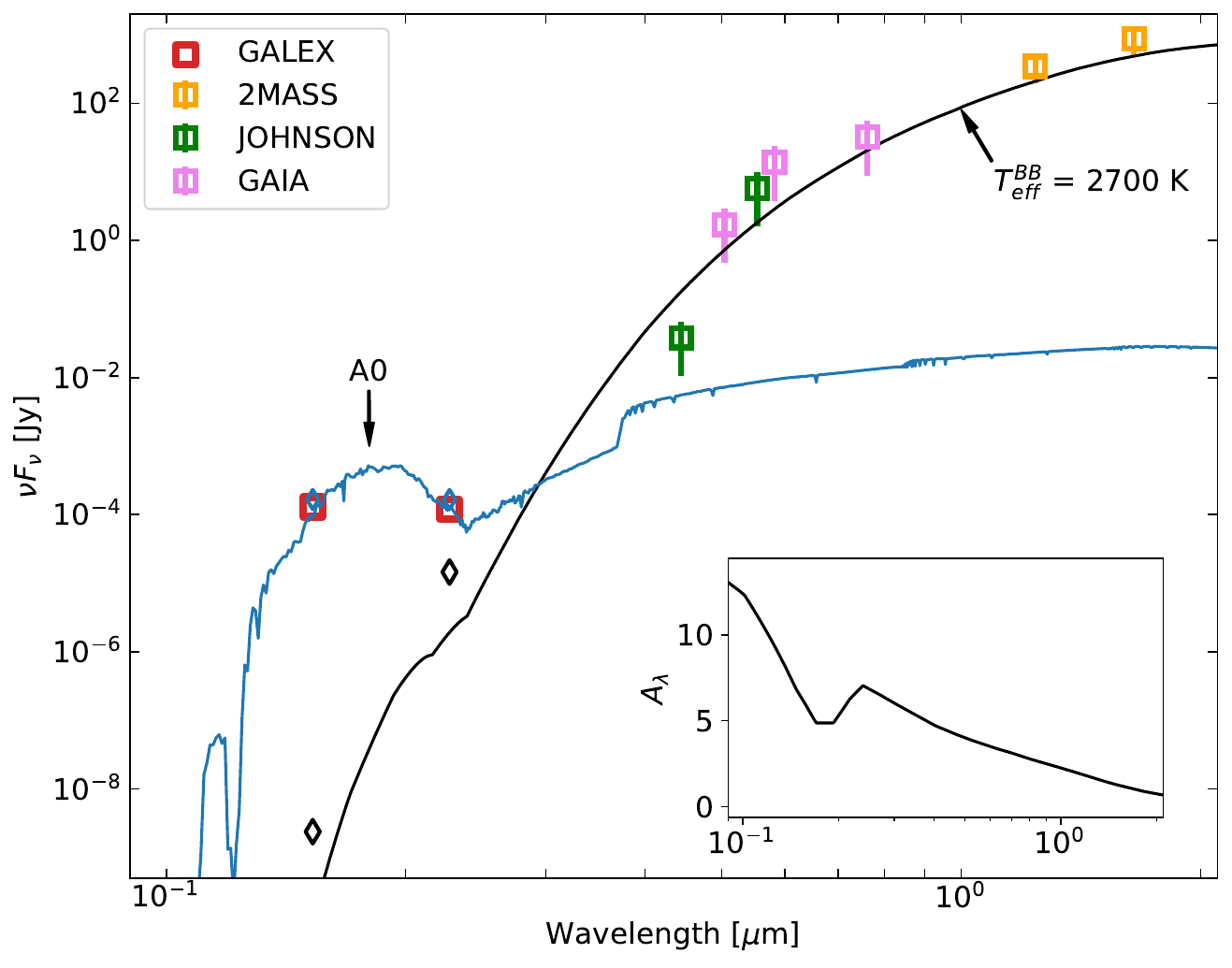}
    \caption{Spectral energy distribution of the V~Hya system from UV to NIR (colored square symbols) and model spectra (black: V Hya as a black body, blue: companion as an A0-type star). The diamonds represent the expected FUV and NUV fluxes from the AGB star (black) and the hot companion (blue). The inset displays the adopted extinction curve from the dust (interstellar + circumstellar).}
    \label{fig:galex}
\end{figure}

To test the nature of the companion obtained from our orbital analysis with the UV excess, the synthetic spectra of the two stars were superimposed on the system photometry from UV to near-IR (NIR). The V Hya B spectrum was modelled as an A0 star with a temperature of 9950~K from  Kurucz models \citep{Kurucz_2003IAUS..210P.A20C} and a radius 1.5~$\rm R_{\odot}$  corresponding to a mass of 2~$\rm M_{\odot}$. The AGB star was modelled by a black-body of 2700~K (inferred temperature from \citealt{Knapp_LC_1999}), as no synthetic spectra for carbon stars exist for $\lambda$~<~300~nm. The SED is displayed in Fig. \ref{fig:galex}. 
To account for the photometric variability induced by the stellar pulsation, the uncertainties of the photometric data points have been set to 2~mag in the optical and 1~mag in the near-infrared.
The spectra are reddened by combining the effect of two sources: the interstellar extinction and the extinction due to the circumstellar envelope (CSE) of the AGB star. 

The visual interstellar extinction was estimated to be $A_{V}$~=~0.09~mag \citep{Lallement_2019A&A...625A.135L},
and the extinction curve of \cite{MW_exctinction_2009ApJ...705.1320G} was adopted. The interstellar contribution is of negligible effect compared to the effect of the circumstellar dust. 
The effect of the dust extinction is obtained assuming a CSE composition made of amorphous carbon, AmC  \citep{AmC_1991ApJ...377..526R}. The dust grains are treated as spherical, with their radius $a$ distributed following the standard MNR distribution ($dn/da = a^{-3.5}$, with $a_{min}$~=~0.005~$\mu$m and $a_{max}$~=~0.25~$\mu$m; \citealt{MNR_1977ApJ...217..425M}). The total extinction curve is displayed in the inset of Fig. \ref{fig:galex}, and the total $A_{V}$ is 3.90 mag. The additional 3.54 mag corresponds to a column density for AmC dust grains of the order of $10^{12}~\rm cm^{-2}$.

The cold nature of the primary does not make it possible to reproduce the observed flux in the UV, whose expected flux in the far-UV (FUV) filter lies several orders of magnitude below its measured value. On the contrary the A0 spectrum, dominating the overall flux for wavelengths below 300~nm, reproduces perfectly the GALEX photometric data in the NUV and FUV filters. The SED confirms the assumption of \cite{GALEX} who attributes the UV excess of V~Hya to the presence of a hot companion. It has to be stressed that the black body radiation curve overestimates the flux at short wavelengths for red stars such as carbon stars, and explains the observed discrepancy between the model and the bluest optical point (Johnson $B$) in Fig. \ref{fig:galex}. Under these circumstances, we could expect to detect signatures of the hot companion even up to 450~nm,~as suggested by \cite{LloydEvansJet}.

\label{sec:nature_of_companion}

\section{Comparison with the light-curve modulations}
\label{sect:comparison_with_LC}
In this section, we compare the RV periodic motion with the light-curve variation. 

\subsection{The pulsation cycle}
To highlight the RV signal associated to the pulsation, the contribution from the Keplerian orbit was removed. For the light-curve variation, the 17 yr signal was fitted by taking a 530-d rolling mean, thus, the signal associated to the Mira variation can be obtained by removing the obtained rolling mean.

\begin{figure}[t]
    \includegraphics[width = 0.5\textwidth]{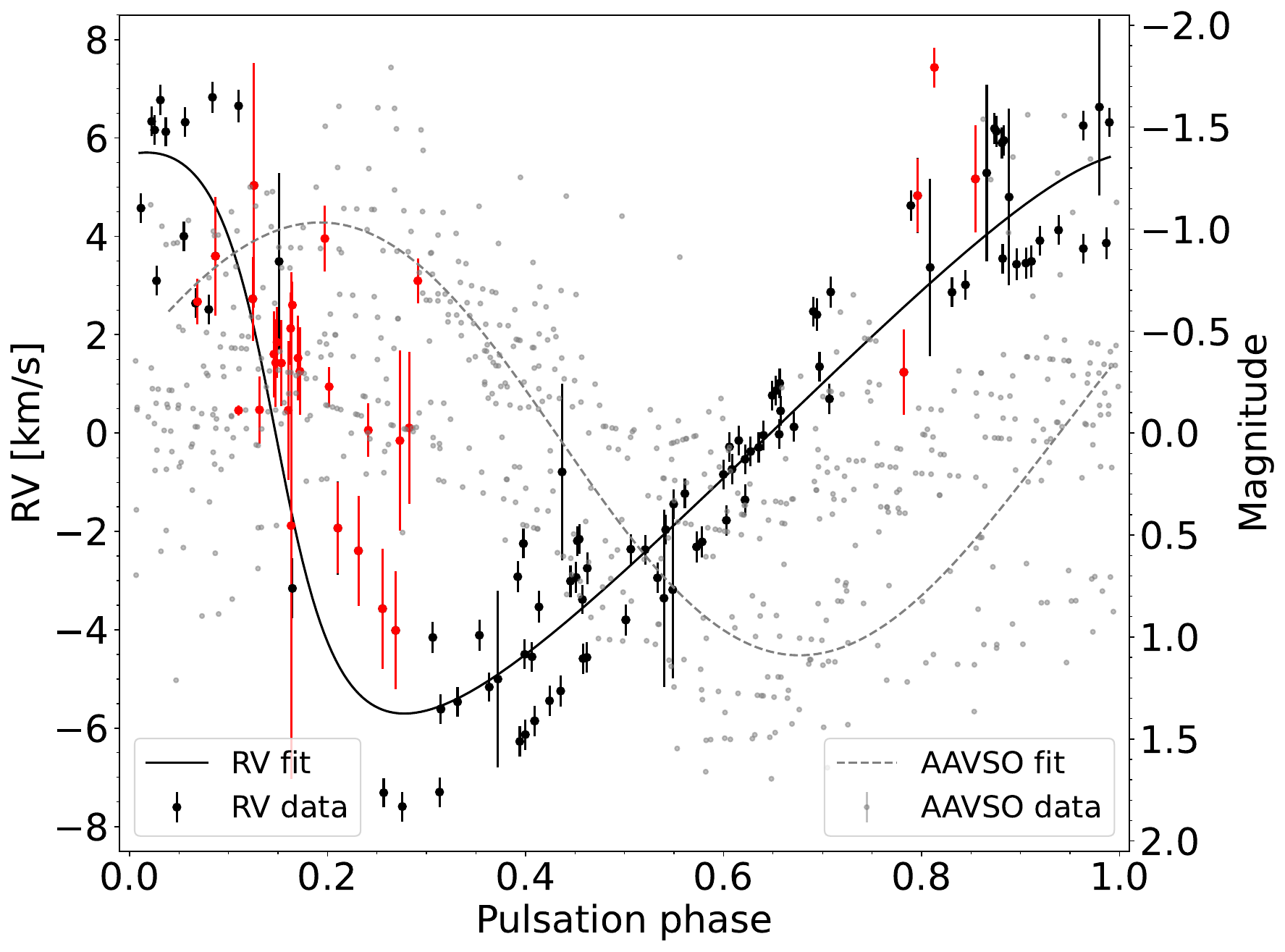}
    \caption{RV curve cleaned from the Keplerian motion, the AAVSO light curve with the 530 d rolling-mean signal removed, and their best-fitting functions phase-folded with respect to the pulsation period. The red dots correspond to the RV data affected by shocks and not used in the analysis (see Sect. \ref{sect:observation}).}
    \label{fig:LC_VR_puls}
\end{figure}

In Fig. \ref{fig:LC_VR_puls}, the RV and light curve signals cleaned from the long variation are plotted as a function of the pulsation phase. The two signals are found to be in quadrature, with the RV signal being in advance with respect to the AAVSO light curve ($\Delta \phi$ = - $\frac{\pi}{2}$). Such a quadrature delay is expected and reflects the change of the AGB photospheric radius: as  the star undergoes expansion of its envelope, its brightness decreases (descending part of the light curve). The expansion leads to blue-shifted RVs for the observer (lower lobe of the RV signal). The same phase delay was already found for the CS Mira star R CMi (see Fig. 5 from \citealt{Chapitre9}).

\subsection{The 17-yr cycle}
\begin{figure*}[t]
    \centering
    \includegraphics[width = \textwidth]{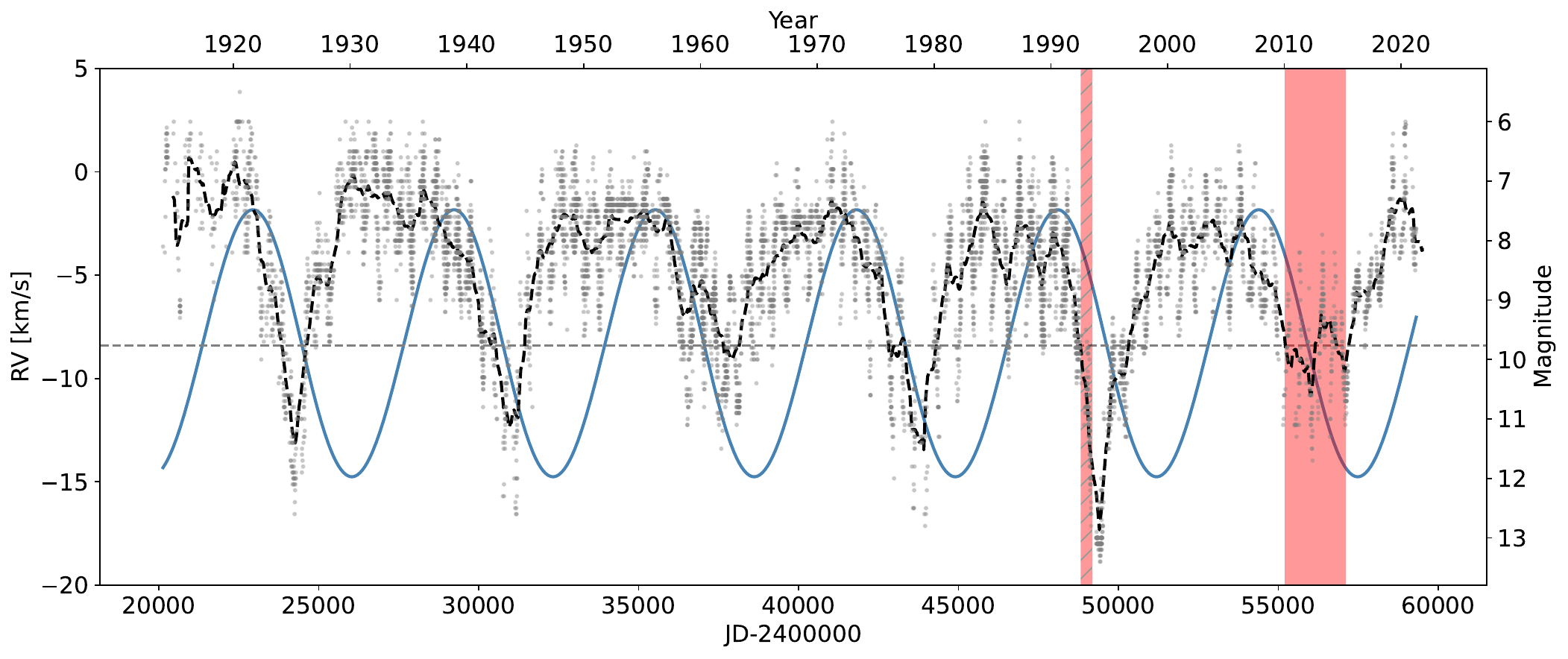}
    \caption{AAVSO lightcurve of V Hya recorded since 1913, displaying six evenly spaced dimming events. The blue curve is the extrapolated spectroscopic orbit. The red regions are the observing epochs of sodium absorption (see Sect. \ref{sodium_doublet_section}) from the HERMES monitoring (filled rectangle), and from the observing dates of \cite{telegram_1993IAUC.5852....2L} (dashed rectangle).}
    \label{fig:LC_VR_tot}
\end{figure*}

In Fig. \ref{fig:LC_VR_tot}, the RV solution (extrapolated to early cycles) is over-plotted over more than a century of photometric observations, covering six dimming events. Their duration and depth vary from cycle to cycle, with a mean amplitude of 3~mag and a mean duration of  30\% of the cycle. The shape of the dimming events is symmetric: the ingress and egress have the same slope, with a value  that varies from 0.001 to 0.003 mag/d, depending on the cycle.

The two long-period signals are again found to be in quadrature. But, in contrast with the short-period signal associated to the pulsation, the RV signal is phase-delayed ($\Delta \phi$~=~$\frac{\pi}{2}$) with respect to the light curve, hinting at a different mechanism than intrinsic variations to explained this phase lag. 
This time, it can easily be explained in the framework of binary motion: the minimum light found by the AAVSO monitoring always corresponds to the superior conjunction, when the primary, namely, V Hya A, is behind its companion and its RVs switch from red-shifted to blue-shifted. A direct consequence is that the obscuration causing the deep minima in the light curve occurs during the companion passage in front of V Hya.

Due to the particular shape of the eclipse, with a long duration and a deep obscuration, and the system inclination, a simple eclipsing scenario by a stellar body has to be rejected. Instead, a more extended and opaque object attached to the companion is needed. 
The inclination angle of about 35$-$40° (see Table \ref{tab:orbital_parameters}) implies that the eclipsing object must have an important extension in the direction perpendicular to the orbital plane. A possible explanation could be that the primary star is eclipsed by a vertical dense outflow.

\section{Time series analysis of spectral lines}

As discussed in the introduction, bipolar outflows have been observed and studied in post-AGB binaries and their signature can be clearly detected in optical spectra. \cite{Dylan_2022A&A...666A..40B} showed that the H$\alpha$ line in the spectra of post-AGB binaries with jets shows a variable profile whose time-series can be modeled to obtain different jet parameters. 
In this section, we report on the temporal variation of particular spectral lines: the sodium and potassium doublets, H$\alpha,$ and the $C_2$ (0,0) vibrational band, and we connect them with the stellar and orbital star periodicity. 

\begin{figure*}[h]
    \centering
    \includegraphics[width = \textwidth]{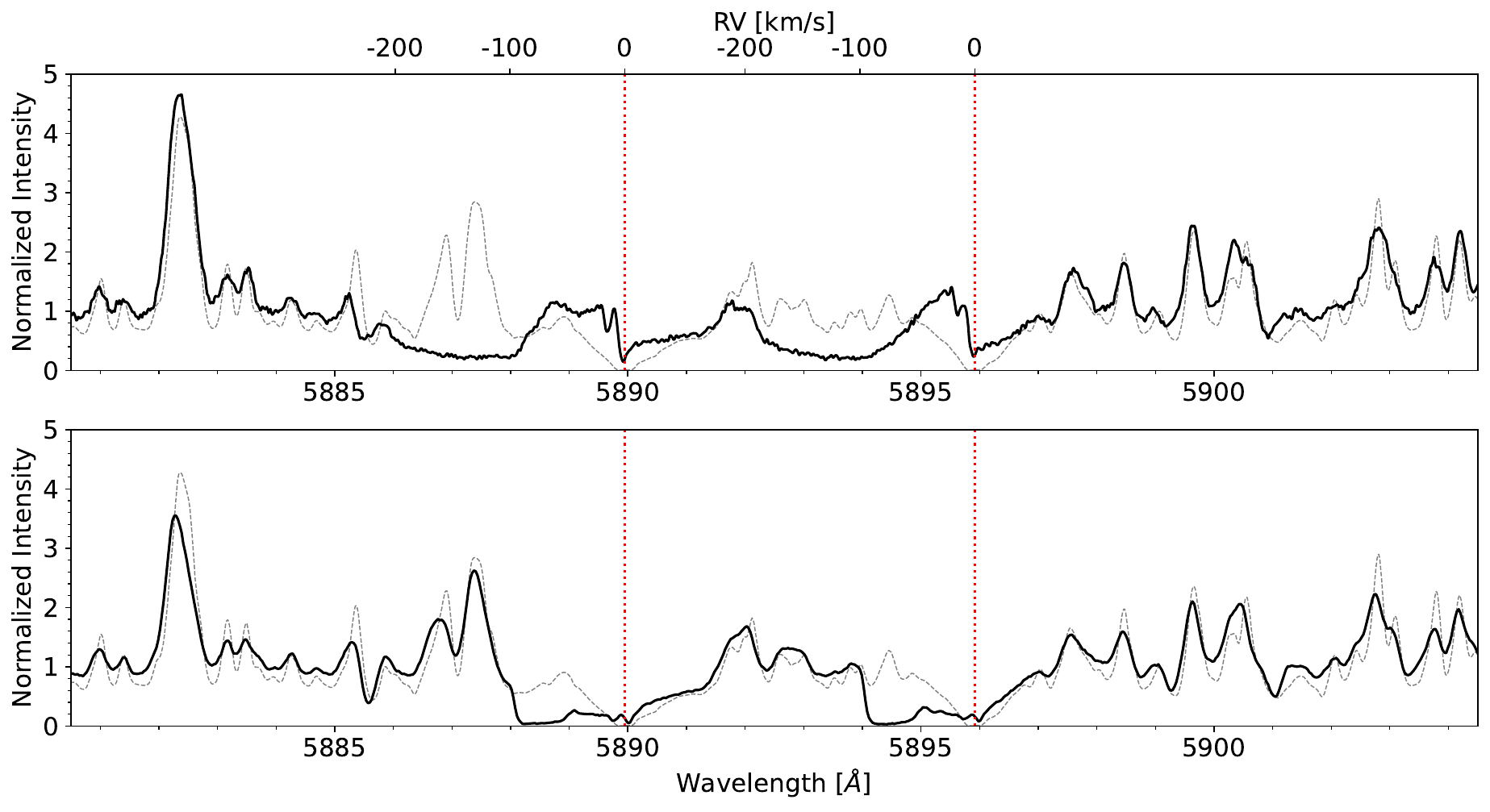}
    \caption{\ion{Na}{I} doublet near superior conjunction (phase 0.46, top) and near inferior conjunction (phase 0.98, bottom). The dashed line is the spectrum of the reference carbon star X Tra used as template for the modelling. The vertical dashed lines indicate the centres of the two components of the doublet. }
    \label{fig:sodium_conjunction}
\end{figure*}

\begin{figure*}
    \includegraphics[width = 0.4\textwidth]{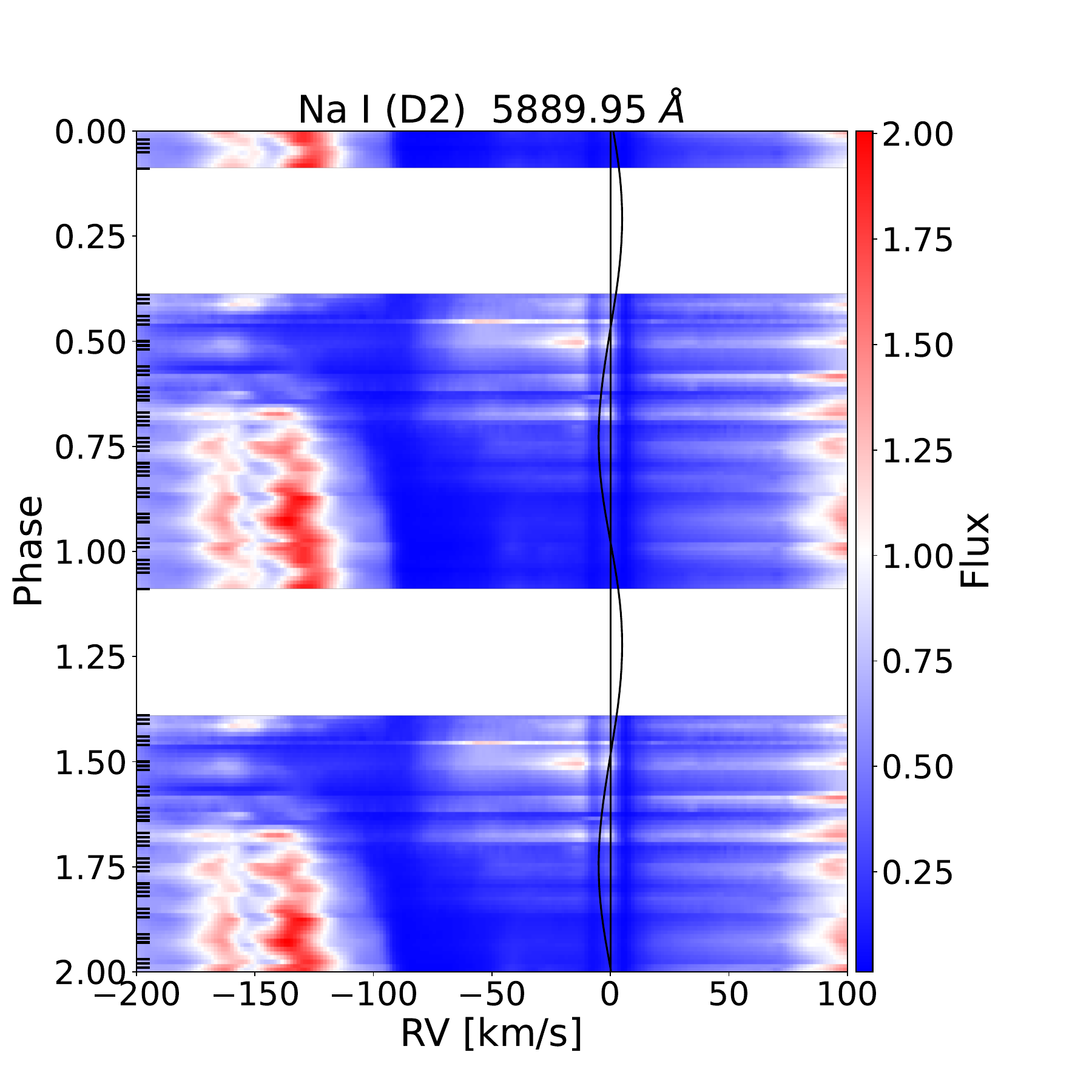}\hfill
    \includegraphics[width = 0.4\textwidth]{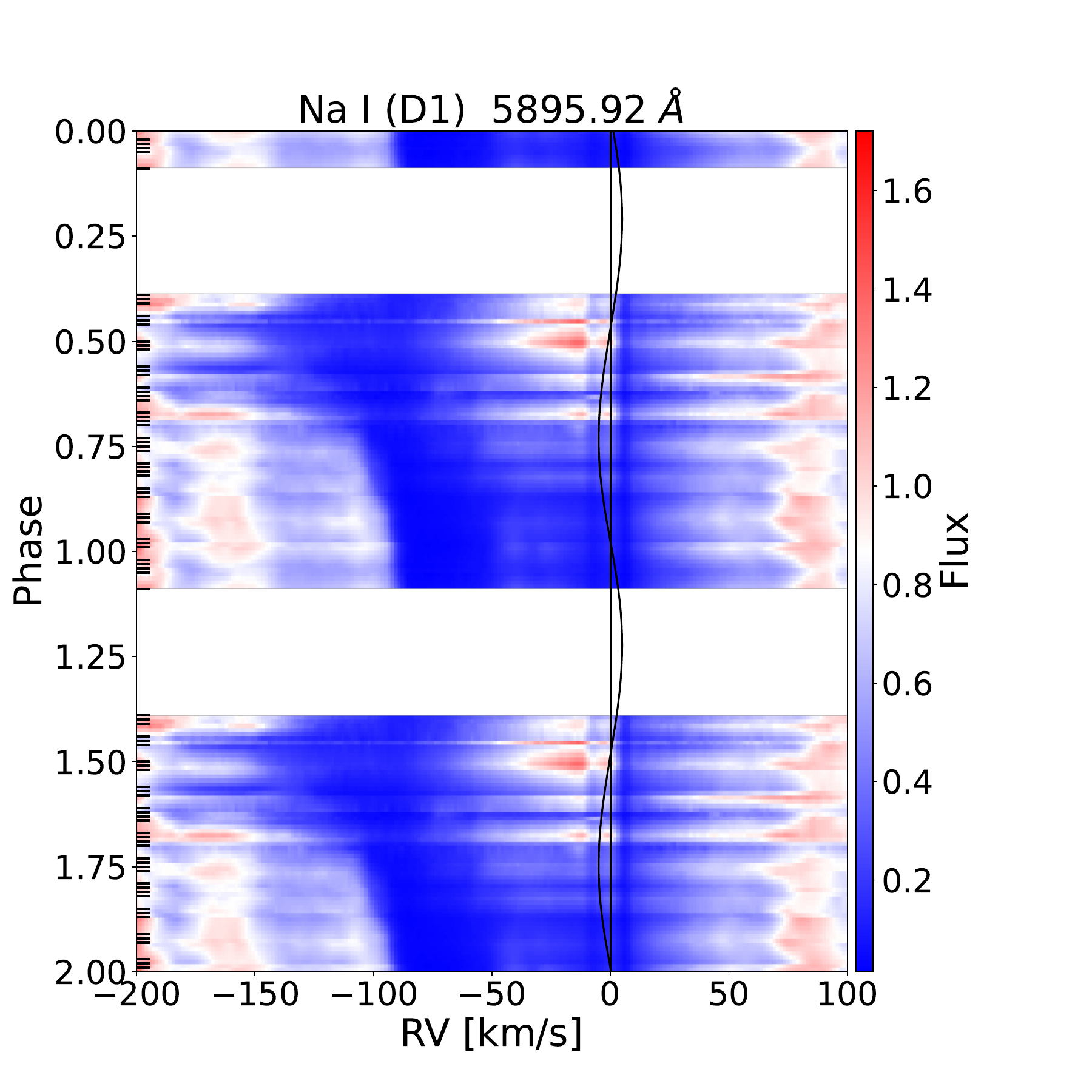}\hfill
    \includegraphics[width = 0.4\textwidth]{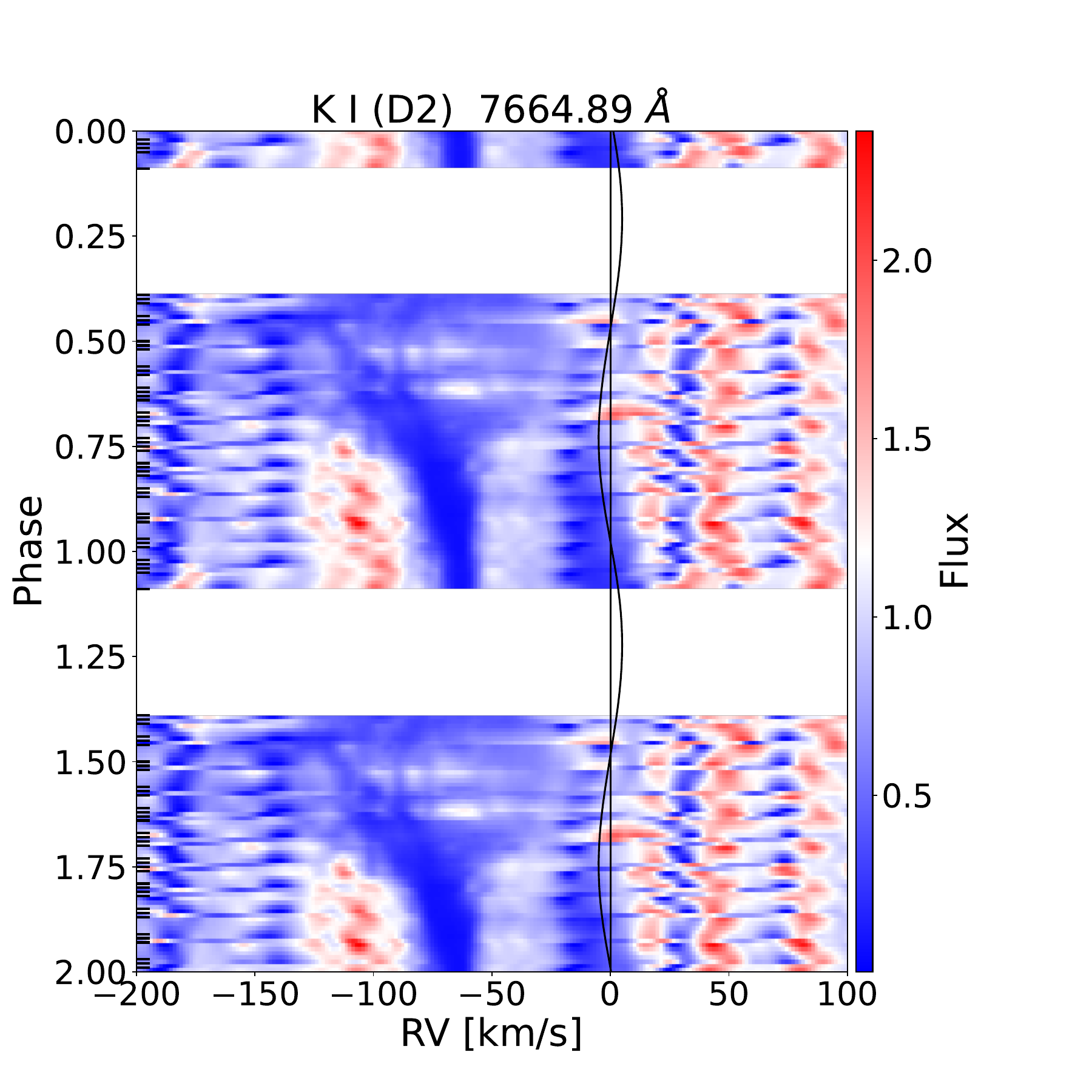}\hfill
    \includegraphics[width = 0.4\textwidth]{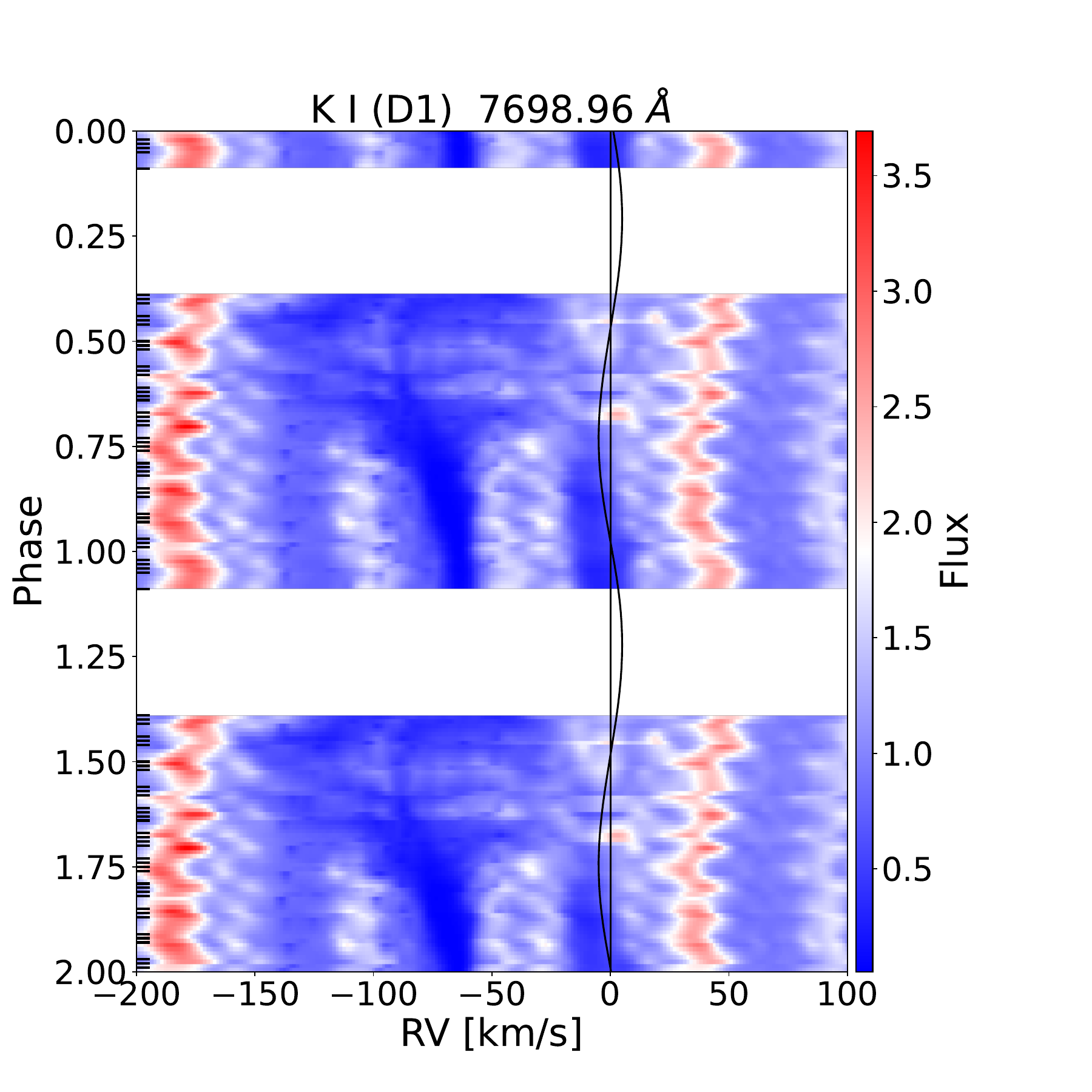}\hfill
    \includegraphics[width = 0.4\textwidth]{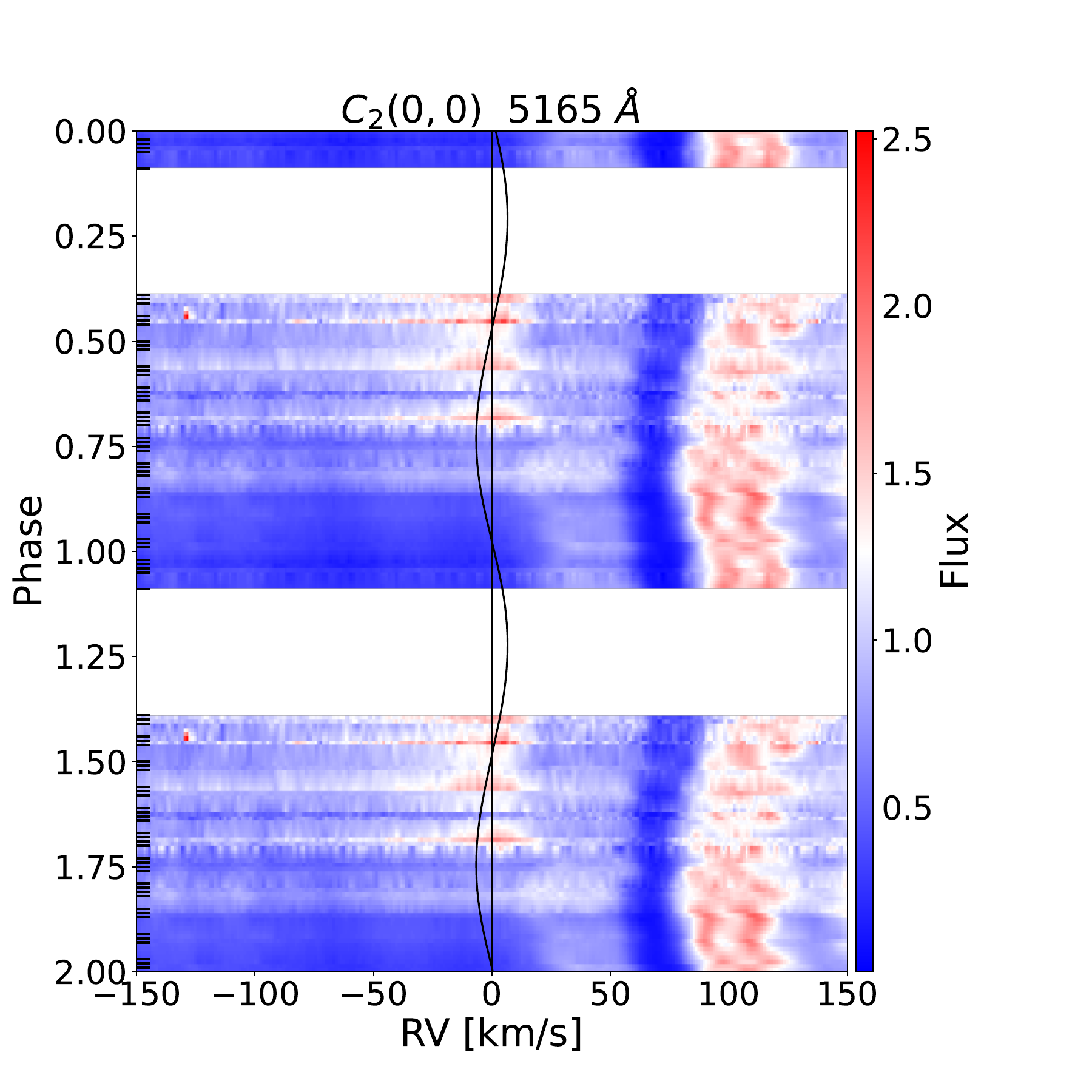}\hfill
    \includegraphics[width = 0.4\textwidth]{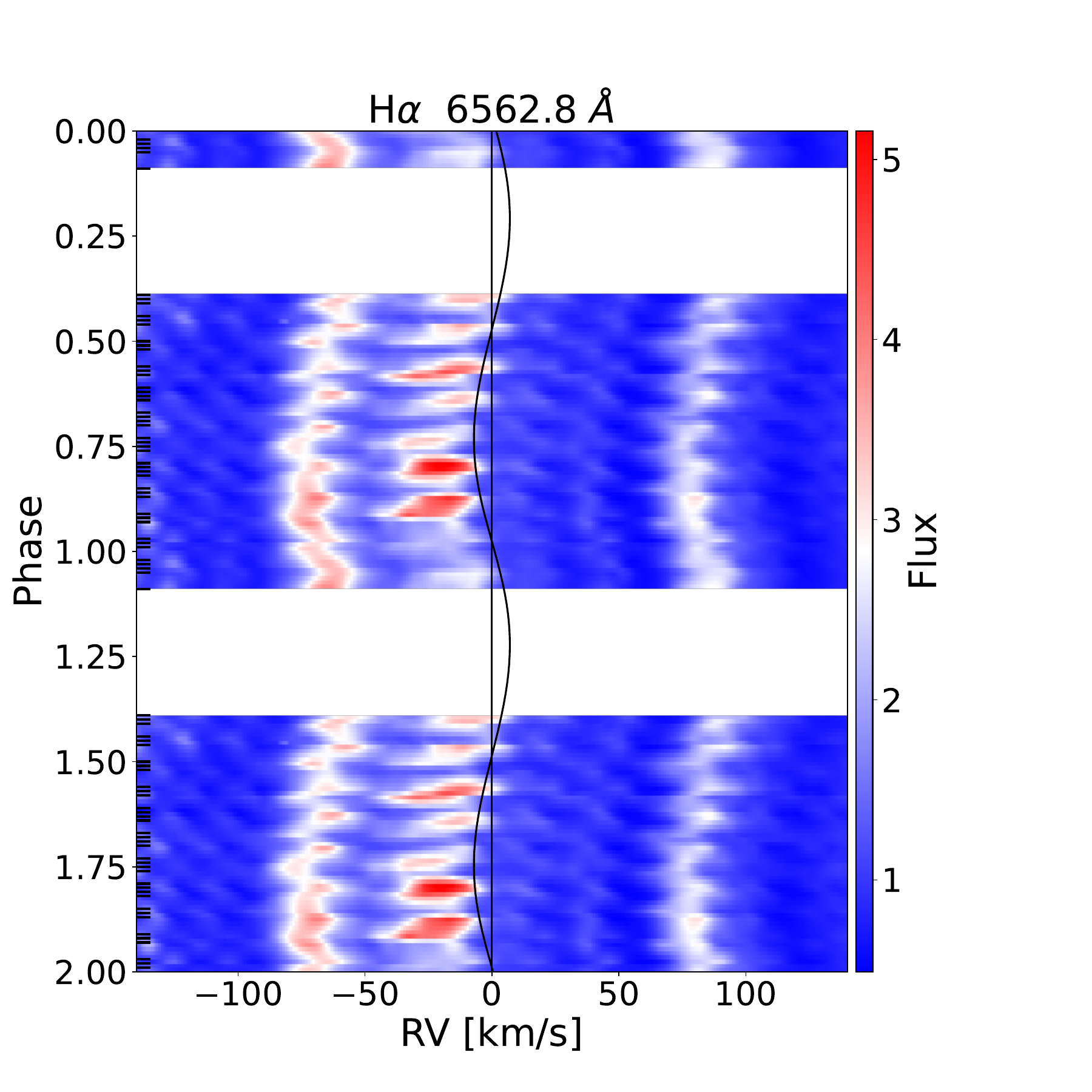}\hfill
    
    \caption{Dynamic spectra of the sodium doublet (top row), potassium doublet (middle row), $\rm C_2$(0,0) (bottom left), and H$\alpha$ (bottom right) of V~Hya as a function of the orbital phase. In each panel, the observed phases are shown twice to guide the eyes, time runs down. The colour represents the pseudo-continuum-normalised fluxes taken as the median value of the spectral window. The black curve represents the primary motion while the vertical line marks the systemic velocity, and the horizontal lines on the left indicate the observed phases.}
    \label{fig:spectra_interpol}
\end{figure*}
\subsection{Sodium doublet}\label{sodium_doublet_section}
\label{sect:sodium_doublet}

The most interesting features are found in the sodium doublet \ion{Na}{I} D-lines. In Fig. \ref{fig:sodium_conjunction}, the \ion{Na}{I} D-profiles at two different orbital phases are plotted, together with a reference spectrum. The phase 0.5 corresponds to the light minimum (superior conjunction, V~Hya behind its companion) and the phase 0 corresponds to the inferior conjunction. A strong and variable blue-shifted absorption is present. This absorption is visible in both lines of the doublet and is most blue-shifted around phases 0 and less blue-shifted around phases 0.5. The profiles of the two sodium lines do not seem to differ.

To highlight the variation of the \ion{Na}{I} D-line profile, the spectra are plotted as a function of the orbital phase and interpolated between the phases to obtain a smooth representation of the variation. This is shown in Fig. \ref{fig:spectra_interpol}. A sample of spectra at different orbital phases can be found in Fig. \ref{fig:sodium_overimposed}.
Since the HERMES monitoring duration is shorter than the orbital phase, the phases between 0.05 and 0.35 are not yet covered and are left blank in the figures.
The 17 year periodicity of the \ion{Na}{I} modulation was confirmed by prior spectroscopic monitoring of \cite{telegram_1993IAUC.5852....2L}. Indeed, a weakening of \ion{Na}{I} absorption followed by an increasing of its emission core is reported at the beginning of the fading event of 1993 (see red rectangle in Fig. \ref{fig:LC_VR_tot}).  

    
The large absorption band seen in Fig. \ref{fig:sodium_conjunction} is also visible in Fig. \ref{fig:spectra_interpol} and reaches its maximum blue-shifted value, -150~$\rm km\,s^{-1}$ at phase 0.5, and its minimum at phase 0. At phase 0.5, an emission (represented in red) is also present on the right side of the blue-shifted absorption. The combination of blue-shifted absorption and emission is reminiscent of P-Cygni profiles.  
The periodicity of the absorption pattern seems to be the same as the orbital one but it is phase-shifted. More precisely, they are in quadrature as the absorption band is at its maximum blue-shifted value at maximum light when the star velocity equals the centre-of-mass velocity. This is easily seen in Fig. \ref{fig:spectra_interpol}, thanks to the sine black line representing the orbital motion.
This phase shift is quite puzzling since one would expect it to be in phase if the absorption was due to the primary star or in anti-phase if it was due to the companion motion. Thus this feature cannot be related to the motion of the stars but must have a different origin. A possible explanation is the presence in the binary system of a conical outflow with a velocity gradient. As the system goes along its orbital motion, the cone obstructs V~Hya light with a varying angle and therefore a varying absorption velocity. 

We can see that even though the shapes are different (\ion{Na}{I} D2 exhibits a double-peak emission whereas \ion{Na}{I} D1 only a left-centred peak), their equivalent widths are quite similar.
The ratio between D2 and D1 equivalent widths range from 0.8 to 1, for a spectral window of 2~\AA. Those values being well below 2 (the expected values for the case of an optically-thin media, and ratio of the oscillator strength), indicating that the media is therefore optically thick. 

\subsection{Potassium doublet}
Peculiarities are also observed in another alkaline metal: the potassium doublet \ion{K}{I} D-lines at $\lambda$ =  7664.89~\AA~and 7698.96~\AA. 
A large blue-shifted absorption band is present and shares the same phase dependency as the sodium doublet, being most blue-shifted at phase 0.5. The absorption is narrower, especially at phase 0 where the bandwidth reaches about 20~$\rm km\,s^{-1}$. 
Pseudo-emission lines following the orbital variation are visible in the red-shifted part of the spectrum (in red in Fig. \ref{fig:spectra_interpol}). Their evolution follows the orbital trend, indicating that those lines are of photospheric origin, contrary to the alkaline resonance lines whose variations are phase-shifted, hinting at a different origin above the photosphere.

\subsection{Swan band}
The Swan bands are vibrational bands of the $\rm C_2$ molecule. Their emission is temperature-dependent and begins to develop above 2450~K \citep{Swan_bands}. The most intense one is the (0,0) transition at 5165~\AA. 
As V~Hya is a cool carbon star, only emission in the (0,0) band is observed in its spectrum. 
In Fig. \ref{fig:spectra_interpol}, the $\rm C_2$ profile seems to change with the orbital phase. The line profile around $\lambda$ = 5165~\(\text{\AA}\) is quite different between superior and inferior conjunctions. Around phase 0.5 a pseudo-emission seems to be present but this feature is absent around phase 0.
Such an abrupt emission of $\rm C_2$ was also reported during the obscuration event of 1993 (\citealt{telegram_1993IAUC.5852....2L}, \citealt{Lloyd_Evans}) confirming the relation between the $\rm C_2$ emission and the star drop of brightness.

\subsection{Balmer lines}
No amplitude modulation correlated with the orbital motion is found in the Balmer lines. Only the H$ \alpha$ emission is strong enough to be observed with a high signal-to-noise ratio (bottom right panel of Fig. \ref{fig:spectra_interpol}) and mainly shows radial-velocity variation that follows the orbital trend with a faint amplitude modulation due to the Mira pulsation. No blue-shifted absorption as for the \ion{Na}{I} or \ion{K}{I} lines is seen. 
Compared to other carbon stars with a Mira pulsation, the intensity of H$\alpha$ for the star V~Hya is quite reduced overall, even outside the eclipse phases, reaching at most five times the continuum.  
As previously reported by \cite{LloydEvansJet}, the H$\gamma$ line is found in absorption and could be associated to the hot companion signature. Indeed, no absorption is found in the Balmer lines of the reference carbon star \object{X~Tra}.

\section{Spatio-kinematic modelling of the outflow}
\label{sect:spatio-kinematic_modelling}
As discussed in Sect \ref{sodium_doublet_section}, the presence of varying blue-shifted absorption in the sodium doublet can be interpreted as the signature of a high-velocity outflow attached to the companion star. \cite{Dylan3} studied the jet imprints on Balmer lines for a sample of post-AGB binaries for a variety of jet configurations. Their modelling approach consists in fitting the model spectral lines created by a parametric jet model to the phase-resolved H$\alpha$ profile using an MCMC routine.

To deduce the jet structure (geometry, velocity, and density profile) we used the same spatio-kinematic modelling engine\footnote{freely available on \url{https://github.com/bollend/spatiokm}} and adapted it for the phase-resolved sodium line at 5895.92 \AA. Hereafter, we summarise the overall fitting routine adopted to fit V~Hya sodium spectral line over the range 0 to -250~$\rm km\,s^{-1}$. We refer to the reference paper \citep{Dylan2} for a more general description of spatio-kinematic jet structures and their numerical implementation.
Among the jet models available, we restrict ourselves to the stellar jet model, which is similar to the model developed by \cite{RedRectangle} for the Red Rectangle Nebula. It has been shown that the fitting routine provide similar results for different configurations, hence we choose the most generic model containing the smallest number of free parameters and assumptions about the launching site of the outflow. The stellar jet model and its different parameters are described in Appendix \ref{appendix:Stellar_jet_configuration}.
As the absorption feature (believed to be the jet imprint) is present at all observed phases which cover nearly 70\% of the orbit, the jet opening angle is expected to be larger than the inclination angle of the system. Therefore, no observed spectrum can be used as template for the unattenuated (background) spectrum in the fitting process. Instead, we used the high-resolution spectrum of the carbon star X~Tra as template for the unattenuated spectrum. The spectrum was modulated in RV to reproduce V~Hya orbital motion and in flux at to reproduce the 530~d intrinsic variability. The dynamic spectrum is shown in Appendix \ref{appendix:Dynamic_model_spectrum}.

The procedure is meant to quantitatively model the absorption of the stellar contribution through the jet. However, in the case of V~Hya, an extra emission pattern is seen in the sodium profile and needs to be modelled and implemented on top of the \cite{Dylan2} method. The modulated emission pattern is not found in other carbon-star spectra, suggesting it to be the result of the star-jet interaction rather than of the star photosphere. We interpreted the blue-shifted emission as re-scattering of V~Hya light in our line of sight.
Additional details on the implementation and normalisation of those two effects are described in Appendix \ref{Appendix:Radiative_transfer}. In summary, the general fitting procedure is a three-step process: first, the synthetic model spectrum is used to fit the stellar-jet model to the phase-dependent observations and to obtain the best-fitting absorbing jet. Then, from the obtained spatio-kinematic parameters, the additional source term is modelled. Lastly, the jet parameters are fine-tuned by fitting the absorbing jet on an updated spectrum consisting in the synthetic model and the modulated source term. 

\subsection{Best-fit model}
\label{section:best_fit_model}

The jet parameters are listed in Table \ref{tab:best_fitting_param}, together with their best-fitting value and their initial range. In the table, $i$ represents the inclination of the orbital plane, $\alpha$ is the jet opening angle, $\theta_{cav}$ is the angle of the jet cavity, $\theta_{tilt}$ is the angle between the jet axis and the normal to the orbital plane, $\varv_0$ is the gas velocity at the jet axis, $\varv_{\alpha}$ is the velocity at the cone edge, $p_v$ and $p_d$ are the power indices governing the density and velocity laws (see Appendix \ref{appendix:Stellar_jet_configuration}), $c_\tau$ is the scaling factor (see Appendix \ref{appendix:Stellar_jet_configuration}), and $R_1$ is the radius of the primary star. Due to a strong correlation between the maximal jet opening angle, $\alpha,$ and the system's minimal inclination, we constrained the inclination angle, $i,$ to its value obtained from the orbital analysis (see Table \ref{tab:orbital_parameters}). For the definitions of all the other parameters, we refer to Appendix \ref{appendix:Stellar_jet_configuration}. The velocity and density structure across the jet is displayed in Fig. \ref{fig:density_velocity}. The velocity follows a decreasing and nearly-linear trend starting from $\varv_{0}$ at the jet axis to $\varv_{\alpha}$ at the jet edge. The maximum deprojected velocity, $\varv_{0}$, is below 200~$\rm km\,s^{-1}$, although such a value is well above the velocity of the wind feeding the AGB circumstellar envelope. 

\begin{figure}[t]
    \centering
    \includegraphics[width= 0.5\textwidth]{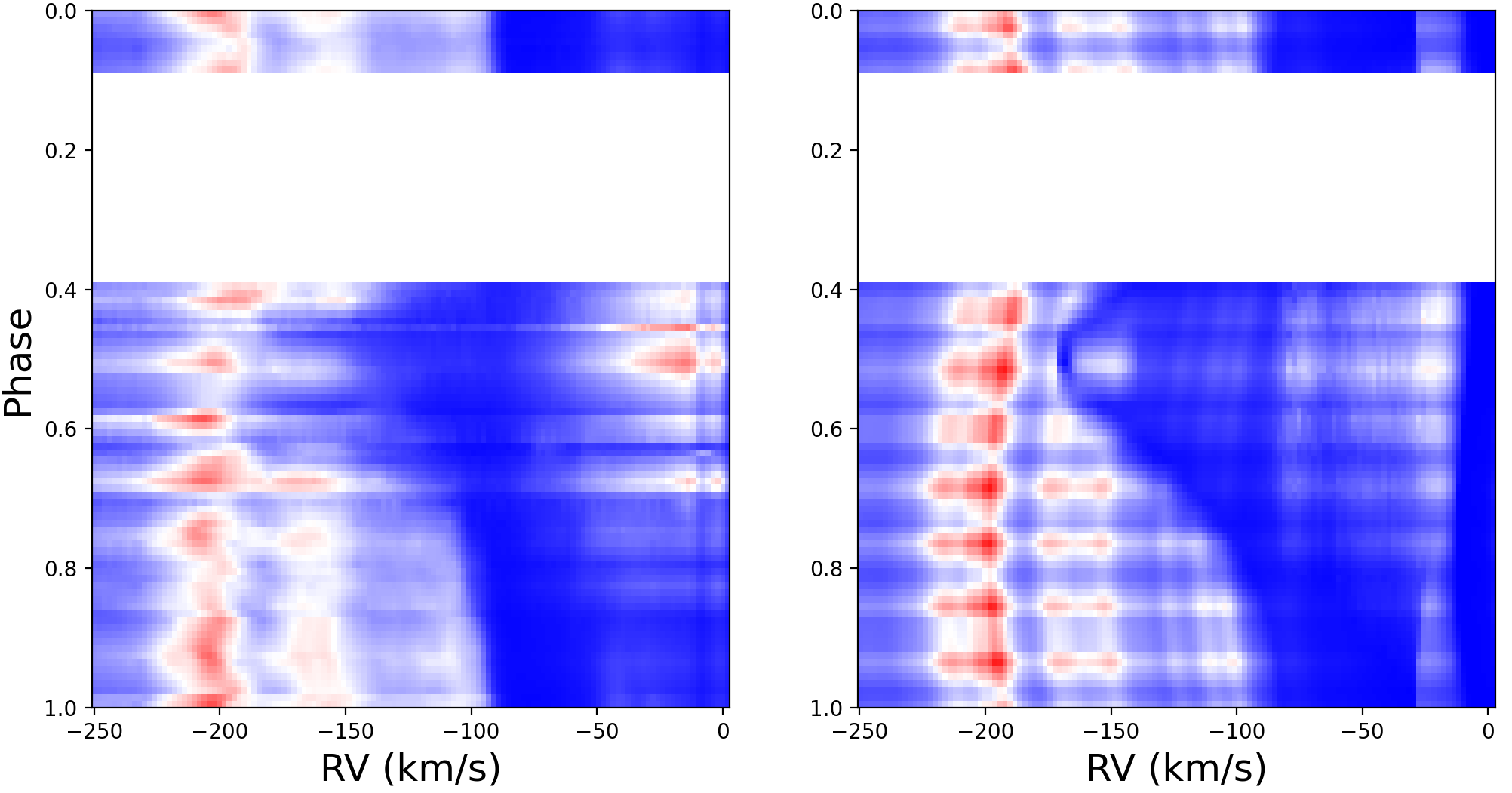}
    \caption{Observed dynamic spectrum of the \ion{Na}{I} D1 line (left) and the model spectrum obtained by the spatio-kinematic fitting of a conical jet (right).}
    \label{fig:sodium2obsmod}
\end{figure}

\begin{figure}[t]
    \centering
    \includegraphics[width= 0.5\textwidth]{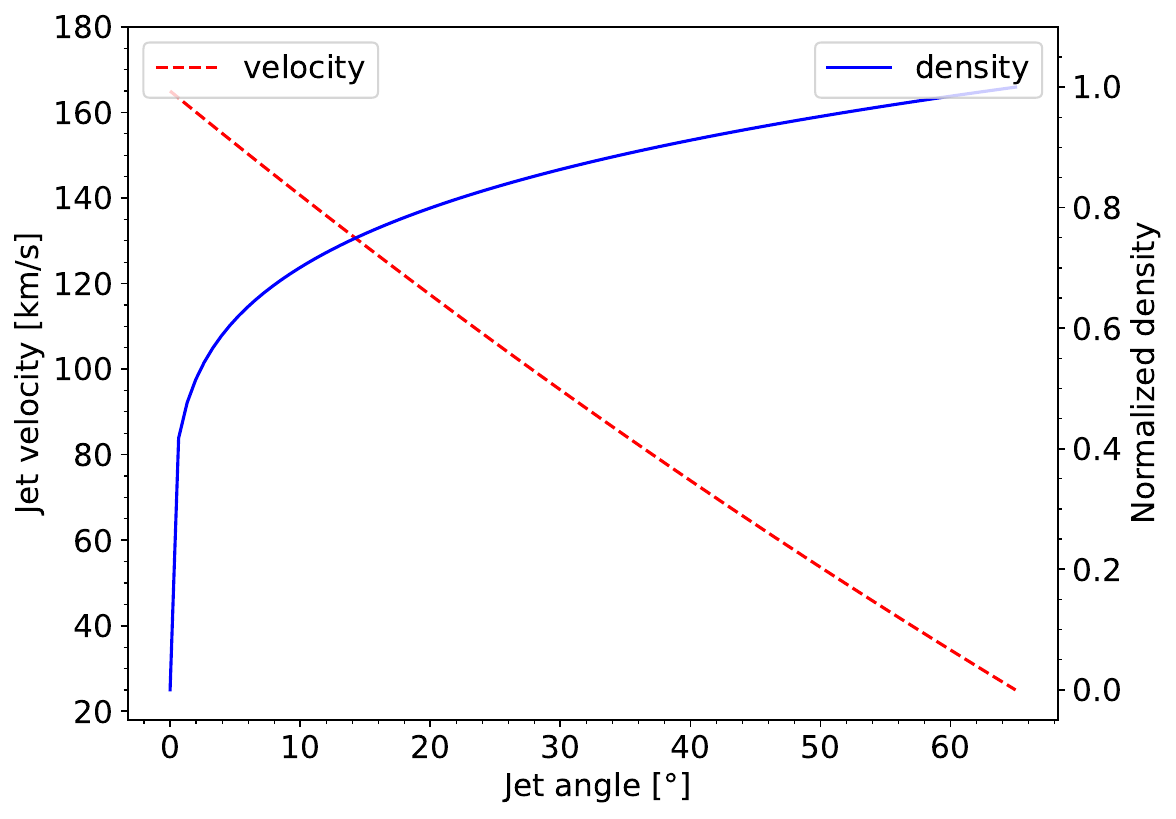}
    \caption{Best-fit model of the velocity and density laws as a function of the jet angle. The density at the edge is set to 1.}
    \label{fig:density_velocity}
\end{figure}

\begin{figure*}[t]
    \includegraphics[width= \textwidth]{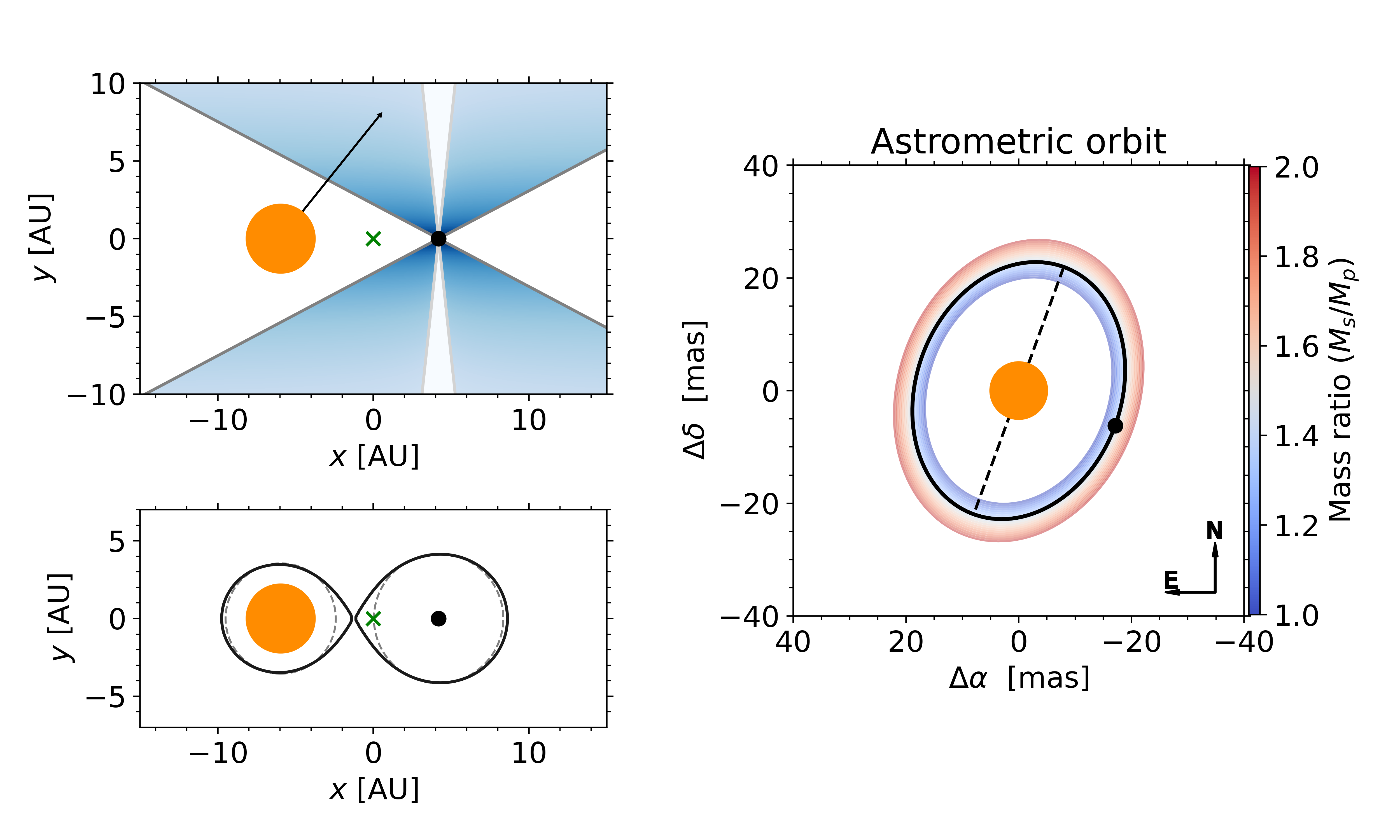}
    \caption{Geometric representation to scale of the binary system where V~Hya~A is represented as an orange filled circle and V~Hya~B as a black dot (to be visible, its radius was increased). V~Hya~A mass was set to 2~$\rm M_{\odot}$, corresponding to a mass of 2.5~$\rm M_{\odot}$ for V~Hya~B. \textbf{Upper left panel}: System depicted edge-on at the superior conjunction. The black line is oriented along the line of sight. The green cross indicates the centre of mass. \textbf{Lower left panel}: System viewed face-on, with the  Roche equipotential in black line and the Roche lobe of the two stars in dashed gray. \textbf{Right panel}: System projected on the sky-plane at the superior conjunction for the counterclockwise solution, assuming a parallax of 2.31~mas. The black line is the best projected orbit, the dashed black line represents the line of nodes. The effect of the mass-ratio uncertainty on the orbital separation is illustrated by the colored ellipse. }
    \label{fig:jet_geometry}
\end{figure*}

The normalised density profile is maximum at the edge of the cone and minimum along its axis, with a cavity angle $\theta_{cav}$. Such a hollow cone is also encountered for post-AGB binaries \citep{Dylan2} and reveals the presence of a cavity along the jet axis. The presence of a cavity in the V~Hya system was already put forward to explain the launching mechanism of the [\ion{S}{II}] jets observed by \cite{Sahai1} and \cite{Scibelli2019}.
\begin{table}[t]
\caption{Jet parameters, range and best-fit values. For the definition of the parameters, see Appendix \ref{appendix:Stellar_jet_configuration}.}
    \label{tab:best_fitting_param}
    \centering
    \begin{tabular}{lrr}
    \hline
    \hline
    Jet parameter & Range & Best fit \\
    \hline
        $i$ [°] &  34 $-$ 40& 38.3 $\pm$ 1.5\\
        $\alpha$ [°] & 30 $-$ 65 & 62 $\pm$ 2\\
        $\theta_{cav}$ [°] & 0 $-$ 10 & 6.9$\pm$ 2.5\\
        $\theta_{tilt}$ [°] & -5 $-$15 & 1.8$\pm$ 2.4\\
         $v_{0}$ [$\rm km\,s^{-1}$] & 100 $-$ 800&194.7 $\pm$ 20.5\\
        $v_{\alpha}$ [$\rm km\,s^{-1}$] & 20 $-$ 200 & 26.4$\pm$ 9.0\\
        $p_v$ & 0 $-$ 8 & 0.32$\pm$0.30 \\
        $p_d$ & -2 $-$10& 0.19$\pm$ 0.17\\
        $c_\tau$ & 0 $-$ 6 & 2.91$\pm$ 0.50\\
        $R_1$~[$R_\odot$]&1.5 $-$ 2.5& 2.19$\pm$ 0.16\\
        \hline
    \end{tabular}
\end{table}
The observed and modelled dynamic spectra of the \ion{Na}{I} D1 line is displayed in Fig. \ref{fig:sodium2obsmod}. It can be seen that the model reproduces the temporal modulation of the blue-shifted absorption 
quite well. The obtained jet geometry and orbital configuration are  sketched in Fig. \ref{fig:jet_geometry}.

\subsection{Limitation of the model}

The above-described analysis allows us to conclude that a high-velocity outflow launched by a companion orbiting V~Hya is able to satisfactorily reproduce the observed spectral modulation. The choice of such a modelling tool was motivated by previous large-scale observations of high-velocity outflows (\citealt{Knapp_CO_1997}, \citealt{Hirano2004}) and by the close proximity of V~Hya evolutionary stage to the post-AGB phase where those structures are widely found. However, no direct evidence of the companion motion is seen in the spectrum as its flux contribution is expected to negligible above 450~nm~(see Sect. \ref{sec:nature_of_companion}), and the exact launch site of the jet remains unknown. 
Further analyses are needed to constrain the temperature and the absolute density structure through a full 3D radiative transfer code. Those ingredients and the putative properties of the companion (see Sect. \ref{sec:nature_of_companion}), are needed to fully characterise the physical mechanism feeding and launching the jet.

\section{Discussion}
\label{sect:discussion}
In this section, we discuss the possible origin of the long-period obscuration event in the framework of the binary model obtained. Alternative explanations not invoking a companion are then discussed. Finally, we confront the obtained model of the V~Hya system with previous large-scale observations to build a multi-scale consistent picture of the system.

\subsection{Obscuration by dust}
As highlighted by the orbital analysis in Sect. \ref{sect:comparison_with_LC},  the dimming episodes of about 3~mag occurring every 17 years are associated with the superior conjunction and the high-velocity absorption of alkaline resonance lines. 
These observations can be explained in the framework of an absorbing gaseous jet passing in front of the line of sight. However, a gaseous jet alone cannot be responsible for the visual dimming event, as such a broad-band attenuation requires particles with a gray absorbing power, such as dust grains. 
A light curve modulation induced by the variation of the dust grains column density through the line of sight with the orbital phase would require a non-uniform distribution of particles above the orbital plane, due to the inclination well below 90 degrees.
For a mean magnitude drop of 3~mag, the observed modulation can only be recovered by assuming the highest dust density at the jet centre, with a dust column density that is three times denser at the superior conjunction when the line of sight crosses the axis of the cone. 
The dust distribution through the jet polar angle would be opposite to what was found for the gas density in the jet, better described by a hollow cone at its centre. 
It raises the question of the physical mechanism behind the dust and gas distributions. Only a better characterisation of the absolute gas density and temperature using a full 3D radiative transfer code would allow us to determine more precisely the dust formation limit, the mass-to-dust ratio, and the exact dust spatial distribution in the system and understand their possible coupling with the gas. However, such a complex investigation is out of the scope of the current paper, as we are focused on the system dynamics.
Interestingly, some post-AGB systems also exhibit a long-term modulation of their light curve, but of a smaller amplitude, (<0.5~mag) associated with the radial-velocity and atomic-line variations. In RV-Tauri stars, those obscurations occur at the inferior conjunction, the opposite situation as encountered here,  explained by the particular orbital configuration of the system where the line of sight grazes the circumbinary disc due to the high orbital inclination \citep{Manick_RV_Tauri_2019A&A...628A..40M}. A few other post-AGB stars with a bipolar morphology show small-amplitude obscuration events together with radial-velocity and H$\alpha$ variation: For instance, \object{V501~Lup} exhibits obscurations of about 0.5 mag in the visible and near-infrared which share the orbital periodicity (2727$\pm$26~days) and could be the result of dust absorption and scattering by the nebula \citep{Manick_V501_Pup_2021MNRAS.508.2226M}. For 89 Her, a long-term obscuration with a periodicity of 5000~d is reported together with modulation of H$\alpha$, \ion{Na}{I} D, and \ion{Ba}{II} lines. With the the long period of the light curve being larger than the orbital one (289~days, \citealt{Dylan3}), two scenarios have been suggested: an orbital wobbling of the circumbinary disc inducing a misalignement with the hour-glass structure or recurrent mass-ejection episodes \citep{Gangi_2021MNRAS.500..926G}.

\subsection{Origin of the long-period variability}
\label{sect:origin_of_long_period-variability}

For the sake of completeness, it ought to be mentioned that alternative scenarios not involving a binary system can be invoked to explain the obscuration of variable stars: a long-period secondary pulsation period that is twelve times longer than the fundamental tone of 530 days or an episodic dust condensation from the star. Below, we describe both of them and argue why we consider them less likely. For the first one, as discussed in Sect. \ref{sect:comparison_with_LC}, the phase shift between the radial velocity and light curves is found to be +90° for the long cycle. Therefore, the usual scenario of radial pulsation induced by the so-called kappa-mechanism (associated instead with a -90° phase shift) cannot be responsible for the 17 yr periodic fading. Besides, this long period would make V~Hya an outlier in the usual period-luminosity diagram, presented by \cite{Wood_2000PASA...17...18W}, being too long to belong to the D-sequence ("long-secondary periods").

The second interpretation, namely, episodic dust formation, is the usual mechanism put forward to account for the abrupt fading by several magnitudes of R Coronae Borealis (R CrB) variables. Those stars exhibit light curves with important dimming events separated by several years \citep{RCorBor_1996PASP..108..225C}. 
As these obscurations are correlated with intrinsic dust formation, their occurrence is far from periodic and their fading episodes show an abrupt decrease followed by a gradual restoration of the star brightness. 
The only exception is the sub-class of DY-Per objects, labeled as cold hydrogen-deficient R CrB stars, whose light curves exhibit a rather regular periodicity in their fading. 
Considering the regularity of V~Hya light curve over a century of observations and previous hints in favour of the binary hypothesis (\citealt{Hirano2004}, \citealt{GALEX}, \citealt{Sahai1}) we consider intrinsic variability scenarios to be less likely to explain the long-term periodicity of the star.

\subsection{Connection to large-scale bipolar outflows}
\label{sect:connection_to_large-scale}
The characterisation of the orbital parameters (including the system inclination) associated with the detection of the high-velocity jet allows us to link the different building blocks of the system;  in particular, the jet that is supposed to be attached to the companion, with its sky-projected orientation. 
A sketch of the system is displayed in Fig. \ref{fig:jet_geometry}. As discussed, the inclination found is consistent with the previous inclination estimates from the bipolar outflow measurement. Additionally, the ascending node being close to 180° (see Table \ref{tab:orbital_parameters}), the jet axis (perpendicular to the orbital plane) points towards the east. Its direction is consistent with previous measurements, at a larger scale, of the high-velocity blue-shifted [\ion{S}{II}] emission and the high-velocity  parabolic outflow found in the radio-emission of CO molecules \citep{Sahai1, Sahai_ALMA_2022ApJ...929...59S}. The line of nodes and jet axis angle have an uncertainty of $\pm$10° due to the inclination degeneracy (see Sect. \ref{sec:Orbital_parameters_with_ORVARA_(step 2)}).  

\begin{figure}[t]
    \centering
    \includegraphics[width = 0.4\textwidth]{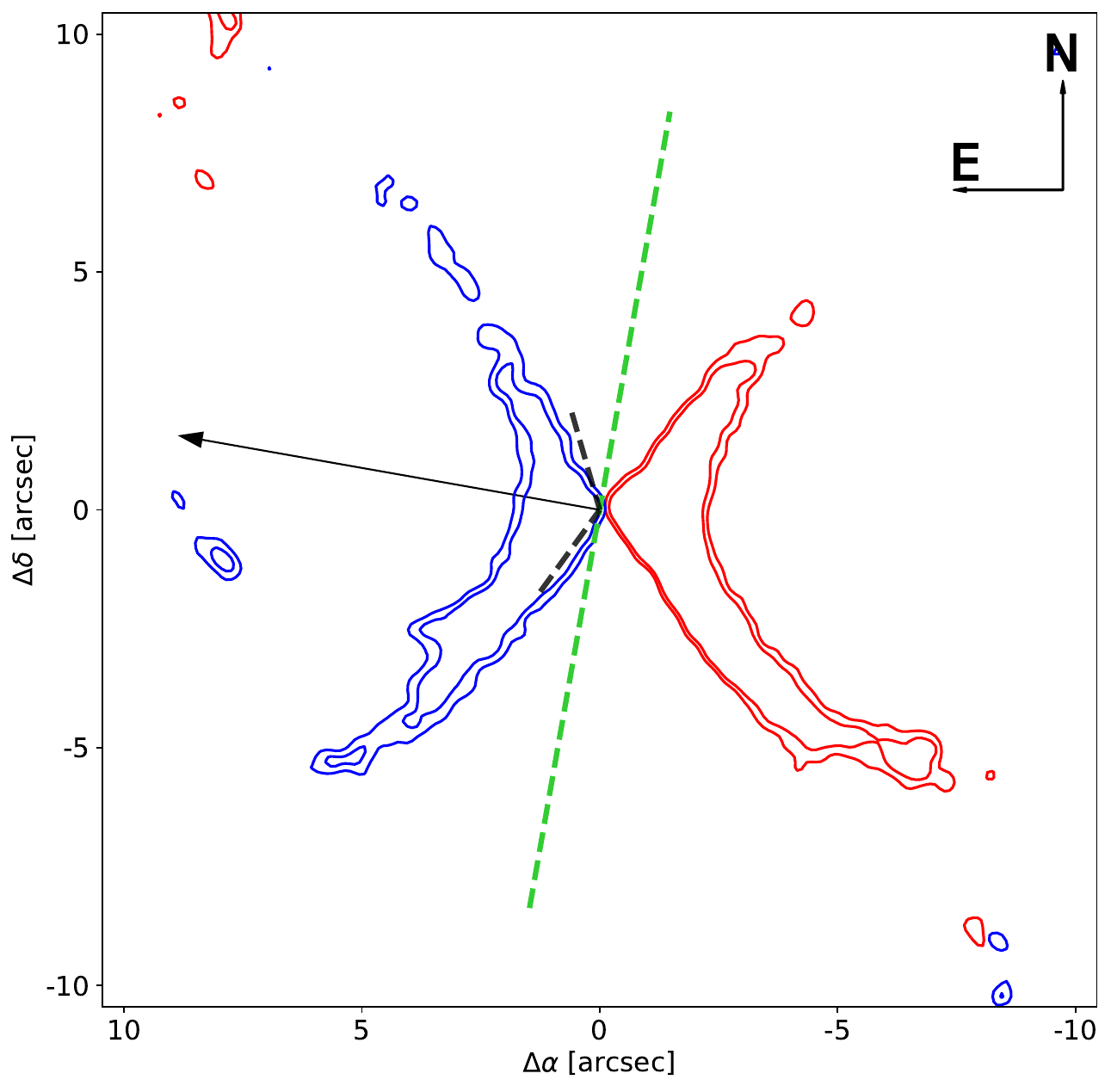}
    \caption{Sky-projected system orientation compared to radio observation. The  contour levels is a stereogram view of the $^{12}$CO~J~=~2$-$1 high-velocity emission. The contour levels are drawn at 6, 8, and 10 mJy/beam for two maps equally shifted by $\pm$100~$\rm km\,s^{-1}$ from the centre-of-mass velocity. The blue-shifted contours are displayed in blue, and the red-shifted in red. The dashed green line represents the orbital line of node, the black arrow the jet axis and the dashed black line sketches the cone edges.}
    \label{fig:ALMA_stereogram}
\end{figure}

Figure \ref{fig:ALMA_stereogram} superimposes the sky-projected jet axis and the system's line of node on the high-velocity emission obtained from ALMA observations (PI: Sahai R., Project ID: 2018.1.01113.S). 
Both geometry and orientation between the two structures match. This result is to be underlined as the two sets of observations, having entirely independent approaches, provide a complementary set of evidence for the system morphology. 
On the one hand, in our study, the conical geometry and orientation of the jet is deduced by fitting phase-dependent blue-shifted absorption of optical resonance lines and compare them with the system orbit. On the other hand, the bipolar jet is mapped using a one-shot observation in the radio-emission of $^{12}$CO~J~=~2$-$1 at larger scale.

This compatibility strongly suggests that they may share a common origin: the modeled conical jet could be seen as the innermost part of a large-scale parabolic structure. This scenario does demonstrate the link between the dynamical binary interaction (as inferred from our spectral monitoring) and the morphology of the large-scale bipolar nebula. 

\section{Conclusion}
\label{sect:conclusion}

Thanks to an unprecedented twelve-year-long spectroscopic monitoring of V~Hya, we performed a temporal study of the dynamical processes at the origin of its two periodic variations. The main results obtained in this paper are two-fold.

In the first step, the radial-velocity curve revealed a short variation of 530~days superimposed on a long trend. Combining the RV data with secular visual photometry and astrometric acceleration, we disentangled the signal associated with the Mira-like pulsation of 530~days from the 17 year cycle. The latter is interpreted as the signature of a Keplerian orbit for a period equal to the long photometric cycle of 6351 days, a low eccentricity ($e = 0.02\pm0.02$), and an inclination of 37°$\pm$2°, compatible with a main sequence companion. 

In a second step, the time-resolved spectroscopy unveiled the presence of a high-velocity gas stream in the system. The resonant lines of sodium and potassium display phase-dependent absorption, blue-shifted up to -150~$\rm km\,s^{-1}$ at the epoch corresponding to the visual minimum, and superior conjunction. We interpret the gaseous jet as the imprint of an interacting binary system. 
A spatio-kinematic modeling of the jet launched from the companion orbiting around V~Hya is fitted to reproduce the observed spectral-line modulation. The obtained jet model is described as a hollow cone with a wide opening angle and a velocity gradient along its radial direction. The highest velocities are oriented along the jet axis, whose projection on the sky plane is pointing toward the east. 
In the meantime, the jet geometry obtained (velocity and direction) is consistent with previous large-scale observations of the high-velocity bullets found in [\ion{S}{II}] lines by \cite{Sahai1},  and high-velocity bipolar flow from the molecular lines of CO (\citealt{Knapp_CO_1997}, \citealt{Hirano2004}, \citealt{Sahai_ALMA_2022ApJ...929...59S}). 

It is, to our knowledge, the first case where the temporal variation of the light curve, high-velocity gas, and the AGB star velocity are reported to be correlated. Our proposed scenario – similar to what is found in many post-AGB binaries (e.g. \citealt{Dylan_2022A&A...666A..40B} and references therein), where the evolved star orbits around the jet launched by its main sequence companion – explains the reported observations in a common framework. 
To conclude, this multi-epoch study reveals the complex behavior of V~Hya, which can be seen as a benchmark case to better understand the  shaping mechanism induced by a binary system at the origin of bipolar planetary nebulae.

Further studies reproducing the radiative transfer inside the circumstellar and circumbinary environments of V~Hya system could help to constrain the connection between the binary interaction and the large-scale bipolar morphology. 
Additionally, enlarging the presented analysis to a larger sample of AGB stars with known companion or bipolar morphology, for instance, TU Tau \citep{TU_Tau_2023arXiv230810972G}, R Crt \citep{R_Crt_2020A&A...635A.200K}, UV Aur \citep{UV_Aur_2009AJ....138.1502H}, RR UMi \citep{RR_UMi_1983A&AS...54..541D} would be needed to obtain a statistical distribution of the orbital parameters of such systems. 

\begin{acknowledgements}
     L.P. is FNRS research fellow. 
A.J. acknowledges support from the {\it Fonds de la Recherche Fondamentale Collective} (FNRS, F.R.F.C.) of Belgium
through grant PDR T.0115.23, and from FNRS-F.R.S. through grant CDR J.0100.21 ('ASTRO-HERMES').

     Use was made of the Simbad database, operated at the CDS, Strasbourg, France, and of NASA’s Astrophysics Data System Bibliographic Services. This research made use of Numpy, Matplotlib, SciPy and Astropy, a community-developed core Python package for Astronomy \citep{Astropy_2018AJ....156..123A}. This work presents results from the European Space Agency (ESA) space mission Gaia. Gaia data are processed by the Gaia Data Processing and Analysis Consortium (DPAC). Funding for the DPAC is provided by national institutions, in particular the institutions participating in the Gaia MultiLateral Agreement (MLA). The Gaia mission website is \url{https://www.cosmos.esa.int/gaia}. The Gaia archive website is \url{https://archives.esac.esa.int/gaia}. 
     This paper makes use of the following ALMA data: ADS/JAO.ALMA\#2018.1.01113.S. ALMA is a partnership of ESO (representing its member states), NSF (USA) and NINS (Japan), together with NRC (Canada), MOST and ASIAA (Taiwan), and KASI (Republic of Korea), in cooperation with the Republic of Chile. The Joint ALMA Observatory is operated by ESO, AUI/NRAO and NAOJ.
     
We acknowledge with thanks the variable star observations from the AAVSO International Database contributed by observers worldwide and used in this research.
\end{acknowledgements}

\bibliographystyle{aa} 
\bibliography{references} 

\FloatBarrier
\begin{appendix} 

\section{Radial velocities}
Tables \ref{tab:RV_table} lists the radial-velocities from the HERMES spectrograph used in Sect. \ref{sec:RV-analysis}.
\begin{table}[h]
\caption{Radial velocities of V~Hya measured with HERMES (barycentric correction applied).}
    \label{tab:RV_table}
\centering
    
    \begin{tabular}{|r|rrr|}
    
    \hline
        N & JD & RV & $\sigma_{RV}$\\
         & [days] & [$\rm km\,s^{-1}$]&  [$\rm km\,s^{-1}$]\\ 
        \hline

        1 &2455204.69 &2.31 &0.02 \\
2&2455222.67 &2.20 &0.02  \\
3&2455259.57& 0.73 &1.63 \\
4&2455568.68& -3.25 &0.08  \\
5&2455611.61& -1.76 &0.08 \\
6&2455656.50& -0.49& 0.08  \\
7&2455657.47& -0.55& 0.08  \\
8&2455660.47& -0.81& 0.08 \\
9&2455661.44& -0.76& 0.08 \\
10&2455932.72& -14.77& 0.08 \\
11&2455935.70& -14.65 &0.08 \\
12&2455940.70& -14.40 &0.08  \\
13&2455940.71& -14.41 &0.08 \\
14&2455948.67& -14.05 &0.08 \\
15&2455954.63& -13.89 &0.08 \\
16&2455966.63 &-13.31 &0.08 \\
17&2455968.70& -13.30& 0.08 \\
18&2455989.56& -12.67& 0.08 \\
19&2455989.57& -12.68& 0.08 \\
20&2456006.54& -11.93& 0.08 \\
21&2456027.48& -11.44& 0.08 \\
22&2456030.46 &-11.35& 0.08 \\
23&2456043.48& -11.00&  0.08 \\
24&2456053.44& -10.65&  0.08 \\
25&2456270.79& -3.90&  0.08 \\
26&2456298.70& -4.01&  0.08 \\
27&2456312.70& -4.27&  0.13 \\
28&2456390.50& -18.69&  0.03 \\
29&2456400.46& -19.03&  0.03 \\
30&2456420.41& -18.85&  0.03 \\
31&2456620.79& -10.12&  0.03 \\
32&2456676.64& -8.03&  0.71 \\
33&2456707.64& -7.83&  0.86 \\
34&2456765.45& -6.99&  0.01 \\
35&2456779.44& -6.98&  0.01 \\
36&2456798.41& -7.22&  0.01 \\
37&2456804.39& -7.27&  0.01 \\
38&2456996.78& -18.56&  0.01 \\
39&2457027.71 &-17.53&  0.09 \\
40&2457053.63 &-16.57&  0.09 \\
41&2457076.60 &-15.71&  0.09 \\
42&2457082.61 &-15.51&  0.09 \\
43&2457106.52& -14.62&  0.09 \\
44&2457111.50& -14.51&  0.09 \\
45&2457129.46& -13.62&  0.09 \\
46&2457131.40& -13.53&  0.09 \\
47&2457133.45& -13.38&  0.09 \\
48&2457153.39& -12.03&  0.14 \\
49&2457482.49& -20.36&  0.02 \\
50&2457482.50& -20.37&  0.02 \\
\hline
    \end{tabular}
\end{table}

\begin{table}[h]
\ContinuedFloat
\caption{continued}

\centering
    
    \begin{tabular}{|r|rrr|}
    
    \hline
        N & JD & RV & $\sigma_{RV}$\\
         & [days] & [$\rm km\,s^{-1}$]&  [$\rm km\,s^{-1}$]\\ 
         \hline
         51&2457491.48& -20.22&  0.02 \\
         52&2457508.47& -19.92&  0.02\\
         53&2457531.37& -19.30&  0.02 \\
         54&2457756.50& -11.66&  0.02 \\
         55&2457763.70& -11.50&  0.02 \\
         56&2457783.63& -10.93&  0.02\\
         57&2457803.59& -10.53&  0.02\\
         58&2457813.62& -10.30&  0.02 \\
         59&2457852.47& -9.774&  0.02 \\
         60&2457875.46& -10.29&  0.02 \\
         61&2457892.40& -10.65&  0.92 \\
         62&2457912.39& -11.47&  0.74\\
         63&2458122.70& -15.92&  0.04 \\
         64&2458133.73& -15.48&  0.04\\
         65&2458164.65& -14.24&  0.04 \\
         66&2458183.60& -13.62&  0.04\\
         67&2458194.57& -13.31&  0.04 \\
         68&2458202.51& -13.12&  0.04\\
        69&2458221.52& -12.48&  0.04 \\
        70&2458261.38& -11.77&  0.77\\
        71&2458488.71& -13.87&  0.90\\
        72&2458499.73& -14.28&  1.00\\
        73&2458512.65& -15.39&  1.02\\
        74&2458519.69 &-15.79&  0.96\\
        75&2458539.60& -15.83&  0.12 \\
        76&2458564.58& -15.65&  0.12 \\
        77&2458596.51& -14.90&  0.12 \\
        78&2458613.41& -14.28&  0.12 \\
        79&2458616.43& -14.19&  0.12\\
        80&2458622.42& -13.97&  0.12\\
        81&2458852.50& -6.45&  0.12\\
        82&2458857.50& -6.40&  0.12\\
        83&2458860.50& -6.35&  0.12 \\
        84&2458900.60& -5.73&  0.12\\
        85&2456341.55& -14.25&  0.53 \\
        86&2459230.68& -8.27&  0.06 \\
        87&2459237.67& -8.02&  0.06 \\
        88&2459240.66& -7.84&  0.06\\
        89&2459247.60& -7.47&  0.05 \\
        90&2459256.72& -6.92&  0.05 \\
        91&2459277.58& -5.89&  0.04\\
        92&2459646.52& -7.96&  0.08 \\
        93&2459649.56& -7.27&  0.03 \\
        94&2459678.50& -7.06&  0.02 \\
        95&2459679.48& -7.01&  0.02 \\
        96&2459649.56& -7.27&  0.03 \\
        97&2459949.70& 0.23&  0.03\\
        98&2459983.62& -0.28&  0.02 \\
        99&2460004.54& -0.65&  0.01 \\
        100&2460011.62& -0.75&  0.01 \\  
\hline
    \end{tabular}
\end{table}
\clearpage
\section{Light-curve periodogram}
\begin{figure}[h]
    \centering
    \includegraphics[width = 0.5\textwidth]{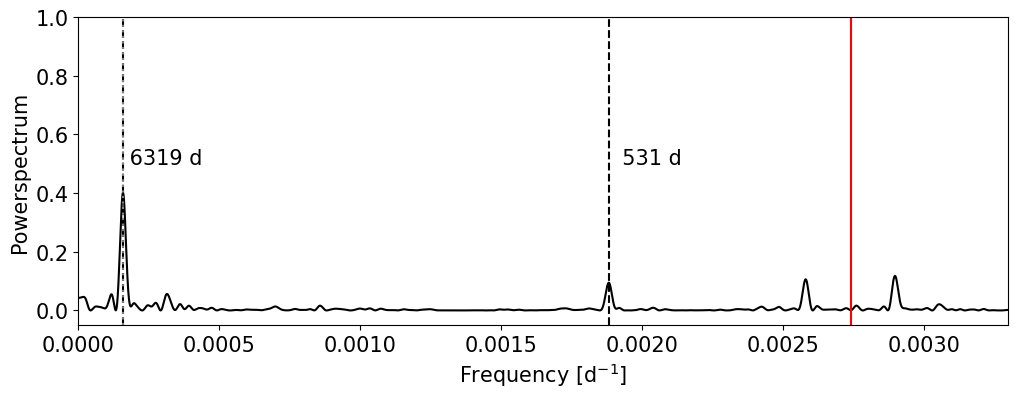}
    \caption{Lomb-Scargle periodogram of the AAVSO light curve}
    \label{fig:Lombscragle}
\end{figure}

Figure \ref{fig:Lombscragle} displays the Lomb-Scargle periodogram of the AAVSO light curve and its four prominent peaks \citep{LombScargle1982ApJ...263..835S}. The two left-most peaks correspond to the observed 6319~d and 531~d periods with a full width at half maximum (FWHM) of 940~days and 6~days, respectively. The two right-most peaks are aliases located at beating frequencies between the 530~day signal, $f_{530}$, and the one-year signal ($f_{365}$, red continuous line):  $f_{alias} = f_{365}\pm f_{530}$.
\section{\textsc{orvara} results}
\subsection{Convergence study}
\begin{figure}[h]
    \centering
    \includegraphics[width = 0.5\textwidth]{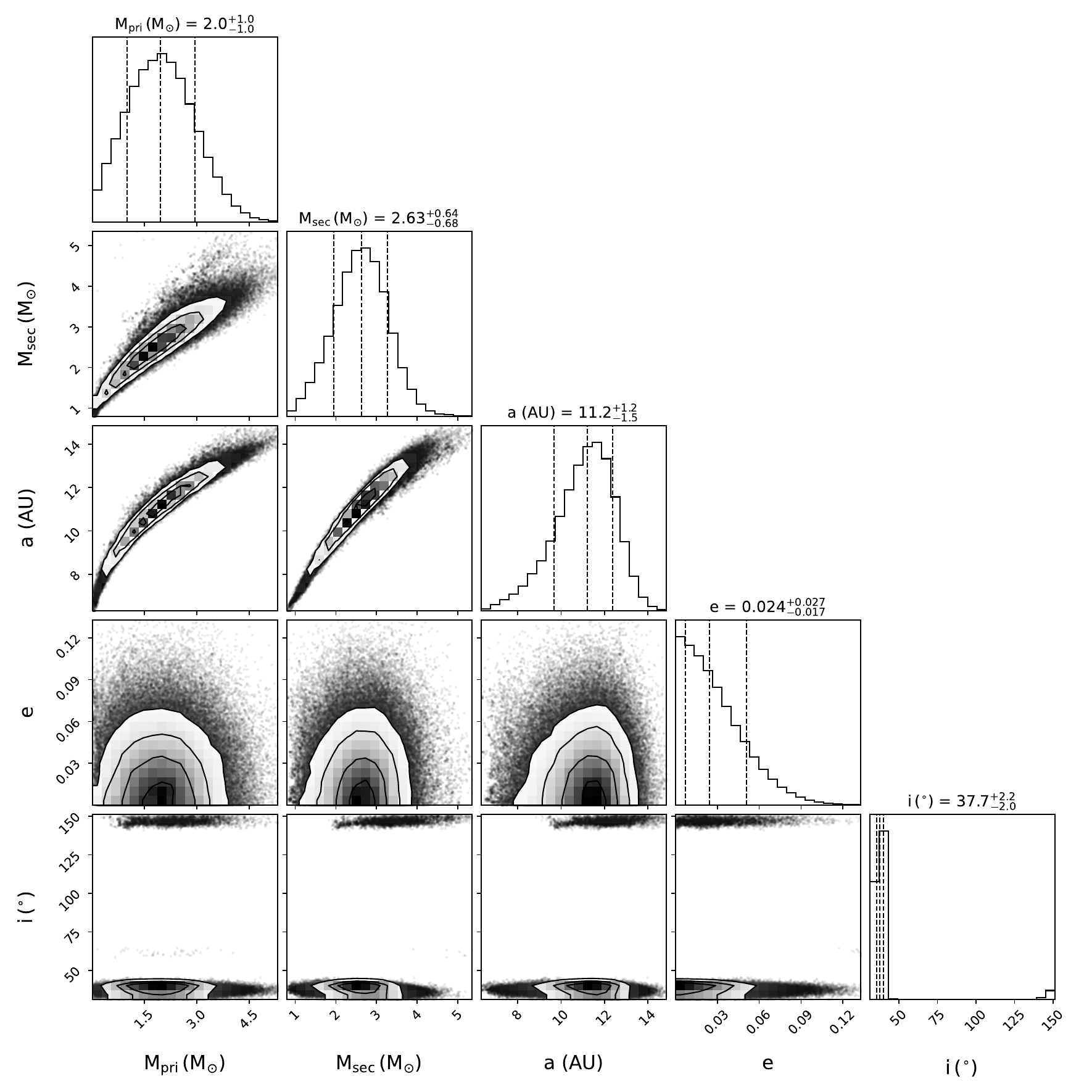}
    \caption{Corner plot of some derived orbital parameters: the mass of the two stars, semi-major axis, eccentricity, and system inclination.}
    \label{fig:orvara_corner}
\end{figure}

Figure \ref{fig:orvara_corner} displays the corner plot from \textsc{orvara}. A strong correlation is present between the mass of the two stars ($M_{pri}$, $M_{sec}$) and, consequently, their separation, $a$. 
The inclination posterior distribution is slightly bimodal, with a main peak located at $i$ = 37°$\pm$2$^\circ$. The secondary faint peak corresponds to 180°~-~37°$\pm$2$^\circ$. The convention is taken such as $i$~<~90$^\circ$ corresponds to counterclockwise rotation of the secondary around the primary while $i$~>~90$^\circ$ is the opposite situation.

The convergence of the chains was assessed by comparing the  variance between chains with the variance within each single chain through the potential scale reduction factor (PSRF) estimator \citep{Gelman_Rubin_MCMC_1992StaSc...7..457G}. The PSRF was <~1.1 for each parameter. 


\subsection{Sensitivity study on the $\mu^H_{\alpha*}$ point}
\begin{figure}[t]
    \centering
    \includegraphics[width = 0.5\textwidth]{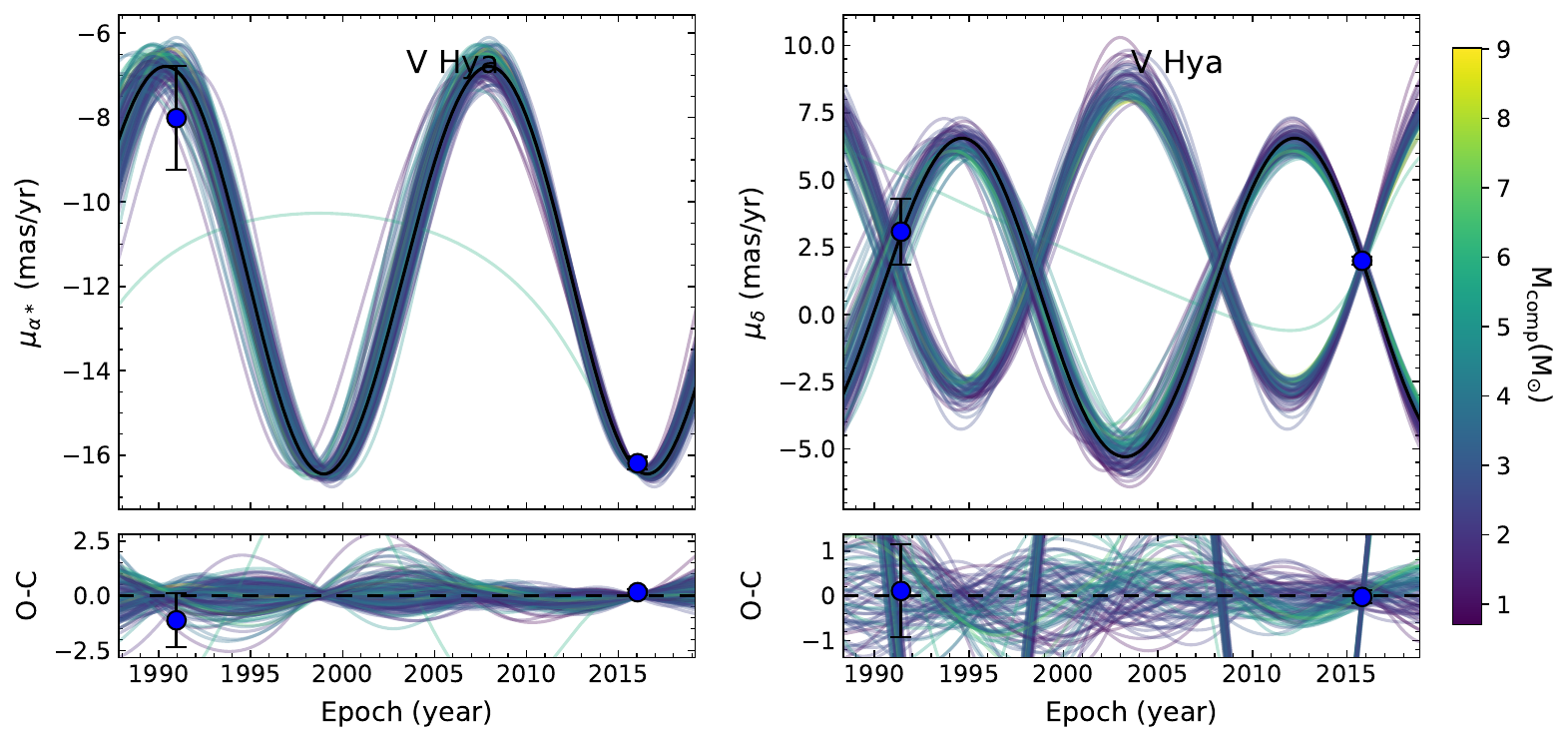}
    \includegraphics[width = 0.5\textwidth]{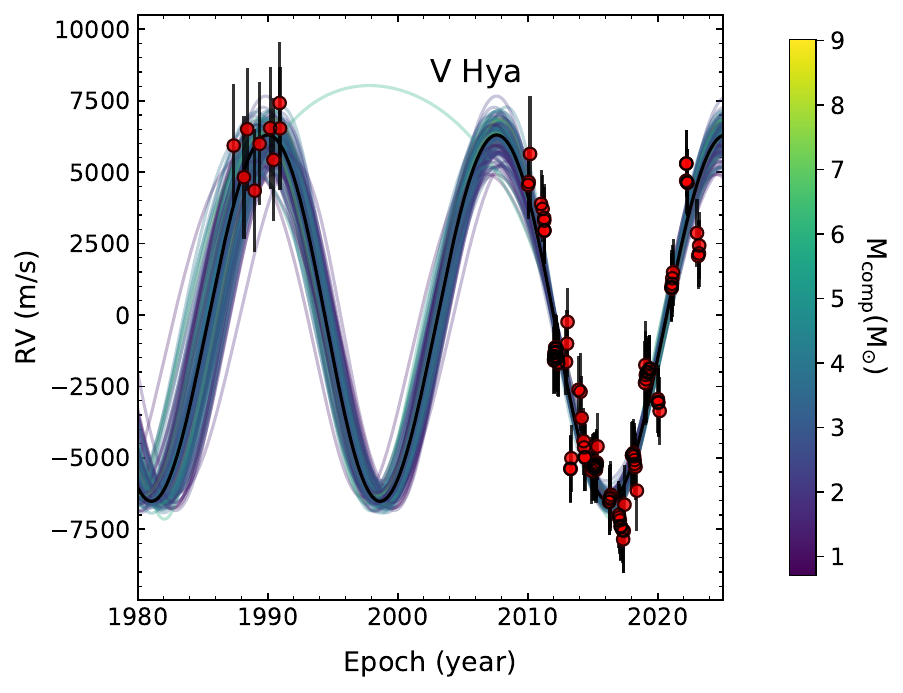}
    \includegraphics[width = 0.5\textwidth]{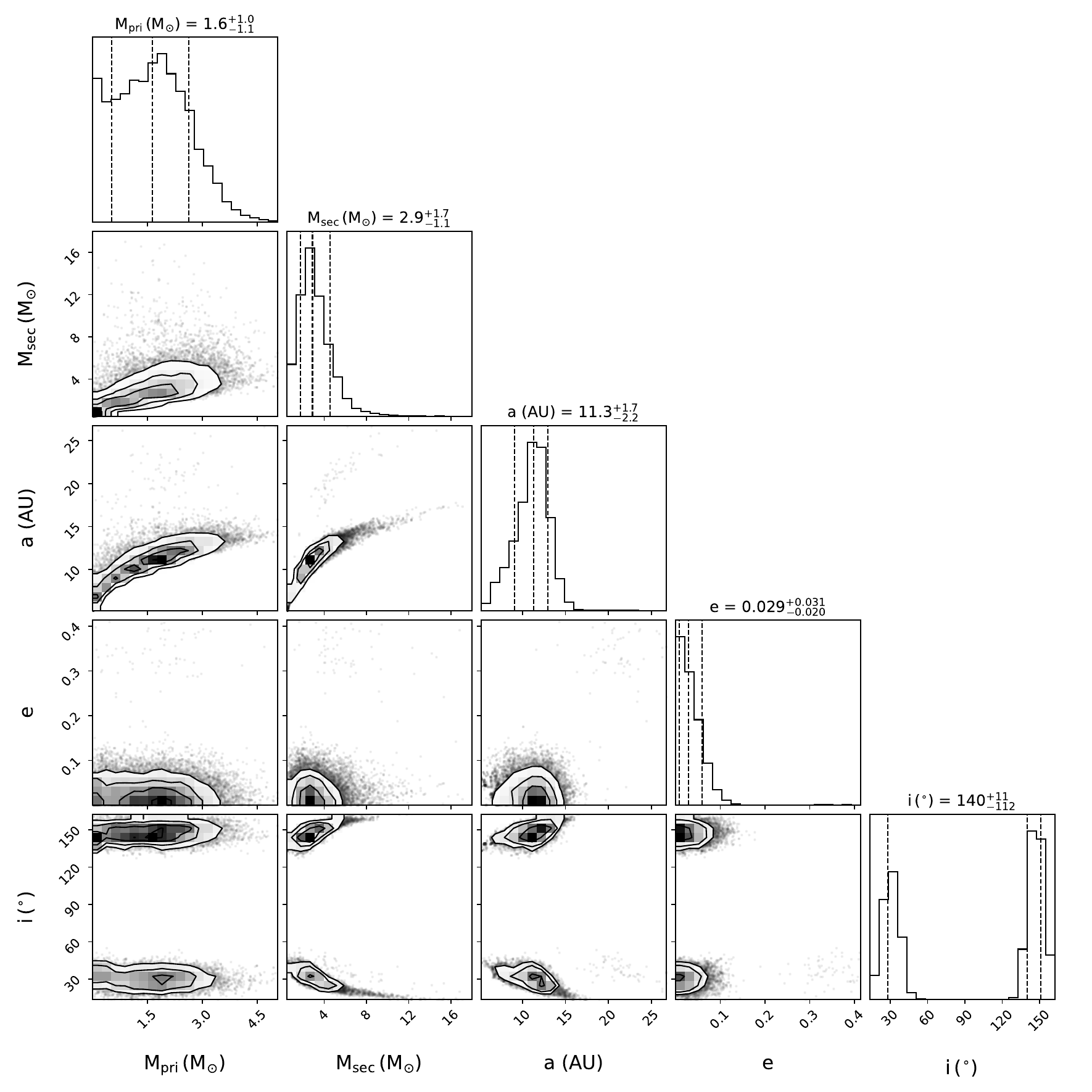}
    \caption{PM curve in right-ascension and declination (top panels), RV curve (middle panel), and corner plot (bottom panel) for $\mu^H_{\alpha*}$ offset by +4~mas/yr.}
    \label{fig:Plus2_orvara_corner}
\end{figure}

\begin{figure}[t]
    \centering
    \includegraphics[width = 0.5\textwidth]{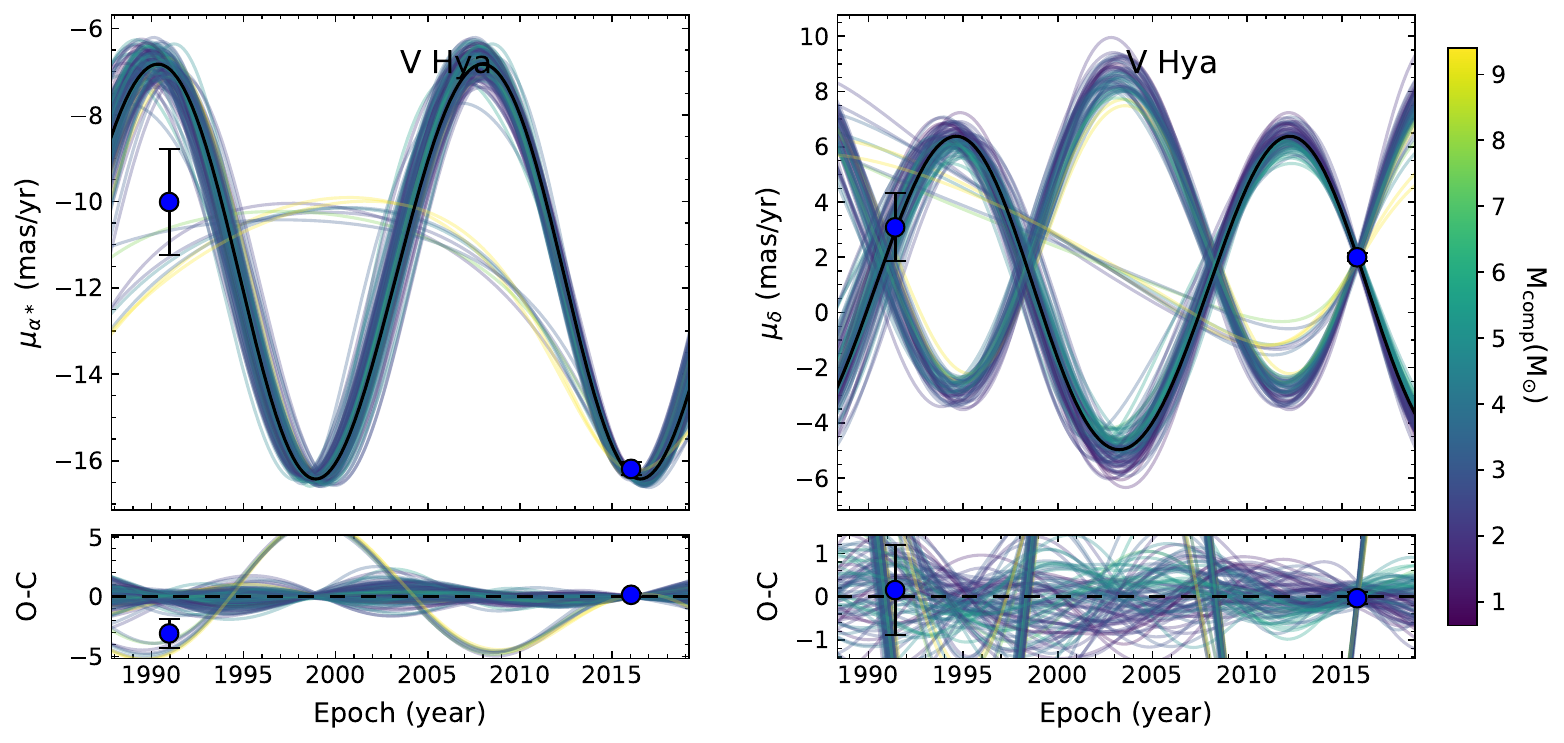}
    \includegraphics[width = 0.5\textwidth]{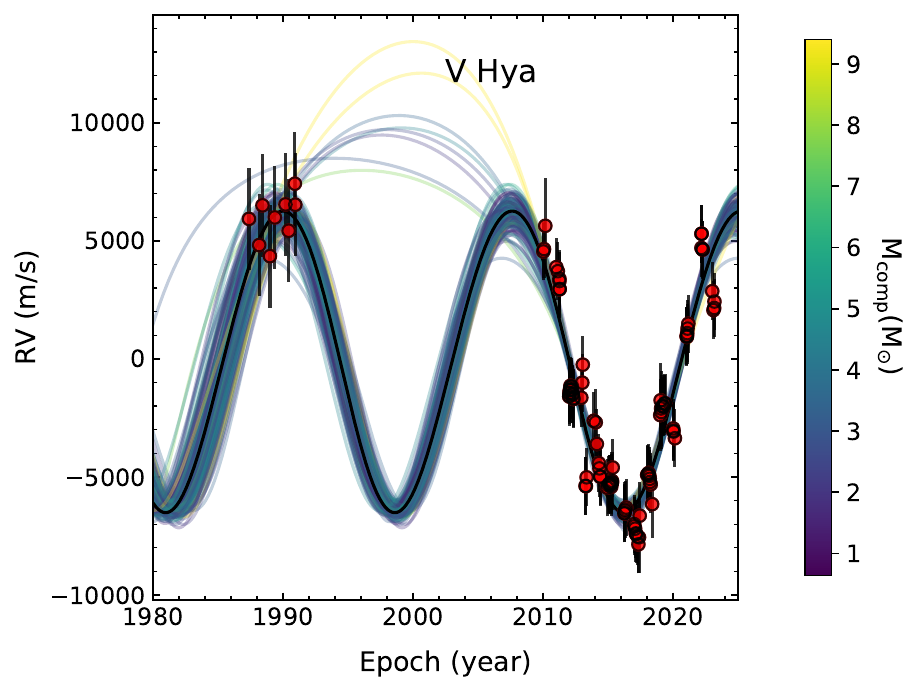}
    \includegraphics[width = 0.5\textwidth]{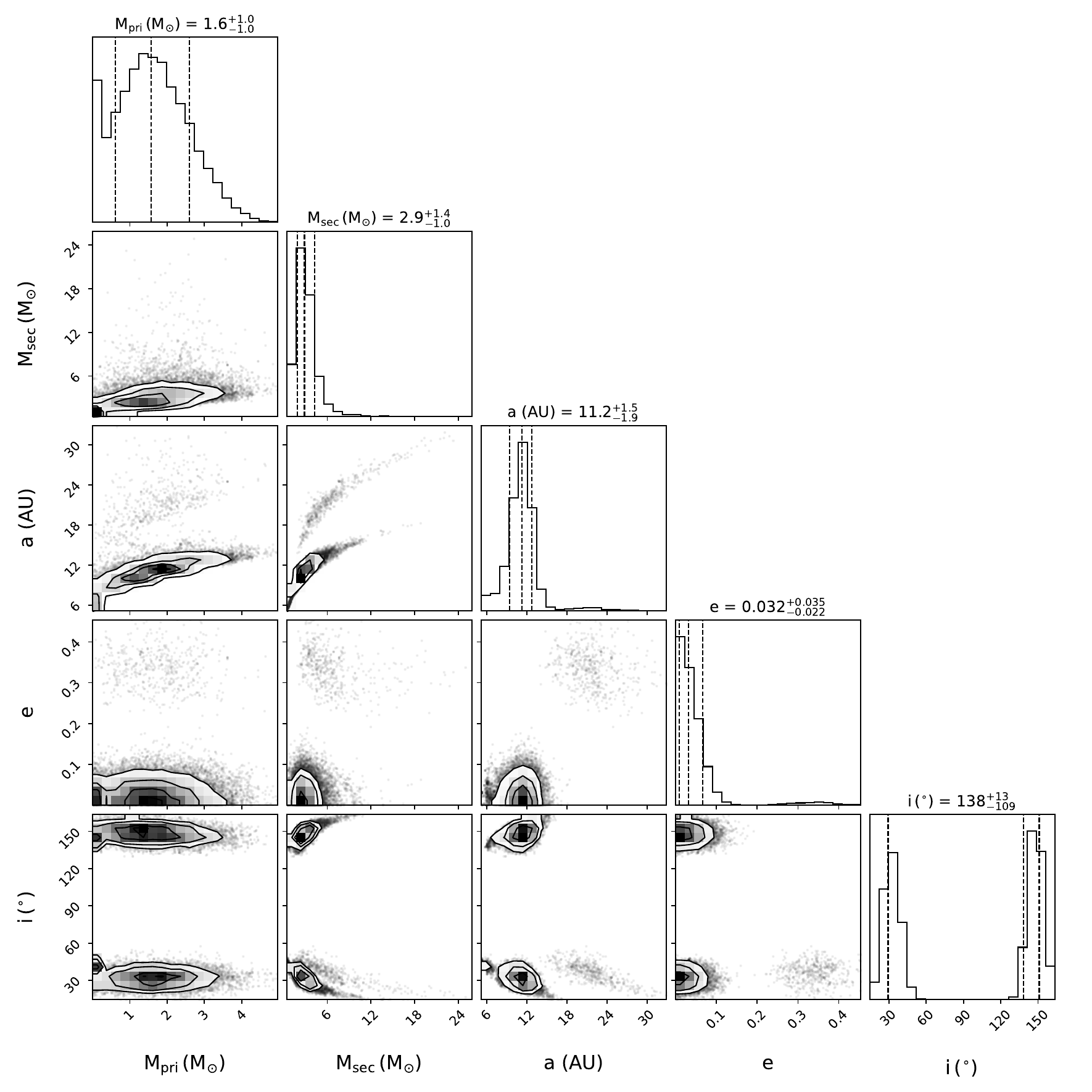}
    \caption{Same as Fig. \ref{fig:Plus2_orvara_corner} but for $\mu^H_{\alpha*}$ offset by +2~mas/yr.}
    \label{fig:Plus1_orvara_corner}
\end{figure}

\begin{figure}[t]
    \centering
    \includegraphics[width = 0.5\textwidth]{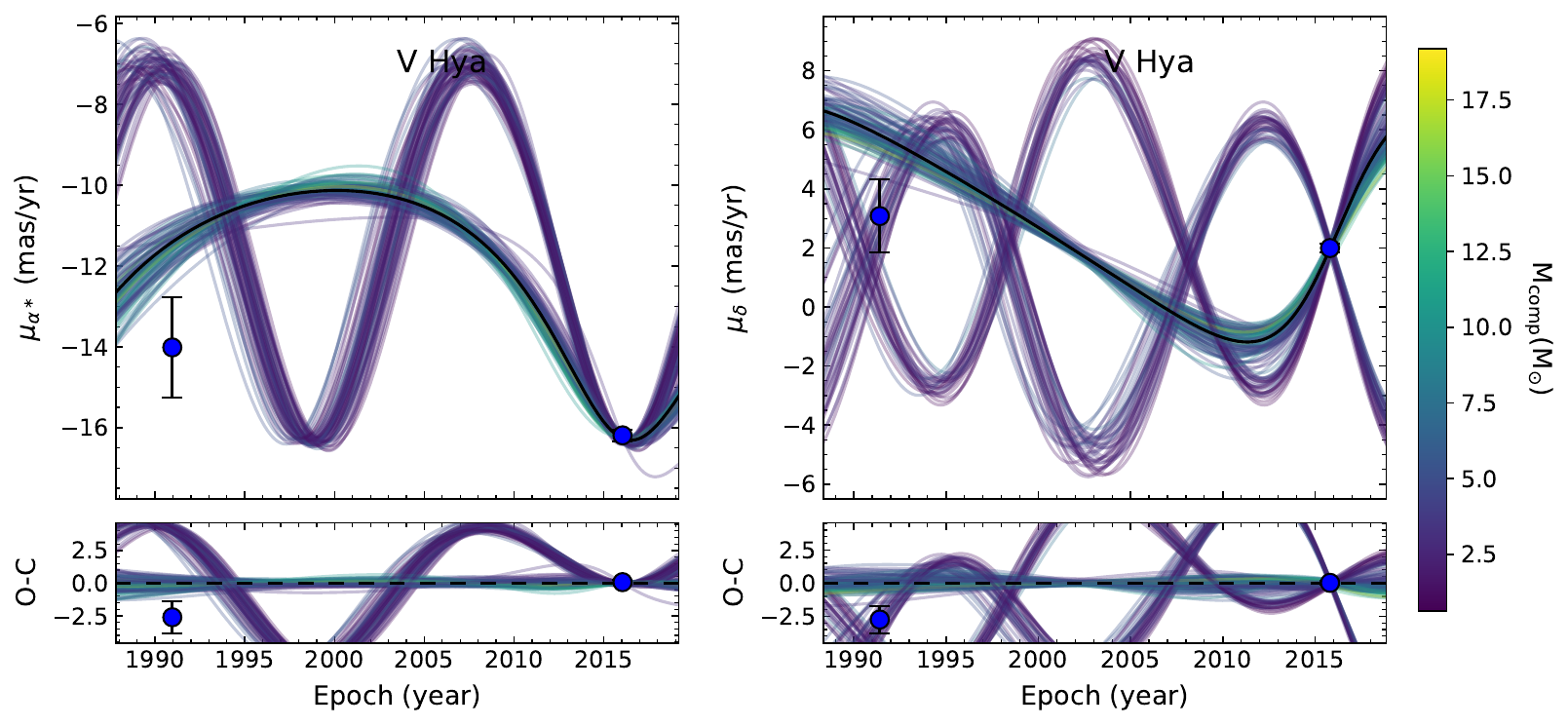}
    \includegraphics[width = 0.5\textwidth]{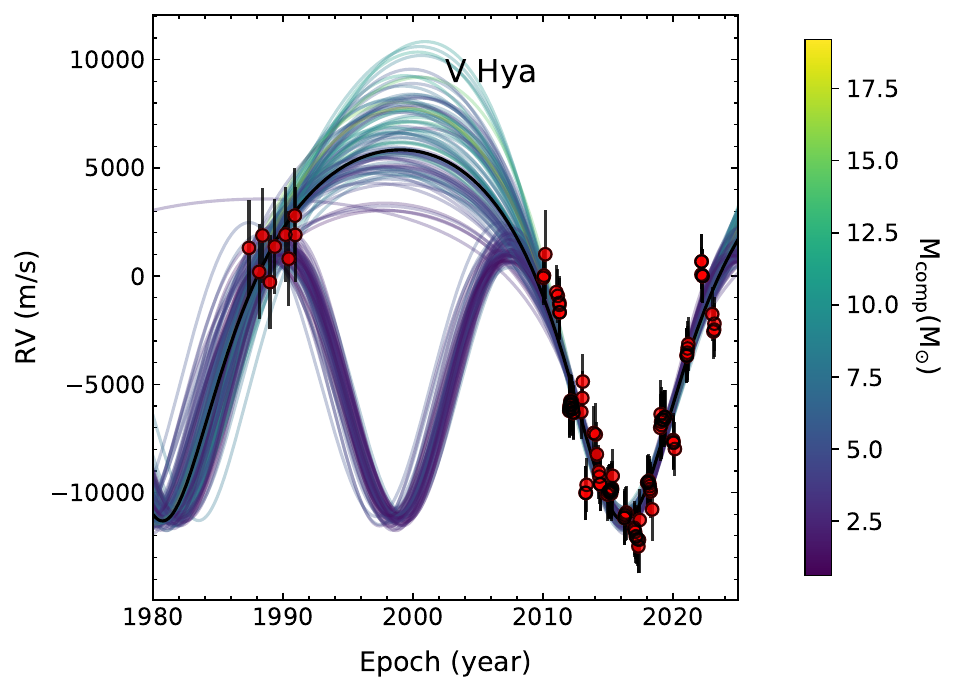}
    \includegraphics[width = 0.5\textwidth]{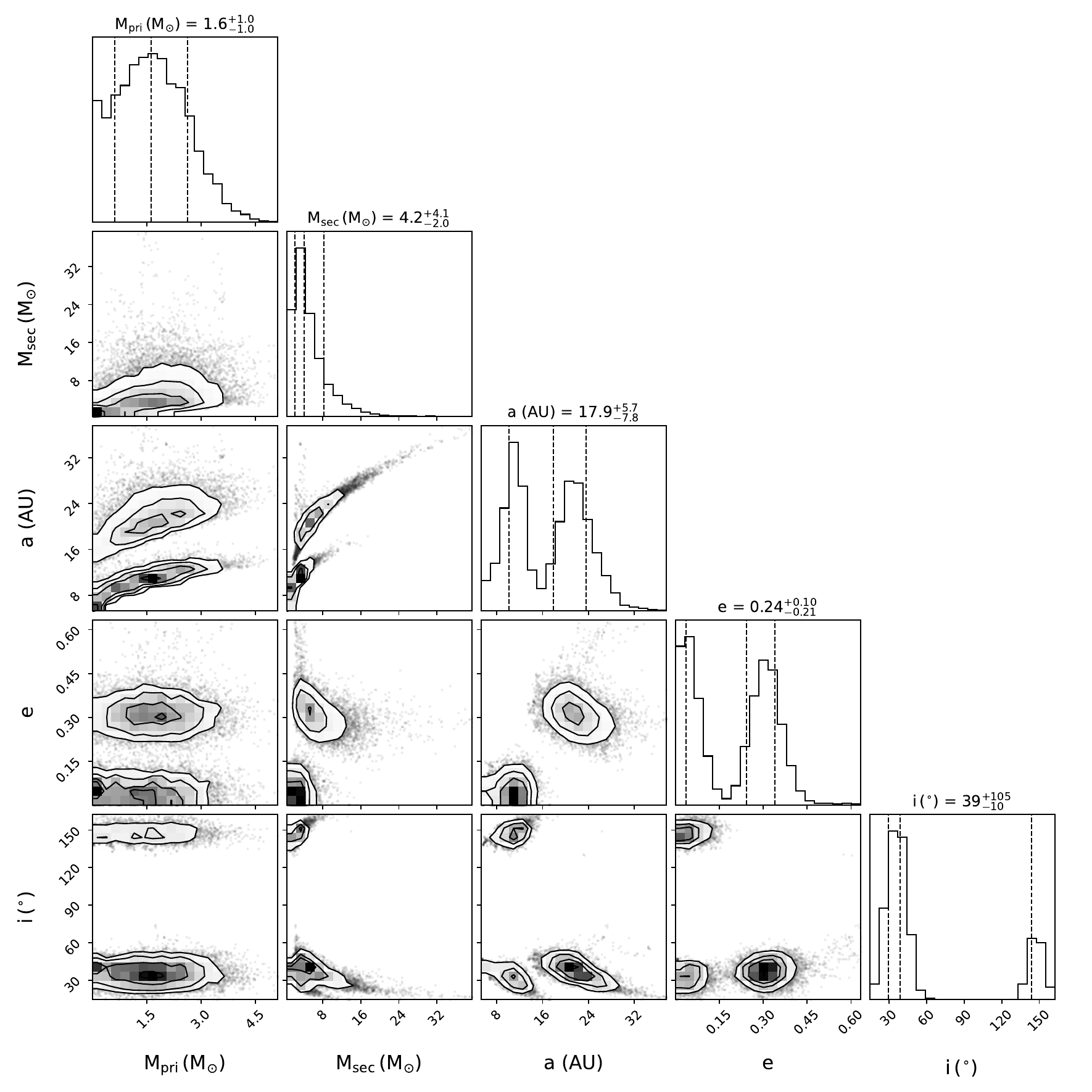}
    \caption{Same as Fig. \ref{fig:Plus2_orvara_corner} but for $\mu^H_{\alpha*}$ offset by -2~mas/yr.}
    \label{fig:Minus1_orvara_corner}
\end{figure}

\begin{figure}[t]
    \centering
    \includegraphics[width = 0.5\textwidth]{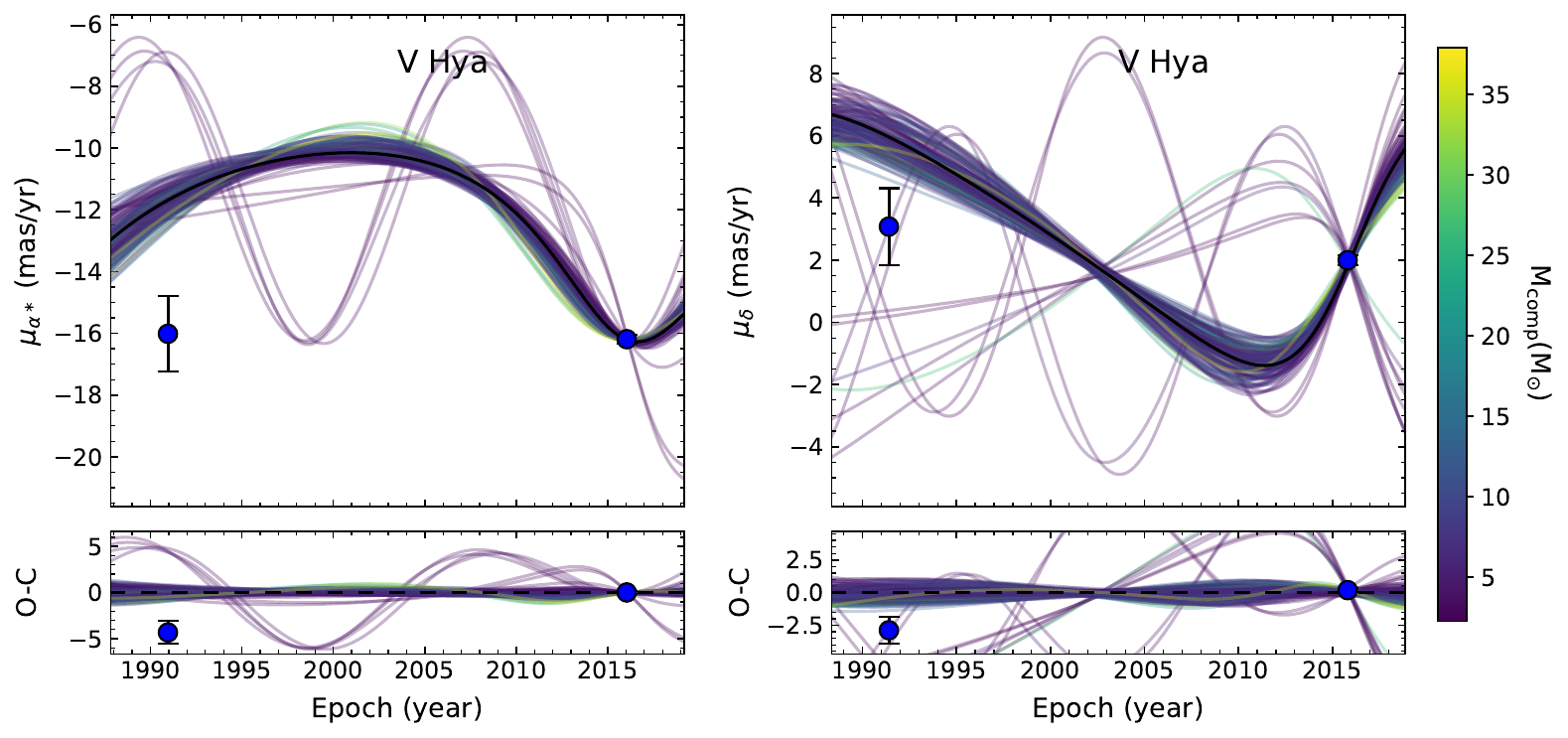}
    \includegraphics[width = 0.5\textwidth]{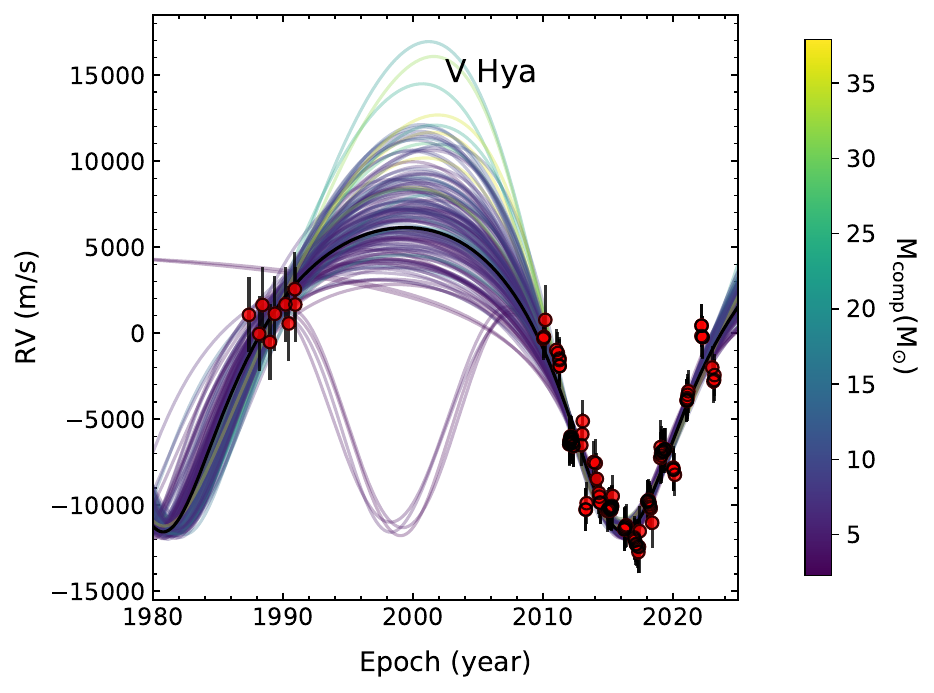}
    \includegraphics[width = 0.5\textwidth]{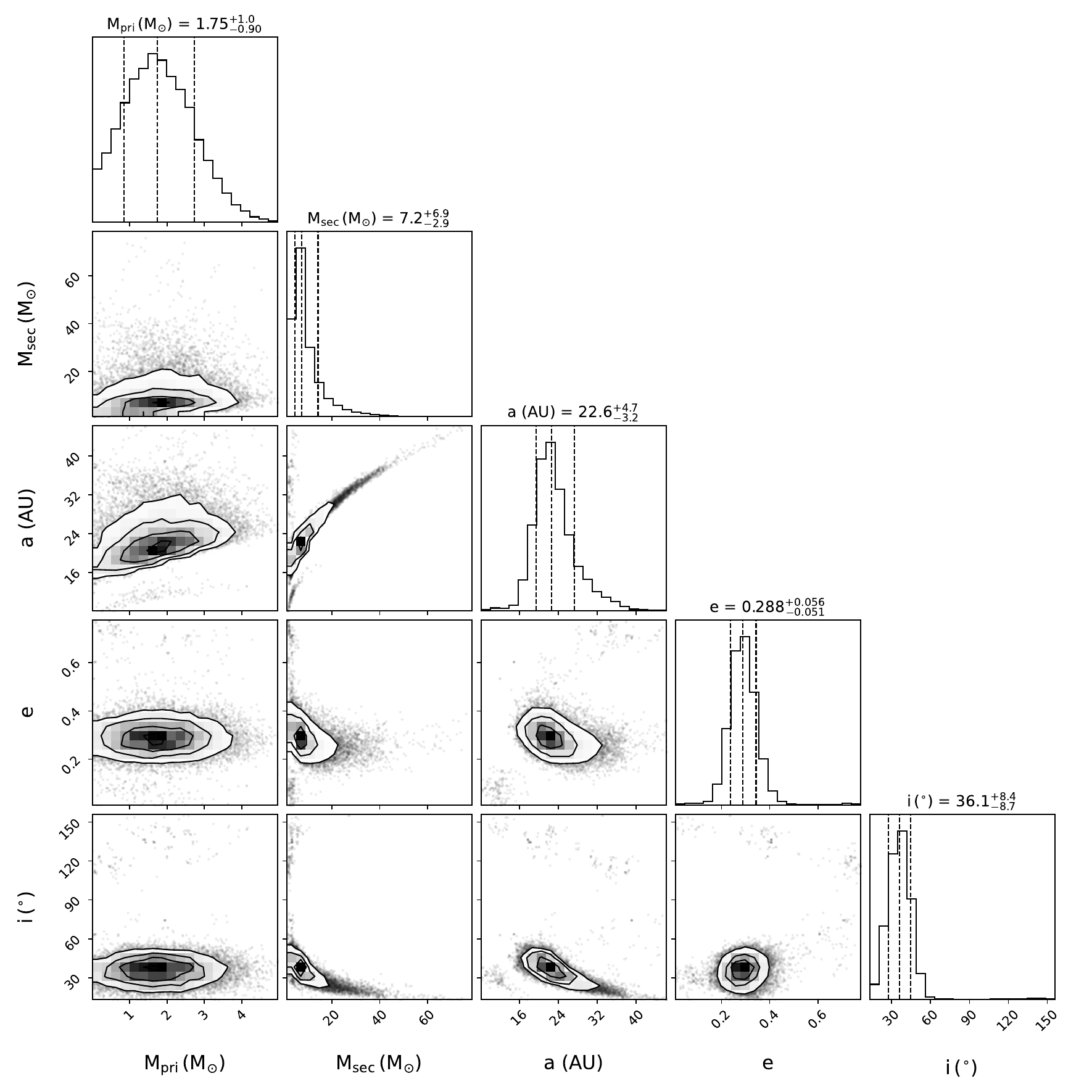}
    \caption{Same as Fig. \ref{fig:Plus2_orvara_corner} but for $\mu^H_{\alpha*}$ offset by -4~mas/yr.}
    \label{fig:Minus2_orvara_corner}
\end{figure}

Figures \ref{fig:Plus2_orvara_corner}, \ref{fig:Plus1_orvara_corner}, \ref{fig:Minus1_orvara_corner}, and \ref{fig:Minus2_orvara_corner} display the proper motion (PM) and radial-velocity (RV) curves, together with the corner plot obtained for four $\mu^H_{\alpha*}$ values, offset from the original value of the catalog of acceleration \citep{acceleration_2021ApJS..254...42B}, by +4, +2, -2, and -4~mas/yr, respectively.

\section{Representative line profiles}
\begin{figure*}[t]
    \includegraphics[width = 0.31\textwidth]{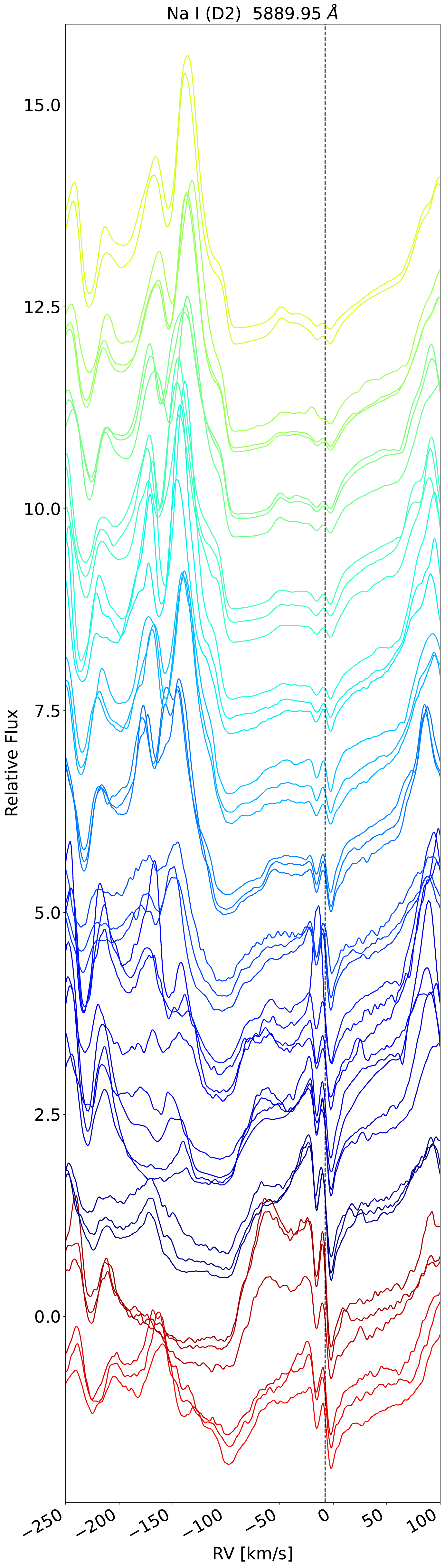}
    \includegraphics[width = 0.31\textwidth]{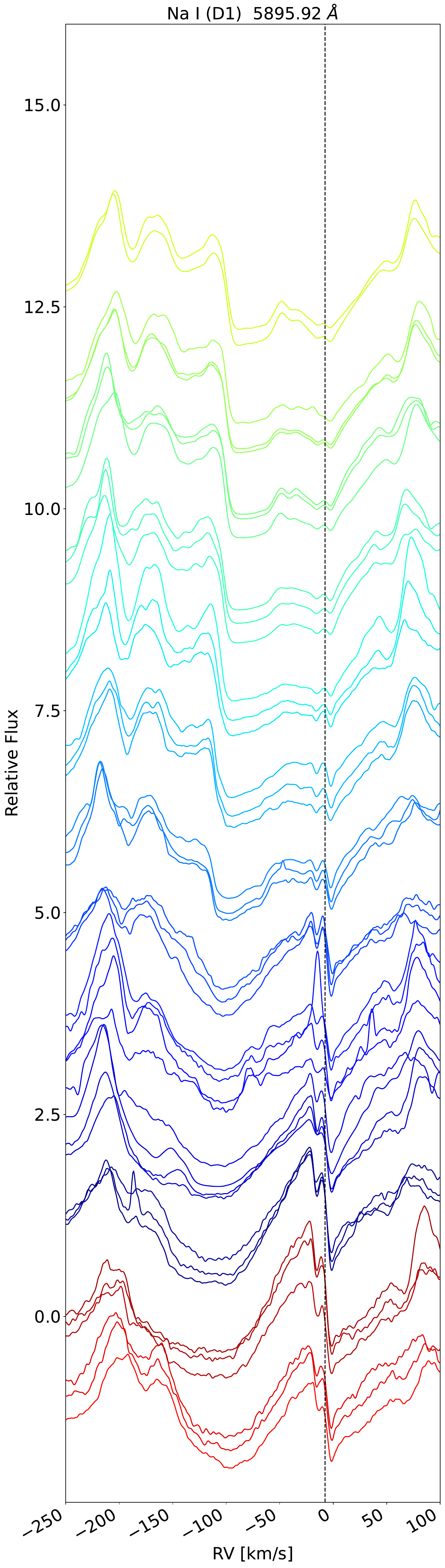}
    \includegraphics[width = 0.3772\textwidth]{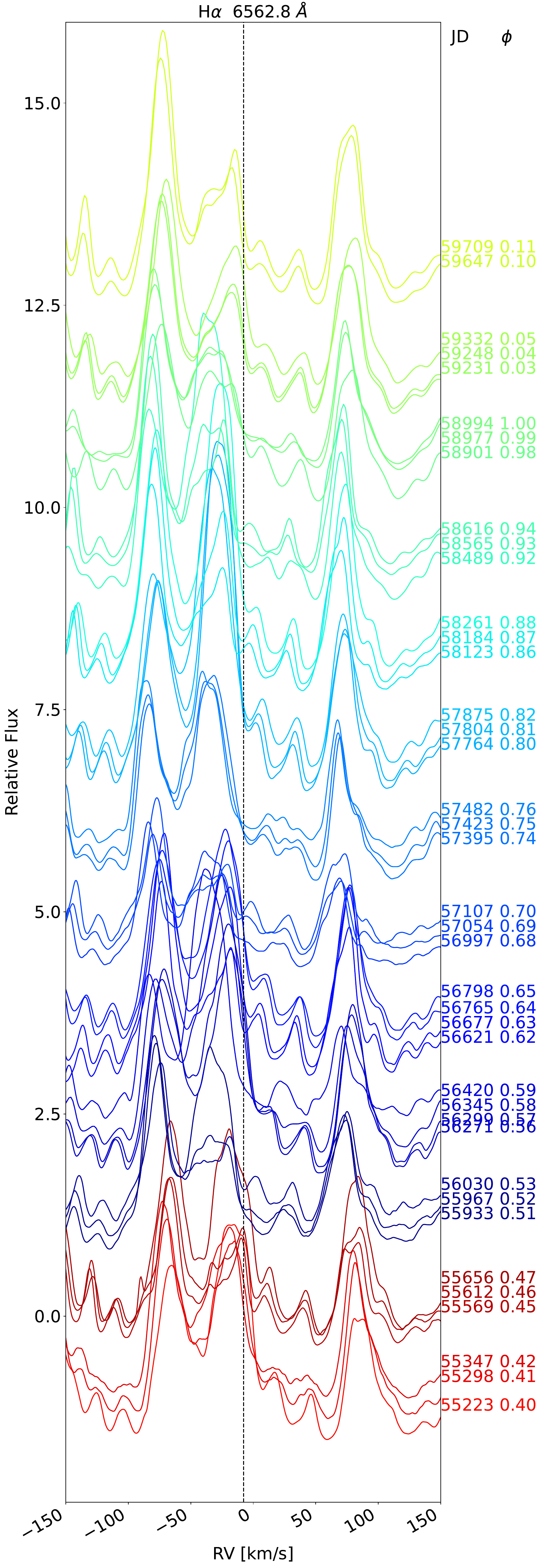}
    \caption{Sodium doublet (leftmost and middle panels) and H$\alpha$ (rightmost panel) lines as a function of the orbital phase. The vertical dashed line marks the systemic velocity of -7.94~$\rm km\,s^{-1}$. The corresponding phase and observing date (expressed in JD-2400000) are displayed on the right-most axis.}
    \label{fig:sodium_overimposed}
\end{figure*}

\begin{figure*}[t]
    \includegraphics[width = 0.31\textwidth]{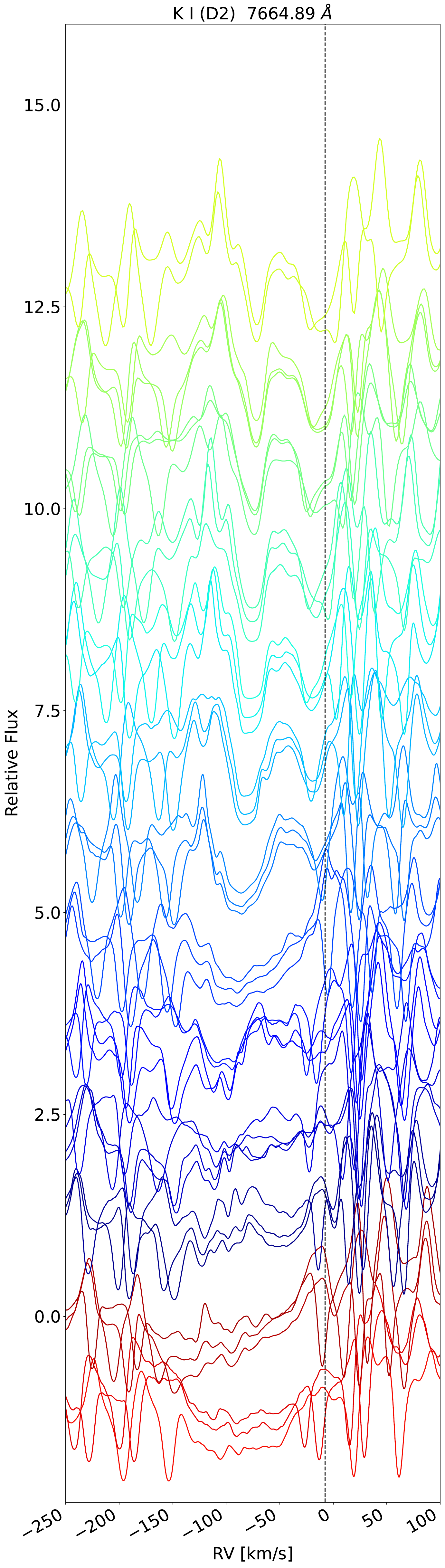}
    \includegraphics[width = 0.31\textwidth]{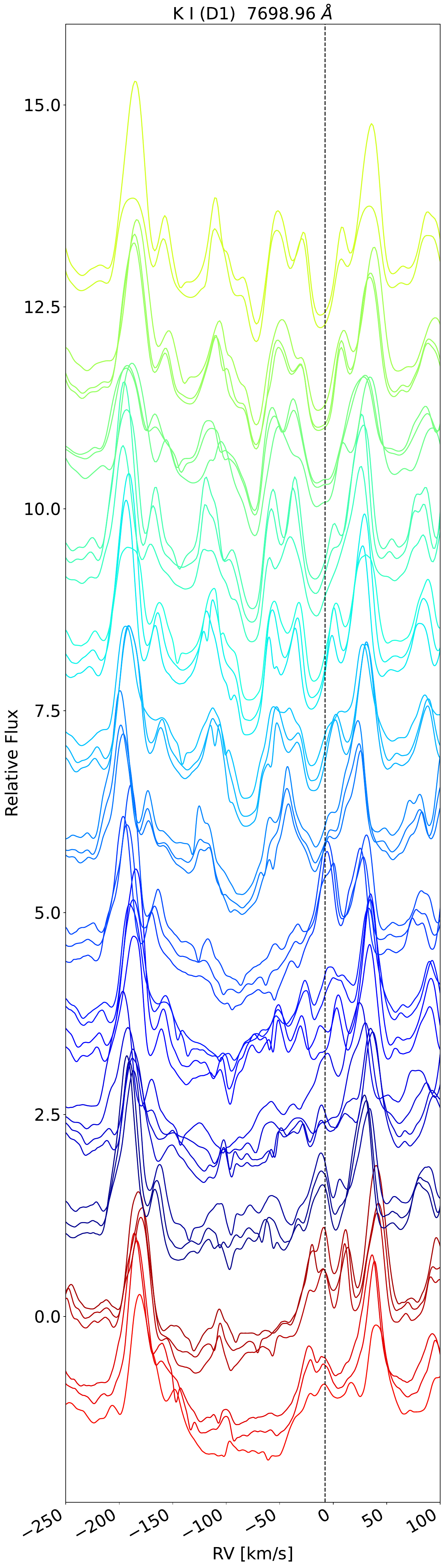}
    \includegraphics[width = 0.3772\textwidth]{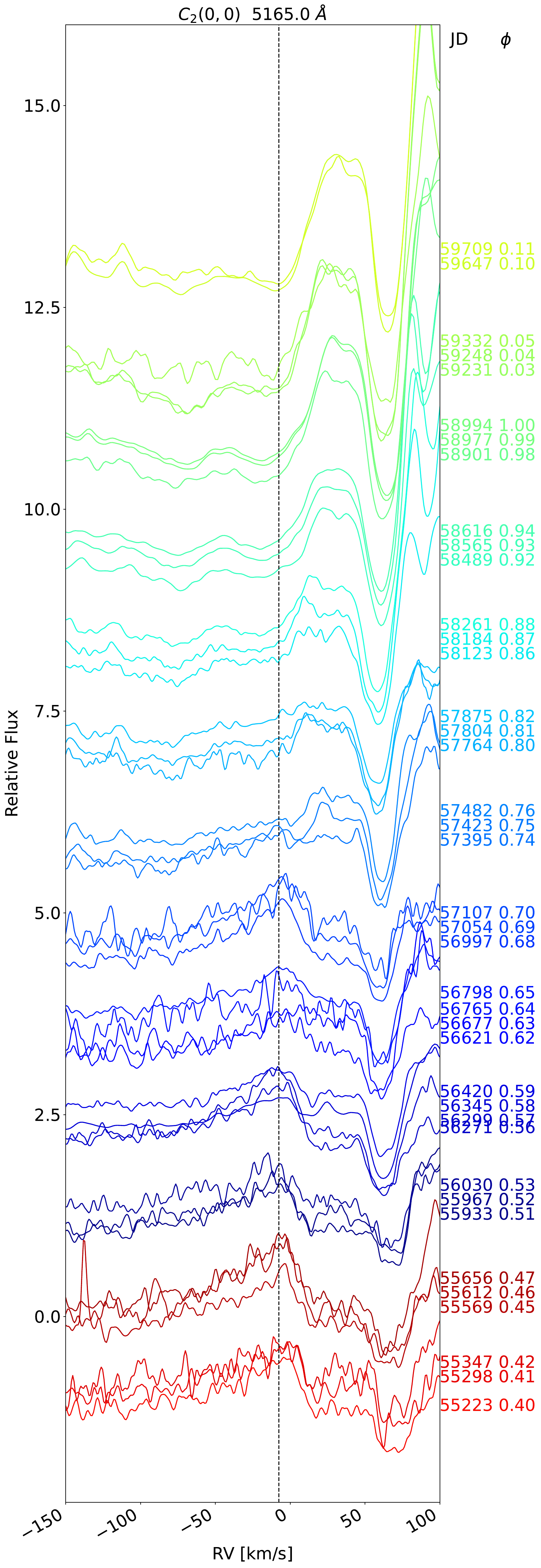}
    \caption{Same as Fig. \ref{fig:sodium_overimposed} but for potassium doublet lines (left-most and middle panels) and the $C_2$ (0,0) line (right-most panel). }
    \label{fig:potassium_overimposed}
\end{figure*}

Figures \ref{fig:sodium_overimposed} and \ref{fig:potassium_overimposed} show the temporal behaviour of spectral lines presented: H$\alpha$, \ion{Na}{I}, \ion{K}{I,} and $C_2$ (0,0).  The latter, being in the bluest part of the spectrum with a low S/N, has been smoothed by a Gaussian filter of FWHM = 2.8~$\rm km\,s^{-1}$.

\clearpage
\section{Spatio-kinematic modelling}
\label{Appendix:spatio-kinematic-modelling}
\subsection{Jet configuration}
In this section we recall the stellar jet configuration used in this study, as introduced in \cite{Dylan_vel_profile_2020A&A...641A.175B}.  
\label{appendix:Stellar_jet_configuration}
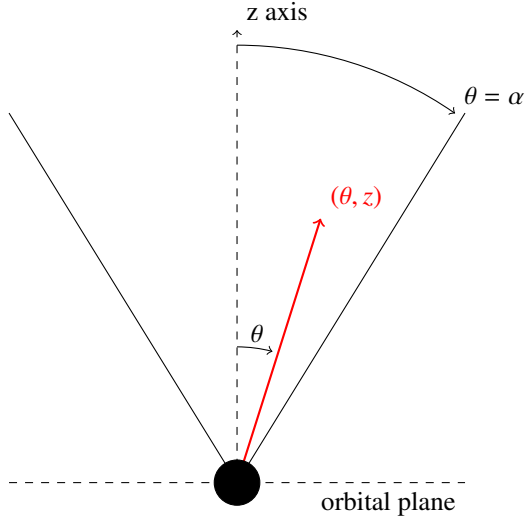
\begin{figure}[ht!]
\centering
\begin{tikzpicture}


\draw (3,0) -- (6,4.9);
\draw[dashed, ->] (3,0) -- (3,6)node[anchor=south west] {z axis};
\draw (3,0) -- (0,4.9);

\draw[ ->] (3,5.8) arc (90:55:5cm) node[anchor=south west] {$\theta = \alpha$};

\draw[red, thick,->] (3,0) -- (4.1,3.5) node[anchor=south west] {($\theta, z$) };
\draw[->] (3,1.8) arc (90:75:1.8cm) node[anchor=south east] {$\theta$};

\filldraw (3,0) circle (0.3cm);

\draw[dashed](0,0) --(6,0) node[anchor=north east] {orbital plane };

\end{tikzpicture}
\caption{Schematic geometry of the stellar jet model and associated parameters. Each point in the jet is described by a coordinate pair: ($\theta$,$z$), where $\theta$ is the angle from the jet axis and $z$ is the vertical distance from the orbital plane.}
\label{fig:tikz_stellar_jet}
\end{figure}

The velocity of the matter inside the jet is described by a latitudinal dependent power law: 

\begin{equation}
    \vec{\varv}(\theta) = [\varv_0 + (\varv_\alpha - \varv_0) f_1(\theta)  ] \vec{1_r},
\end{equation}
with 
\begin{equation}
    f_1(\theta) = \frac{e^{- |\theta/\alpha|p_\varv} - e^{-p_\varv}}{1 - e^{-p_\varv}}.
\end{equation}
Here, $\vec{1}_r$ is the radial vector, $\theta$ the polar latitudinal coordinate, $\alpha$ the half-opening angle of the jet,  $p_v$ is a power index to fit,  $v_0$ is the velocity at the centre, and $v_{\alpha}$ the velocity at the cone edge.

The density profile is also latitude-dependent and decreases with the jet height $z$ as follows:

\begin{equation}
    \rho(\theta, z) = \Bigg ( \frac{\theta}{\alpha} \Bigg )^{p_d} z^{-2},
\end{equation}
where $p_d$ is a power index to fit and the $z$ dependence is fixed to satisfy the continuity equation along the cone height.  

The stellar jet model is composed of 10 free parameters: the inclination of the system, $i,$ which we constrain by the astrometric results: the jet opening angle, $\alpha$; the velocity at the edge, $v_{\alpha}$, and at the centre, $v_0$; the power index of the velocity law, $p_v$; the power index of the density, $p_d$; $\theta_{cav}$, the angle of the cavity within the jet; $\theta_{tilt}$ is the angle misalignment with respect to the orbital axis ($z$ coordinate); $R_1$ is the radius of the emitting surface; and $c_{v}$, a general scaling factor for the optical depth. 

\subsection{Dynamic model spectrum}
\label{appendix:Dynamic_model_spectrum}
Archive spectra of carbon stars from the UVES Paranal Observatory Project (POP) database \citep{UVES_POP_2003Msngr.114...10B} have been compared to select a template spectrum for V~Hya; among them HD134453 (X Tra), HD92055 (U Hya), and HD32088. The three tested stars show a similar behavior around the sodium doublet lines, confirming the unusual nature of V~Hya spectral profile. We kept X Tra as a template model. In Fig. \ref{fig:sodium_conjunction}, the spectrum of V~Hya is superimposed to the one of X Tra for two opposite orbital phases.

\begin{figure}[t]
    \centering
    \includegraphics[width = 0.5\textwidth]{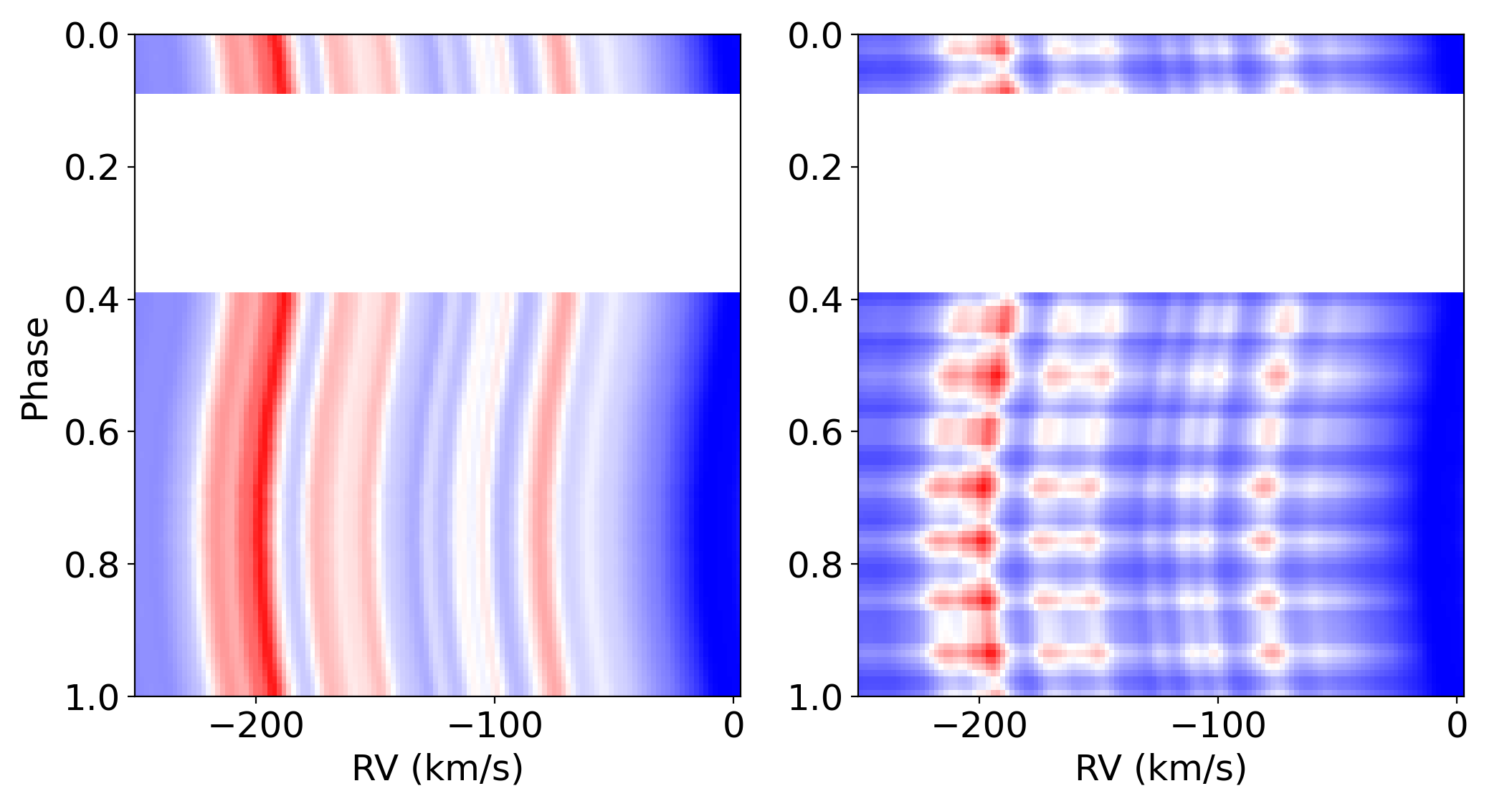}
    \caption{Dynamic model spectrum of X Tra for the sodium doublet at 5895.92 \AA\ (left). Dynamic model spectrum with an additional 530 d-modulation (right).}
    \label{fig:sodium2osbmod_template}
\end{figure}

To obtain the dynamic synthetic spectrum at 5895.92 \AA, the observed X Tra spectrum was Doppler-shifted according to V~Hya orbital motion (see Fig.  \ref{fig:sodium2osbmod_template}, left). A 530-d intensity modulation is added to mimic the effect of the intrinsic scatter due to the Mira-like pulsation. The final dynamic spectrum used as template for the fitting routine is displayed in the right panel of Fig. \ref{fig:sodium2osbmod_template}.

\subsection{Radiative transfer through the jet}
\label{Appendix:Radiative_transfer}
\begin{figure}[t]
    \centering
    \includegraphics[width = 0.3\textwidth]{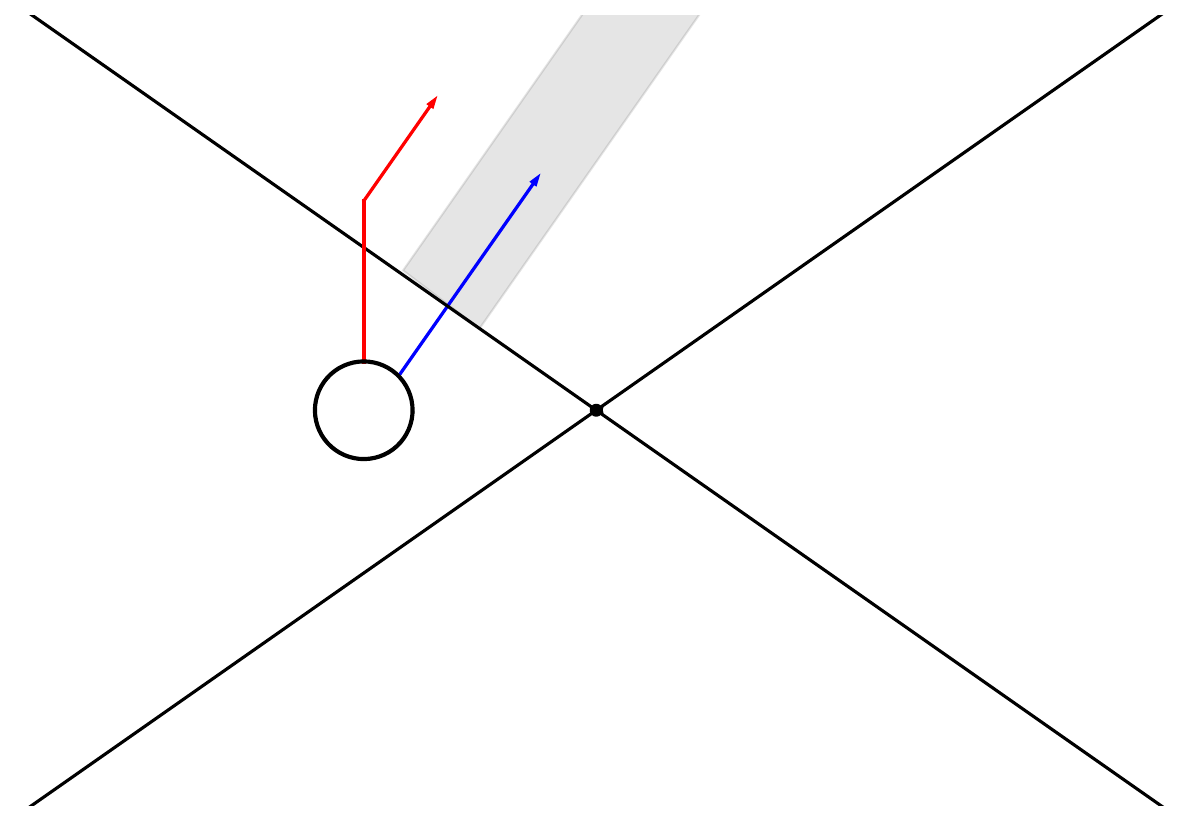}
    \caption{Illustration of the stellar light propagating through the cone: the light passing through the gray shaded area is attenuated by the jet cone (see the blue line tracing the ray). The light emitted in another direction but passing through the rest of the cone can be re-scattered or re-emitted into the line of sight (red arrow).}
    \label{fig:cone_scattering}
\end{figure}

At phase $\phi$, the total intensity at a specific wavelength, $I_\lambda$. is obtained by summing the contribution of two terms:
\begin{equation}
I_\lambda(s) = I_{\lambda}^0 e^{-\tau(s)} + \beta S_\lambda^{scat}.
\end{equation}
The first term describes the attenuation of the photospheric light due to the absorption of photons along the line of sight by the neutral atoms present the jet.
The region attenuating the incoming stellar light is delimited by a cylinder normal to the effective stellar surface. The situation is illustrated in Fig. \ref{fig:cone_scattering}. 
$I_{\lambda}^0$ represents the unattenuated initial intensity and $\tau(s)$ is the optical depth at a specific position $s$ along the line of sight in the jet. This term is proportional to the density at that point $\tau(s) \propto c_v \rho(s) \Delta s$, with $\Delta s$ the distance between two consecutive points on the line of sight and $c_v$ a scaling parameter to fit. We refer to the reference paper \citep{Dylan2} for additional details on the numerical implementation of the radiative transfer equation and the spatio-kinematic modelling.

The second term is a scattering source. Photons emitted from the star in a direction different from the line of sight may also intersect the cone and interact with gaseous media. The interaction is modelled by a change in the direction of propagation. A fraction of the photons is bounced back in the line of sight and contributes to the scattering source of the gaseous cone - $S_\lambda^{scat}$ (illustrated with the red arrow in Fig. \ref{fig:cone_scattering}).
Knowing the velocity field inside the cone, the source term is modelled by performing a first-order ray-tracing of a grid of scattering points evenly spaced in the cone. The final source function is obtained by summing the contribution of each scattered photon depending on their final projected velocity and weighted by their probability to be re-scattered in the direction of the observer. This probability only depends on the deflection angle between the photon initial direction and the line of sight.  

\begin{figure}[t]
    \centering
    \includegraphics[width = 0.5\textwidth]{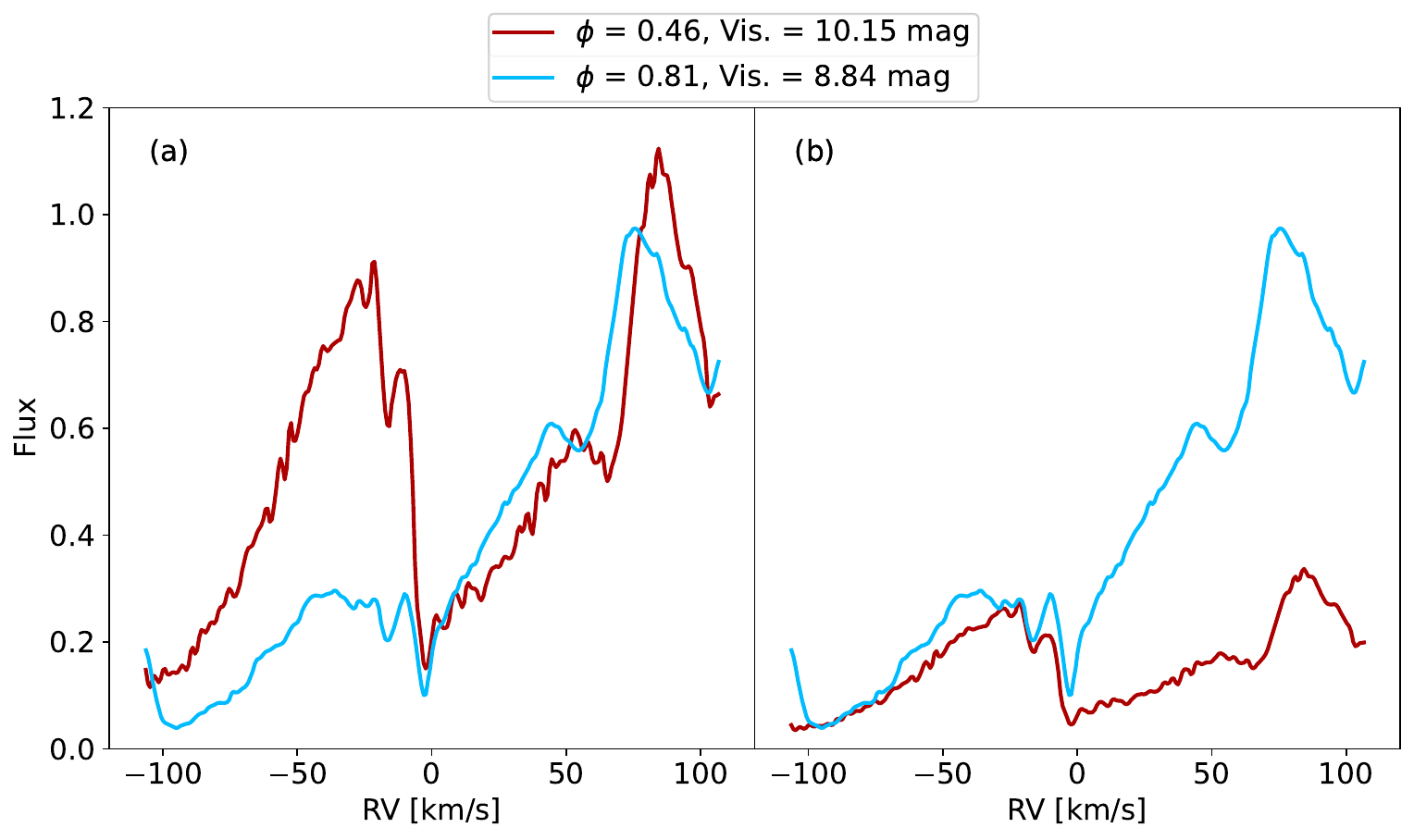}
    \caption{Normalisation effect on the \ion{Na}{I} D1 line profile for spectra taken at two different phases. (a) The normalisation is obtained by dividing each spectrum by its mean over a 100~\AA~ window. This procedure is incorrect (see text). (b) The normalisation between the two spectra takes into account the relative flux ratio of 0.33 between the two phases corresponding to a drop of 1.31~mag. The RV zero point is set to $\lambda$ = 5895.92~\AA.}
    \label{fig:NaD1_normalized}
\end{figure}

The orbital modulation,  $\beta$, accounts for the normalisation correction due to the different origins of two emitting sources unresolved by the instrument. The scattering source being emitted from an extended region offset from the star, it does not follow V~Hya orbital motion and is therefore not affected by the periodic obscuration event. 
The spectra displayed in the analysis are always normalised by averaging over a given spectral window dominated by photospheric pseudo-emission. The obtained window-normalised spectra can therefore be compared, independently of their flux modulation. The side effect of such a normalisation is that any extrinsic emission, unaffected by V~Hya obscuration, is artificially modulated by a factor inversely proportional to the stellar flux modulation: appearing brighter when the star fades and nearly vanishing at the opposite phase. In practice, such extrinsic emission will vary in strength with the orbital phase, $\phi$, according to a correction factor, expected to be related to the light-curve modulation, equal to: 
\begin{equation}
    \beta_\phi = 10^{0.4\Delta m_\phi},
\end{equation}
where $\Delta m_\phi$ is the visual light-curve modulation expressed in magnitude. Figure \ref{fig:NaD1_normalized} illustrates the effect of both types of normalisation on the relative strength of the \ion{Na}{I} D1 line. The two spectra displayed are taken near light maximum regarding the pulsation cycle but at two different orbital periods. The magnitude difference in the visual between the two epochs is equal to 1.31, corresponding to a flux attenuation factor of about 3.31.

\end{appendix}

\end{document}